\documentclass[twocolumn,showpacs,preprintnumbers,amsmath,amsfonts,amssymb,aps,pra,groupedaddress,floatfix]{revtex4-1}
 \usepackage{geometry}
\usepackage{todonotes}
\usepackage[caption=false]{subfig}
\usepackage{graphicx}% Include figure files
\usepackage{dcolumn}% Align table columns on decimal point
\usepackage{bm}% bold math

\usepackage{amsfonts}
\usepackage{amsmath}
\usepackage{amssymb}
\usepackage{mathtools}

\usepackage{siunitx}
\usepackage[%dvips,
bookmarks=true,% Lesezeichen erzeugen
bookmarksopen=false,% Lesezeichen ausgeklappt
bookmarksnumbered=true,% Anzeige der Kapitelzahlen am Anfang der Namen der Lesezeichen
breaklinks=true,% ermöglicht einen Umbruch von URLs
colorlinks=true,% Einfaerbung von Links
linkcolor=blue,% Linkfarbe: schwarz
anchorcolor=black,% Ankerfarbe: schwarz
citecolor=black,% Literaturlinks: schwarz
filecolor=black,% Links zu lokalen Dateien: schwarz
menucolor=black,% Acrobat Menü Einträge: schwarz
urlcolor=black,% URL-Farbe: schwarz
pdfpagemode=UseOutlines,
pdftitle=Scattering Theory for Floquet States,
pdfsubject=Scattering Theory for Floquet States,
pdfauthor=Thomas Bilitewski
]{hyperref}

\newcommand{\vac}{\vert {\rm vac} \rangle} 

\DeclarePairedDelimiter\bra{\langle}{\rvert}
\DeclarePairedDelimiter\ket{\lvert}{\rangle}

\DeclarePairedDelimiterX\braket[3]{\langle}{\rangle}%
{#1\,\delimsize\vert\,#2\,\delimsize\vert\,#3}

  \DeclarePairedDelimiterX\prodt[2]{\langle}{\rangle}%
{#1\,\delimsize\vert\,#2}

 \DeclarePairedDelimiterX\prodtt[2]{\langle \langle}{\rangle \rangle}%
{#1\,\delimsize\vert\,#2}
 
\DeclarePairedDelimiterX\brakett[3]{\langle \langle}{\rangle \rangle}%
{#1\,\delimsize\vert\,#2\,\delimsize\vert\,#3}

\DeclareSymbolFont{bbold}{U}{bbold}{m}{n}
\DeclareSymbolFontAlphabet{\mathbbold}{bbold}

\makeatletter
\newlength{\sfp@hseplen}\newlength{\sfp@vseplen}
\define@cmdkey{subfigpos}[sfp@]{pos}[ul]{}% \sfp@pos
\define@cmdkey{subfigpos}[sfp@]{font}[\small]{}% \sfp@font
\define@cmdkey{subfigpos}[sfp@]{vsep}[2\baselineskip]{\setlength{\sfp@vseplen}{\sfp@vsep}}% \sfp@vsep
\define@cmdkey{subfigpos}[sfp@]{hsep}[10pt]{\setlength{\sfp@hseplen}{\sfp@hsep}}% \sfp@hsep
\newcommand{\subfigimg}[3][,]{%
  \setkeys{Gin,subfigpos}{pos,font,vsep,hsep,#1}% Set (default) keys
  \setbox1=\hbox{\includegraphics{#3}}% Store image in box
  \ifnum\pdfstrcmp{\sfp@pos}{ul}=0% UPPER LEFT placement of subfig label
    \leavevmode\rlap{\usebox1}% Print image
    \rlap{\hspace*{\sfp@hsep}\raisebox{\dimexpr\ht1-\sfp@vsep}{\sfp@font{#2}}}% Print label
    \phantom{\usebox1}% Insert appropriate spacing
  \else\ifnum\pdfstrcmp{\sfp@pos}{ur}=0% UPPER RIGHT placement of subfig label
    \leavevmode\usebox1% Print image
    \llap{\raisebox{\dimexpr\ht1-\sfp@vsep}{\sfp@font{#2}}\hspace*{\sfp@hsep}}% Print label
  \else\ifnum\pdfstrcmp{\sfp@pos}{lr}=0% LOWER RIGHT placement of subfig label
    \leavevmode\usebox1% Print image
    \llap{\raisebox{\sfp@vsep}{\sfp@font{#2}}\hspace*{\sfp@hsep}}% Print label
  \else% Assume LOWER LEFT placement of subfig label
    \leavevmode\rlap{\usebox1}% Print image
    \rlap{\hspace*{\sfp@hseplen}\raisebox{\sfp@vsep}{\sfp@font{#2}}}% Print label
    \phantom{\usebox1}% Insert appropriate spacing
  \fi\fi\fi
}
\makeatother

\begin{document}

\title{Scattering Theory for Floquet-Bloch States}

\author{Thomas \surname{Bilitewski}}
\email{tb494@cam.ac.uk}
\affiliation{T.C.M. Group, Cavendish Laboratory, J.J. Thomson Avenue, Cambridge CB3 0HE, United Kingdom}
\author{Nigel R. \surname{Cooper}}
\affiliation{T.C.M. Group, Cavendish Laboratory, J.J. Thomson Avenue, Cambridge CB3 0HE, United Kingdom}

\date{February 11, 2015}
\begin{abstract}

  Motivated by recent experimental implementations of artificial gauge
  fields for gases of cold atoms, we study the scattering properties
  of particles that are subjected to time-periodic Hamiltonians.
  Making use of Floquet theory, we focus on translationally invariant
  situations in which the single-particle dynamics can be described in
  terms of spatially extended Floquet-Bloch waves. We develop a
  general formalism for the scattering of these Floquet-Bloch
  waves. An important role is played by the conservation of Floquet
  quasi-energy, which is defined only up to the addition of integer
  multiples of $\hbar\omega$ for a Hamiltonian with period $T=
  2\pi/\omega$.  We discuss the consequences of this for the
  interpretation of ``elastic'' and ``inelastic'' scattering in cases
  of physical interest.  We illustrate our general results with
  applications to: the scattering of a single particle in a
  Floquet-Bloch state from a static potential; and, the scattering of
  two bosonic particles in Floquet-Bloch states through their interparticle
  interaction. We analyse examples of these scattering processes that
  are closely related to the schemes used to generate artifical gauge
  fields in cold-atom experiments, through optical dressing of
  internal states, or through time-periodic modulations of
  tight-binding lattices. We show that the effects of scattering
  cannot, in general, be understood by an effective time-independent
  Hamiltonian, even in the limit $\omega \to \infty$ of rapid
  modulation.  We discuss the relative sizes of the elastic scattering
  (required to stablize many-body phases) and of the inelastic
  scattering (leading to deleterious heating effects). In particular,
  we describe how inelastic processes that can cause significant
  heating in current experimental set-up can be switched off by
  additional confinement of transverse motion.

\end{abstract}

% \pacs{}
%\keywords{}

\maketitle

%\tableofcontents

\makeatletter
% \let\toc@pre\relax
% \let\toc@post\relax
% \makeatother 

% \newpage
% \listoffigures
%  \newpage
%  \listoftables
%  \newpage

% \newpage
% \listoffigures
% \newpage
% \listoftables
% \newpage
% \tableofcontents
 \newpage

\section{\label{sec:intro}Introduction}

There have been long-standing research efforts to find ways to cause neutral atoms to experience artificial gauge fields, extending the capabilities of ultracold gases as simulators of quantum many-body systems\cite{fetter, advances,Bloch2008, dalibardreview,Gauge_fields_Review}. Such gauge fields can mimic the orbital effects of magnetic fields on a charged particle, and can generate topological energy bands, which could lead to novel many body phases of degenerate fermonic or bosonic atoms. A wide variety of theoretical proposals for how to generate artificial gauge fields has been put forward, and there have now been several successful experimental implementations of artificial gauge fields using schemes inspired by these proposals\cite{MadisonCWD00,abos01,SchweikhardCEMC92,Spielman_Experiment2,MagneticFields_1,Struck,Jimenez,munichuniform,mitharper,esslingerhaldane,munichchern}. These advances open up possibilities of future studies of strongly correlated physics, analogous to (fractional) Quantum Hall physics, in cold gas setups with topological bandstructures.

All of the methods used to generate these artificial gauge fields exploit periodic time-dependent forcing\footnote{The case of rotation is special in settings where any potentials
that are static in the laboratory frame have perfect cylindrical symmetry.
Then, in the frame of reference co-rotating with any stirring potential
the Hamiltonian is time-independent, so energy is absolutely conserved. In
general, without cylindrical symmetry of the static potentials, rotation
at angular frequency $\omega$ also leads to a time-periodic Hamiltonian
and only conservation of quasi-energy.}.  Quantum mechanical systems with a periodic time-dependence allow a treatment within Floquet theory \cite{Shirley,Hanggireview}, the analogue of Bloch's theorem for time-periodicity instead of space-periodicity. 
This allows the reformulation of the time-periodic Schr\"odinger equation in terms of an eigenvalue problem which enables the use of well-known methods from time-independent quantum theory in contrast to the general time-dependent case in which such methods are generally not applicable.
Such time-periodic systems are more easily accessible to a theoretical treatment while still going beyond equilibrium physics and showing a range of novel phenomena. They can be used to simulate otherwise inaccessible static Hamiltonians as well as to show genuinely non-static behaviour.

One class of methods to generate artificial magnetic fields makes use of the internal structure of the atomic species. By coupling different internal states by optical (two-photon) transitions and by using a specific spatial dependence of the laser fields one can engineer a geometric phase equivalent to the motion of charged particles in a magnetic field \cite{dalibardreview, Gauge_fields_Review}. Such methods can be implemented for particles in the continuum \cite{Spielman_Theory,Spielman_Experiment1,Spielman_Experiment2,Dudarev2004,Ruseckas2005} or very shallow optical lattices\cite{Flux_Lattice_Cooper_2011,Flux_Lattice_Cooper_Dalibard_2011,FQH_Cooper}, for particles in deep species-dependent optical lattices \cite{Jaksch_2003,Mueller_2004,Dalibard_2010}, and for collective ``spin'' degrees of freedom in lattices\cite{yaofci}. 

Another class of methods uses periodic shaking or modulation of species-independent optical lattices. 
Notable effects include the tuning of magnitude and sign of the tunnelling strength leading to the phenomenon of coherent destruction of tunnelling in two-level systems \cite{DrivenTunneling_Review} and in tight-binding models to dynamic localisation \cite{Dynamic_Localization}, a tunable superfluid to Mott transition \cite{SFM_SL} as well as the possibility to invert bands.
More sophisticated driving protocols to generate artificial gauge fields with high fluxes have been proposed and experimentally realised \cite{Sias_SL,TopologicalProperties,Kolovsky,Creffield,NAGF_TI,Edge_States,SOC,MagneticFields_1,MagneticFields_2,baurssc}. 
These approaches do not depend on any internal structure of the atoms and are thus applicable to a wide range of species.
One can divide these proposals into two classes: either the lattice position is periodically changed \cite{Sias_SL,Struck,Struck_SL_Ising}, referred to as shaking in the following; or the on-site energies in a static lattice are modulated by additional time-dependent fields in such a way as to resonantly restore tunnelling between different sites in the lattice \cite{MagneticFields_1,MagneticFields_2,munichuniform,mitharper}.
In all of these systems the time-periodic driving effectively modifies the tunnelling elements, allowing these to become complex with site-dependent phases. Interpreting these as arising from a Peierls substitution the tunneling phases correspond to the presence of an artificial gauge field.

With the experimental achievement of artificial gauge fields using these techniques, as demonstrated by at least the resulting single-particle phenomena, it is of great interest now to consider the consequences for systems of many interacting quantum particles. Much work has been done in exploring the effective interactions between particles in the dressed-state bands \cite{Williams2011,Spielman_Experiment3,Julia-Diaz2013,Zhang2013,Zhang2013a,Jiang2011,ZHAI2012,Stanescu2008}, which leads to many novel features. However, this work has largely ignored the aspects relating to the periodic time-dependence giving conservation only of the Floquet quasi-energy, but see \cite{Wannier_Stark1,WannierStark2,Stability_Floquet_condensate,phi4_RG} for notable exceptions. In \cite{Stability_Floquet_condensate} the stability of a Bose-Einstein condensate in a shaken lattice was analysed and the system was shown to have both stable and unstable phases. 

Generally, time-periodic driving with characteristic angular frequency $\omega$ allows the absorption of energy quanta $\hbar \omega$ from the field. At the single particle level, there is a periodic energy transfer, as typical of a Rabi oscillation in a dressed state, which does not lead to entropy generation or other heating processes. However, in the presence of inter particle interactions, one can anticipate that 
this energy absorption can cause forms of ``inelastic'' 
scattering that can have a heating effect. Possible processes include the excitation of particles into higher bands or even particle loss, or absorption of energy into transverse directions also resulting in heating and loss of particles from the experimental region. These processes necessitate investigation.

Scattering through time-periodically modulated potentials has been studied previously in other contexts \cite{Li1999a,Fedirko2000,Emmanouilidou2002,Pimpale1991}. In particular, the transport properties of time-driven mesoscopic systems have been investigated \cite{Platero2004,KOHLER2005} and formulations been given within the Keldysh formalism \cite{Jauho1994} as well as in the Floquet framework \cite{Camalet2003}. A general formulation for oscillating scatterers in terms of a Floquet scattering matrix was developed in \cite{Moskalets2002}. The study of the transport in periodically driven systems and the associated scattering properties is still an active field of research \cite{Moskalets2014,Moskalets2014a,Kibis2014,FoaTorres2014,Dasenbrook2014}.
The novelty of the systems considered here is that the non-interacting Hamiltonian will be periodically driven and the scattering will be caused by static two-body interactions, not by an external oscillating one-body scattering potential. Due to the Floquet structure of the single-particle states the interactions will effectively become time-dependent. Importantly, this also means that for the systems we consider the asymptotic states will be time-dependent, in contrast to the situation usually assumed in transport measurements in which the time-dependence is restricted to the scattering region. Thus, while the underlying theory describing the scattering processes and the resulting phenomena (like the absorption of quanta $\hbar \omega$ from the photon field during scattering) is similar, the physical origin of those processes is very different. Such systems have been studied in the context of atom-atom and atom-electron collisions in intense laser fields and a perturbation theory in the particle-interactions has been established \cite{Joachain2009,Byron1984,Francken1988,Doerr1994}.

In this paper we shall study the scattering processes that arise in these time-driven systems. The focus will be on elucidating the role of two-body elastic scattering processes, required for the realisation of strongly-correlated many-body phases, as compared to inelastic two-body processes which will limit experimental lifetimes or the temperatures achievable in experiments.
We will begin with a discussion of the Floquet theory framework in section~\ref{sec:floquet} which will be used to introduce the concepts and notation required for the following treatment of two model systems in section~\ref{sec:app}. Firstly, section~\ref{subsec:raman} will present a two-level system coupled by laser fields which will serve to illustrate effects of time-periodic driving on the scattering properties in the continuum case. Secondly, in section~\ref{subsec:lattice} we will consider a lattice with time-modulated on-site energies to highlight similarities and differences to the continuum.
We present the results of two-particle scattering in a model relevant to current and future cold gas experiments realising artificial gauge fields in this fashion.

%==========================================================================
\section{\label{sec:floquet}Floquet theory}
%==========================================================================

Our studies rely on the application of Floquet theory \cite{Shirley,Hanggireview,DrivenTunneling_Review}, as is relevant for quantum mechanical systems with periodic time-dependence. We shall make use of an inner product in an extended space which allows the usage of the language and many of the techniques developed for time-independent Hamiltonians \cite{sambe_pra_7_2203_1973}, the basis of which we now review.

Consider a time-varying Hamiltonian $H(t)$ that is periodic in time, $H(t+T)=H(t)$, where $T=2 \pi /\omega$ is the oscillation period and $\omega$ the associated frequency. The corresponding time-dependent Schr\"odinger equation
\begin{equation}
 \left[H(t) - i \hbar \frac{\partial}{\partial t} \right] |\Psi(t)\rangle =0
\end{equation}
allows solutions of a specific form called Floquet states. These may be written as
\begin{equation}
| \Psi_{\alpha}(t)\rangle = \exp[-i \epsilon_{\alpha} t/\hbar] |\Phi_{\alpha}(t)\rangle, 
\end{equation}
where $|\Phi_{\alpha}(t)\rangle$ is called the Floquet mode which has the same time-periodicity as the Hamiltonian, i.e. it satisfies $|\Phi_{\alpha}(t+T)\rangle= |\Phi_{\alpha}(t)\rangle$, and $\epsilon_{\alpha}$ is called the quasi-energy which is only defined up to multiples of $\hbar \omega$. Due to the time-periodicity of the Floquet modes one may expand them as 
\begin{equation}
|\Phi_{\alpha}(t)\rangle= \sum_m e^{i m \omega t}\ket{\phi_{\alpha}^{m}} .
\end{equation}
Clearly, the same physical state is obtained from
 \begin{equation}
 \begin{split}
| \Psi_{\alpha}(t) \rangle &= \exp[-i (\epsilon_{\alpha}+m \hbar \omega ) t/\hbar] \exp[i m \omega t]|\Phi_{\alpha}(t)\rangle\\
                  &=\exp[-i \epsilon^{m}_{\alpha} t/\hbar] |\Phi^{m}_{\alpha}(t)\rangle, 
 \end{split}
\end{equation}
where the shifted states are defined as $|\Phi^{m}_{\alpha}(t) \rangle= \exp[i m \omega t]|\Phi_{\alpha}(t)\rangle$ with quasi-energy $\epsilon^{m}_{\alpha}=\epsilon_{\alpha}+m \hbar \omega $ for any integer number $m$.

The formal analogy to Bloch's theorem is now evident: just as the
(discrete) spatial translational invariance of a lattice Hamiltonian
leads to Bloch functions with an exponential plane wave part and a
periodic part labelled by the crystal or quasi-momentum $k$, so too
does the invariance of the Hamiltonian under discrete time translation
$t \rightarrow t+T$ cause the Floquet states to consist of an
exponential part and a periodic part labelled by the quasi-energy
$\epsilon$. Just as the Bloch quasi-momentum $k$ is only defined up to
reciprocal lattice vectors and conventionally taken to lie in the
first Brillouin zone (BZ), so too the Floquet quasi-energy is defined
only up to addition of $\hbar\omega$. One may then define
$\epsilon^0_{\alpha}£$ to lie in the range $ - \hbar \omega/2 <
\epsilon^0_{\alpha} \le \hbar \omega /2$.  However, as we discuss
below, other conventions for the Floquet energies may be more suitable
and physically transparent: for example one might
choose $\epsilon_{\alpha,0}$ in such a way as to most closely
correspond to the eigenenergies of a static Hamiltonian.  

Defining the Hermitian operator $\mathcal{H} = H -i \hbar \frac{\partial}{\partial t}$, one sees that the Floquet modes themselves satisfy an eigenvalue equation
\begin{equation}
 \mathcal{H} \Phi_{\alpha}(t) = \epsilon_{\alpha} \Phi_{\alpha}(t),
\end{equation}
with the additional restriction that $\Phi_{\alpha}(t)$ must be a periodic function of period $T$. Moreover, clearly the shifted Floquet modes $\Phi^m_{\alpha}(t)$ satisfy the same equation, but with their shifted eigenvalues $\epsilon^m_{\alpha}$. As eigenfunctions of a Hermitian operator they form a complete and orthogonal basis set with respect to a suitably extended inner product. This is achieved via\cite{Hanggireview,sambe_pra_7_2203_1973}
\begin{equation}
 \prodtt{\Phi^{n}_{\alpha}}{\Phi^{m}_{\beta}} = \frac{1}{T} \int_0^{T} dt \prodt{\Phi^{n}_{\alpha}(t)}{\Phi^{m}_{\beta}(t) }= \delta_{\alpha,\beta} \delta_{n,m},
\label{eq:extendedinner}
\end{equation}
where $\prodt{\Phi^{n}_{\alpha}(t)}{\Phi^{m}_{\beta}(t)}$ denotes the usual inner product between state vectors at equal times. Moreover, at equal times the Floquet modes form a complete set for the Hilbert space of $H$,
\begin{equation}
 \sum_{\alpha} \ket{\Phi_{\alpha}(t)} \bra{\Phi_{\alpha}(t)}=\hat{\openone}
\end{equation}
where $\hat{\openone}$ denotes the identity in the Hilbert space of $H$ only, i.e. not including the space of time-periodic functions on which $\mathcal{H}$ acts.
This formulation allows one to carry over many of the techniques known from time-independent quantum mechanical systems and extend them to the time-periodic case by use of the scalar product (\ref{eq:extendedinner}).

\subsection{Scattering theory\label{subsec:scat_theo}}

The theory of scattering within the Floquet framework \cite {Joachain2009} is most conveniently described in the interaction picture of quantum mechanics. 
The conceptional difference in the scattering of Floquet states arises from the fact that the interaction picture will be defined with respect to a time-periodic non-interacting Hamiltonian $H_0(t)$ in contrast to the more conventional case of a static non-interacting Hamiltonian. The special properties of the associated propagator, encoding the fact that energy is only conserved modulo $\hbar\omega$, will ultimately lead to inelastic scattering processes.

We consider a Hamiltonian of the form $H(t) = H_0(t) +V$, which is split into a non-interacting part $H_0$ that describes free particle motion, and an interaction $V$ that will describe the particle scattering. The non-interacting Hamiltonian $H_0(t)$ consists of a static part and a time-periodic single particle coupling term that is strong and thus must be treated in a non-perturbative way, whereas the  interaction $V$ will be treated in the framework of perturbative scattering theory. We assume that $V$ is time-independent, as this will be relevant in the physical situations discussed later, but the analysis can be readily extended to general time-dependent $V(t)$. 

We define the unitary time-evolution operator $U_0(t,t^{\prime})$ associated with $H_0(t)$ by 
\begin{equation}
 \ket{\Psi_0(t^{\prime})} = U_0(t,t^{\prime}) \ket{\Psi_0(t)} .
\end{equation}
Due to the time-dependence of $H_0(t)$ this operator depends on both start and end times and due to the time periodicity has a special structure. Specifically, it has a spectral representation as
\begin{align}\label{eq:Floquet_Propagator}
 U_0(t,t^{\prime})&=\sum_{\alpha} e^{-i \epsilon_{\alpha} (t^{\prime}-t)/\hbar} \ket{\Phi_{0,\alpha}(t^{\prime})} \bra{\Phi_{0,\alpha}(t)}\\
                  &=\sum_{\alpha,n,m} e^{-i \epsilon_{\alpha} (t^{\prime}-t)/\hbar} e^{-i (n \omega t -m \omega t^{\prime}) }\ket{\phi^{m}_{0,\alpha}} \bra{\phi^{n}_{0,\alpha}}
\end{align}
We define the states in the interaction picture in the usual way via
\begin{equation}
 \ket{\Psi^{\rm I}(t)} = U_0(t,t_0) \ket{\Psi(t)}
\end{equation}
which then satisfy the Schr\"odinger equation
\begin{equation}
  i \hbar \partial_{t} \ket{\Psi^{\rm I}(t)} = V^{\rm I}(t) \ket{\Psi^{\rm I}(t)}
\end{equation}
with
\begin{equation}\label{eq:V_interaction_picture}
 V^{\rm I}(t)= U_0(t,t_0) V U_0(t_0,t) .
\end{equation}
The corresponding time-evolution operator $U^{\rm I}(t_0,t)$ then satisfies the differential equation
\begin{equation}
 i \hbar \partial_t U^{\rm I}(t_0,t) =  V^{\rm I}(t) U^{\rm I}(t_0,t) 
\end{equation}
with the initial condition $U^{\rm I}(t_0,t_0)= \hat{\openone}$. Rewritten as an integral equation
\begin{equation}
 U^{\rm I}(t_0,t) = \hat{\openone} - \frac{i}{\hbar} \int_{t_0}^{t}dt^{\prime} \, V^{\rm I}(t^{\prime}) U^{\rm I}(t_0,t^{\prime})
\end{equation}
it allows the usual iterative solution in the Dyson series
\begin{equation}\label{eq:Dyson_Series}
 U^{\rm I}(t_0,t) = \hat{\openone} - \frac{i}{\hbar} \int_{t_0}^{t}dt^{\prime} \, V^{\rm I}(t^{\prime}) + \mathcal{O}(V^{2}) \,.
\end{equation}
The full unitary evolution operator is then given by
\begin{equation}
 U(t_0,t)=U_0(t_0,t) U^{\rm I}(t_0,t)\,.
\end{equation}
This treatment clarifies in what way the usual scattering theory can be applied to Floquet states. The only difference arises via the use of the propagator $U_0(t,t_0)$, Equation~(\ref{eq:Floquet_Propagator}), whose structure therefore determines the differences to the standard case of a time-independent Hamiltonian. Since every Floquet state generically contains components that evolve with phases $e^{-i(\epsilon +m \hbar \omega t)/\hbar}$ for all integer $m$, $V^{\rm I}(t)$ will most generally contain time-dependent terms oscillating with $e^{-i(\Delta \epsilon +m \hbar \omega t)/\hbar}$ where $\Delta \epsilon=\epsilon_{\alpha}-\epsilon_{\beta}$ is the quasi-energy difference of any two Floquet states. Therefore, it is immediately apparent that generically a transition between an initial state with quasi-energy $\epsilon_{\rm i}$ and a final state with quasi-energy $\epsilon_{\rm f}=\epsilon_{\rm i} + m \hbar \omega$ for any integer $m$ can be induced by a static interaction $V$ due to the structure of the Floquet states.
Keeping only the term that is first order in $V$ leads to the Born approximation for the scattering, which reduces to the application of the Floquet Fermi golden rule for transition rates. Including higher orders in the scattering potential will lead to a further mixing of the Floquet states at the same quasi-energy, but will not change the qualitative picture described based on the Floquet Fermi golden rule.
\subsection{Floquet Fermi golden rule\label{subsec:FFGR}}
The extension of Fermi's golden rule to the Floquet framework, called the Floquet Fermi golden rule (FFGR) was presented in Ref.\cite{FFGR}. Since it will be central to the applications described below, here we present a simple derivation of the FFGR following from the above propagator. 

We consider a case in which $V$ is switched on at $t=0$ and compute transition rates from an initial state $\ket{\Psi_{\rm i}}$ to final states $\ket{\Psi_{\rm f}}$. We take the initial state at $t=0$ and the final state at all times to be Floquet eigenstates of the unperturbed Hamiltonian $H_0(t)$, i.e. $\ket{\Psi_{\rm i}(t=0} = \ket{\Phi_{0,{\rm i}}(t=0)}$ and  $\ket{\Psi_{\rm f}(t}=e^{-i \epsilon_{\alpha}t} \ket{\Phi_{0,\alpha}(t)}$. For notational simplicity we drop this subscript indicating the states of the unperturbed Hamiltonian in the following.

Thus, the relevant amplitude is
\begin{align}
A({\rm {\rm i} \rightarrow {\rm f}},t)&=\braket{\Psi_{\rm f}(t)}{U(0,t)}{\Psi_{\rm i}(t=0)}\\
&= \braket{\Psi_{\rm f}(t)}{U_0(0,t) U^{\rm I}(0,t)}{\Psi_{\rm i}(t=0)}\\
  &=\braket{\Psi_{\rm f}(t=0)}{U^{\rm I}(0,t)}{\Psi_{\rm i}(t=0)} .
\end{align}
Using the expansion up to first order of the time-evolution operator $U^{\rm I}(0,t)$ we obtain for the transition amplitude in the case of ${\rm i} \ne {\rm f}$
\begin{align}
  &A({\rm i} \rightarrow {\rm f},t)=\frac{-i}{\hbar} \int_{0}^{t}dt^{\prime}\braket{\Psi_{\rm f}(0)}{V^{\rm I}(0,t^{\prime})}{\Psi_{\rm i}(0)}\label{eq:FFGR_derivation_1} \\
  &\quad=\frac{-i}{\hbar} \int_{0}^{t}dt^{\prime}\braket{\Psi_{\rm f}(0)}{U_0(t^{\prime},0) V U_0(0,t^{\prime})}{\Psi_{\rm i}(0)} \label{eq:FFGR_derivation_2}\\
  &\quad=\frac{-i}{\hbar} \int_{0}^{t}dt^{\prime} e^{-i(\epsilon_{\rm i} - \epsilon_{\rm f}) t^{\prime}/\hbar}\braket{\Phi_{\rm f}(t^{\prime})}{V }{\Phi_{\rm i}(t^{\prime})}\label{eq:FFGR_derivation_3}\\
  &\quad=\sum_{n.m}\frac{-i}{\hbar} \int_{0}^{t}dt^{\prime} e^{-i(\epsilon_{\rm i} - \epsilon_{\rm f}-(n-m)\hbar \omega) t^{\prime}/\hbar}\braket{\phi^{m}_{\rm f}}{V }{\phi^{n}_{\rm i}}\label{eq:FFGR_derivation_4}\\
  &\quad= \sum_{n.m} \frac{e^{-i(\epsilon_{\rm i} - \epsilon_{\rm f}-(n-m)\hbar \omega) t/\hbar} -1}{(\epsilon_{\rm i} - \epsilon_{\rm f}-(n-m)\hbar \omega)}   V^{mn}_{\rm fi} \label{eq:FFGR_derivation_5}\\
  &\quad= \sum_{l.m} \frac{e^{-i(\epsilon_{\rm i} - \epsilon_{\rm f}-m\hbar \omega) t/\hbar} -1}{(\epsilon_{\rm i} - \epsilon_{\rm f}-m \hbar \omega)}   V^{ll+m}_{\rm fi}\label{eq:FFGR_derivation_6}
\end{align}
where to get from Equation~(\ref{eq:FFGR_derivation_2}) to Equation~(\ref{eq:FFGR_derivation_3}) we used the spectral representation of the propagator Equation~(\ref{eq:Floquet_Propagator}) and the orthogonality of the Floquet modes at equal times. In Equation~(\ref{eq:FFGR_derivation_5}) we have defined the matrix-element $V^{mn}_{\rm fi}=\braket{\phi^{m}_{\rm f}}{V }{\phi^{n}_{\rm i}}$ of the perturbation $V$ between the $n$-th Fourier component $\ket{\phi^{n}_{\rm i}}$ of the initial Floquet mode $\ket{\Phi_{\rm i}(t)}=\sum_n e^{i n \omega t} \ket{\phi^{n}}$ and the $m$-th Fourier component $\ket{\phi^{m}_{\rm f}}$ of the final Floquet mode $\ket{\Phi_{\rm f}(t)} = \sum_m e^{i m \omega t}  \ket{\phi^{m}_{\rm f}}$. 

From the transition probability, $P({\rm i} \rightarrow {\rm f},t)= \left\vert A({\rm i} \rightarrow {\rm f},t)\right\vert^2$, one derives the Floquet Fermi golden rule by computing the rate $\gamma_{\rm i\rightarrow \rm f}= \lim_{t\rightarrow \infty} P({\rm i} \rightarrow {\rm f},t)/t$. In contrast to the derivation of the usual Fermi golden rule, the amplitude contains {\it two} sums over the Fourier components of the Floquet modes. The sum over $m$ allows the emission/absorption of energy quanta $m \hbar \omega$ during the scattering process. In computing the transition rates $m$ is fixed by the resonance condition $\epsilon_{\rm i}-\epsilon_{\rm f}= m \hbar \omega$.
With this in mind the standard steps lead to
\begin{equation}
 \gamma_{\rm i\rightarrow \rm f} =\sum_{m,l,n}  \frac{2 \pi}{\hbar} \delta(\epsilon_{\rm i}-\epsilon_{\rm f}-m\hbar \omega)   V^{nn+m}_{\rm fi}   V^{l+ml}_{\rm if} . 
\end{equation}
As a final step we may rewrite this in a more convenient form as
\begin{equation}
 \gamma_{\rm i\rightarrow \rm f} =\sum_{m}  \frac{2 \pi}{\hbar} \delta(\epsilon^0_{\rm i}-\epsilon^0_{\rm f}-m\hbar \omega) |\brakett{\Phi_{\rm f}^{m}}{V}{\Phi_{\rm i}^{0}}|^2 
\label{eq:ffgr}
\end{equation}
where $|\Phi_{\rm f}^{m}(t)\rangle= e^{i m \omega t}|\Phi_{\rm f}^{0}(t)\rangle$ are the shifted Floquet modes introduced above. Written in this way the rate has the same form as the conventional Fermi's golden rule apart from the additional summation over $m$ and the use of the extended scalar product. The explicit sum over $m$ justifies the remarks that only quasi-energy is conserved or equivalently that energy is only conserved up to quanta of $\hbar \omega$. 
Transitions with the absorption/emission of $m \hbar \omega$ will occur within the FFGR if $V^{nn+m}_{\rm fi}=\braket{\phi^{n}_{\rm f}}{V }{\phi^{n+m}_{\rm i}} \ne 0$ for some $n$, i.e. if the interaction couples different Fourier components of the Floquet modes. 

\subsection{Inelastic Scattering}
In light of the fact that for a time-periodic Hamiltonian only the quasi-energy is conserved, it is important to consider the definition of ``inelastic scattering'' in these circumstances. One obvious choice for the quasi-energies is to reduce all to a BZ $-\hbar \omega/2  < \epsilon_\alpha^0 \le \hbar \omega/2$. 
However, this choice may be inconvenient and even hide some of the relevant physics. 

We illustrate this by describing two simple examples.

\begin{figure}
    \subfloat%[First sub-figure\label{subfig-1:dummy}]
    {%
      \includegraphics[width=0.45\textwidth]{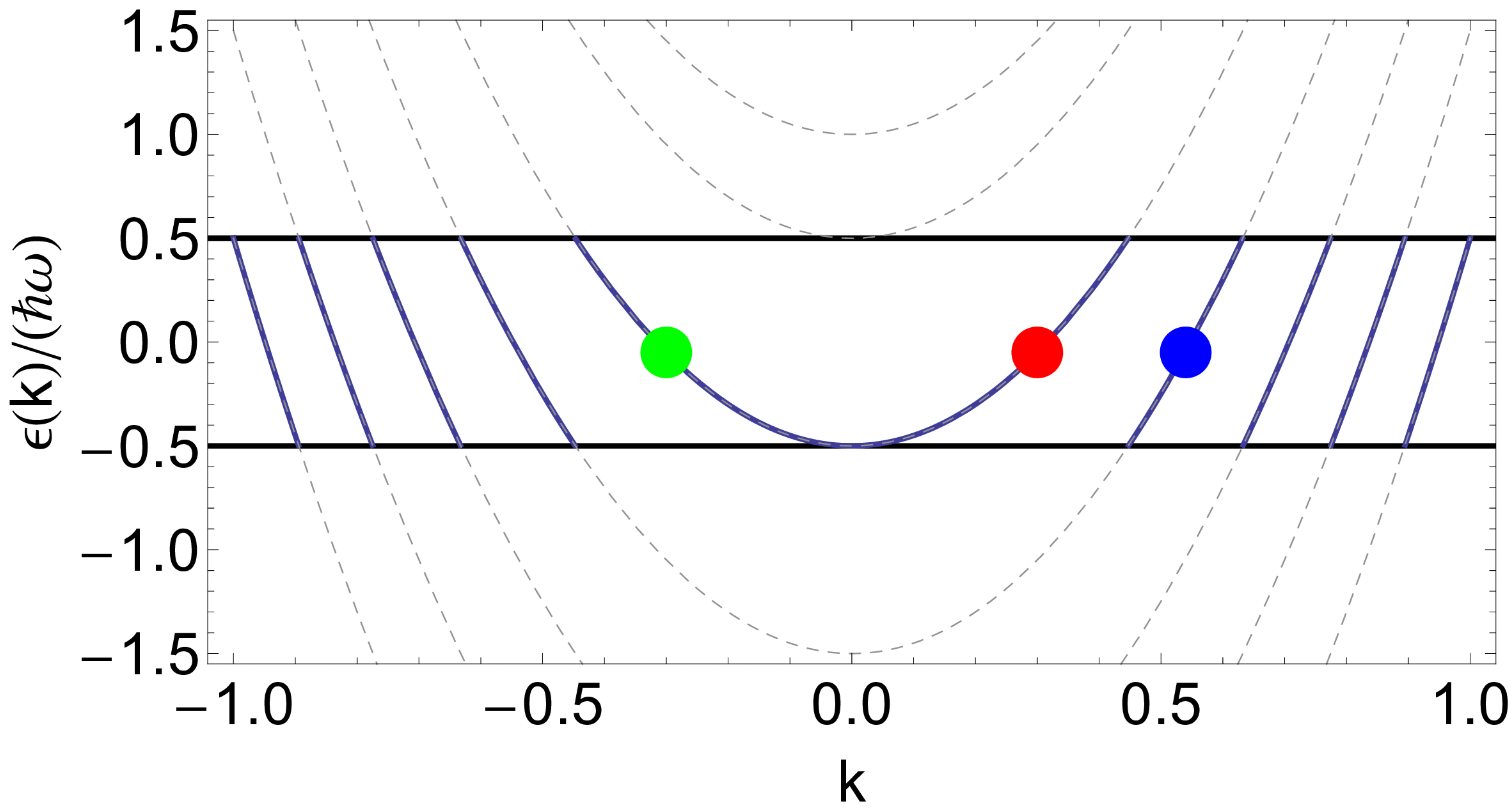}
    }
    \caption{(Color online) Unbounded single-particle dispersion $\epsilon^0_k/(\hbar \omega) \propto k^2 $ as a function of dimensionless wavevector $k$ reduced to the first BZ (bold) and continuous dispersion with periodically repeated images (dashed). Scattering an initial state (red/middle) to final state (green/left) would be considered elastic and scattering from initial state (red/middle) to a different branch (blue/right) would be considered inelastic. When regarding these processes with respect to the continuously defined dispersions, elastic scattering corresponds to no change in the Floquet index $m$ whereas inelastic scattering changes $m$.}
    \label{fig:floquet_disp2}
\end{figure}

As a first example consider a single particle for which the Floquet
energy spectrum of $H_0(t)$ consists of Floquet-Bloch waves with a
parabolic energy dispersion as a function of the dimensionless wavevector $k$, with
$\epsilon^0_k = \hbar\omega (k^2 -1/2)$, see
Fig.~\ref{fig:floquet_disp2}. (The dimensions and energy offset are chosen for presentational convenience. We shall present a model leading to
a similar case in Sec.\ref{subsec:toy_model} below.) The
single-particle states of this unperturbed time-periodic Hamiltonian
can be fully described by an energy dispersion that is a continuous
function of wavevector $k$.  Within the Floquet framework, one can
equally well choose to reduce the quasi-energies to a BZ, shown as the
set of bold curves in Fig.~\ref{fig:floquet_disp2}. In terms of the
reduced quasi-energies the dispersion is discontinuous and there is a
discrete infinite set of quasi-energetically degenerate momenta. When
a potential $V$ that breaks translational symmetry is introduced, it
can cause one-body scattering from an initial state (e.g. the red dot)
to the final states of the same quasi-energy (e.g. the green or blue
dots). Given the simple nature of the parabolic energy dispersion for
the unperturbed single particle, it is natural to call the transition
to a different branch ``inelastic'' (red to blue) and the transition
staying within the same branch ``elastic'' (red to green). This may be conveniently
achieved by defining $\epsilon^{0}_{k} = \hbar\omega (k^2 -1/2)$ to
depend continuously on the dimensionless wavevector $k$ (i.e. not to be restricted to $-\hbar/\omega/2<  \epsilon^{0}_{k}\leq \hbar\omega/2$), and by considering the whole
family of periodically repeated dispersions (the dashed lines in Fig~\ref{fig:floquet_disp2}): ``inelastic'' scattering  (i.e. between different branches) 
then
corresponds to a change in the Floquet index $m$.

\begin{figure}
    \subfloat%[First sub-figure\label{subfig-1:dummy}]
    {%
      \includegraphics[width=0.22\textwidth]{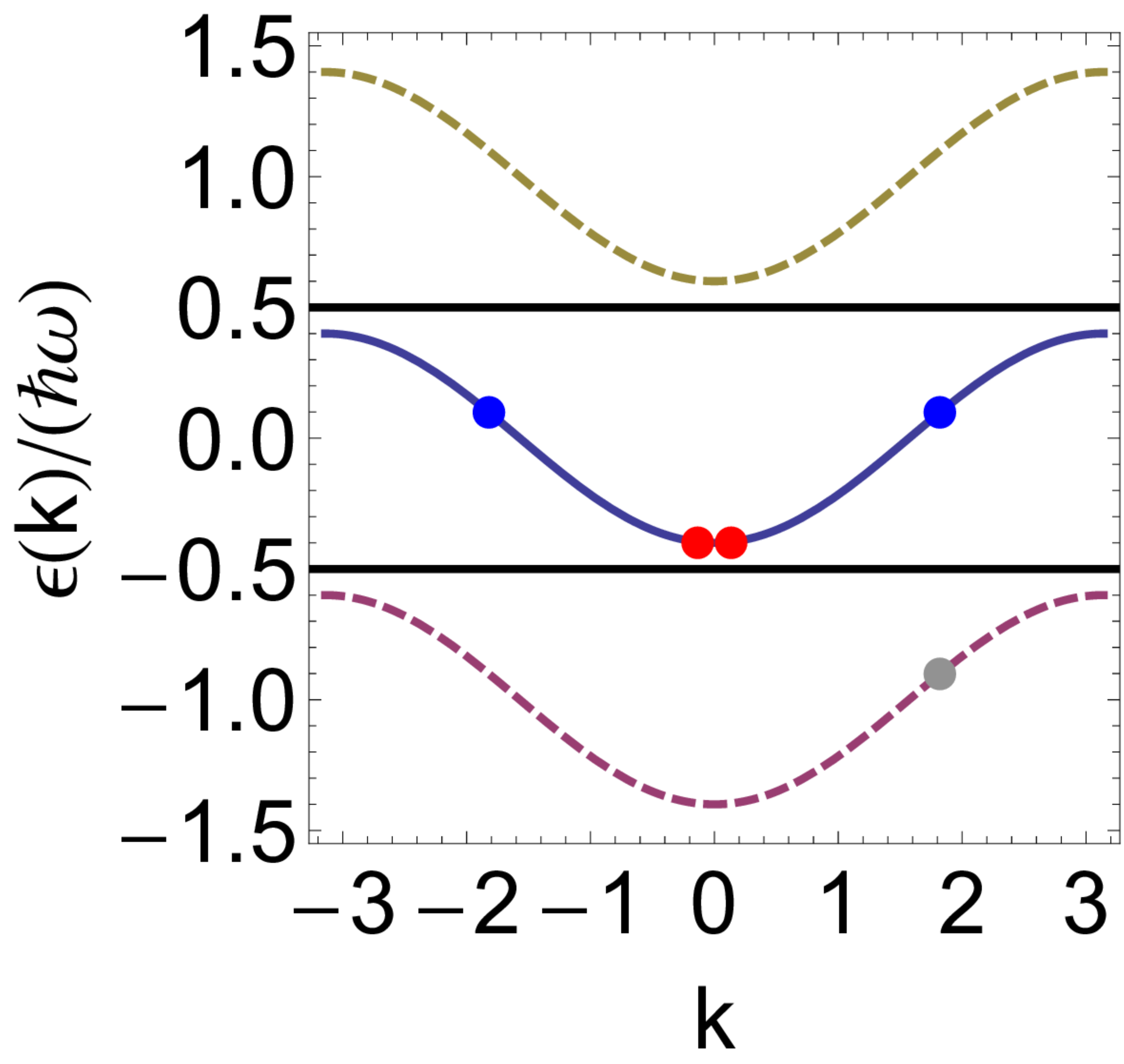}
    }
    \hfill
    \subfloat%[First sub-figure\label{subfig-2:dummy}]
    {%
      \includegraphics[width=0.22\textwidth]{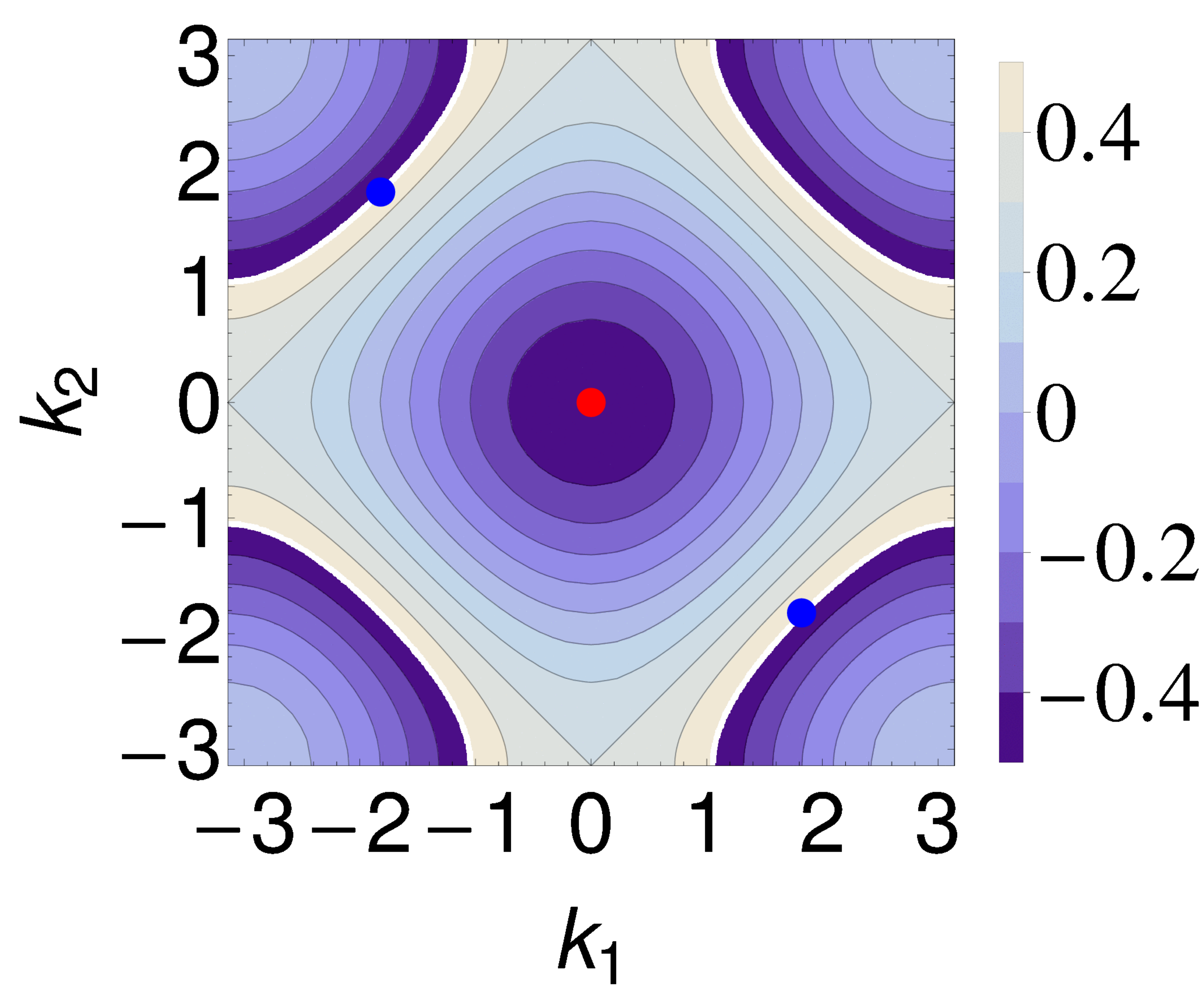}
    }
    \caption{(Color online) Bounded single-particle dispersion $\epsilon(k)/(\hbar \omega)=-0.4  \cos(k) $ as a function of dimensionless wavevector $k$ in the energy BZ with periodically repeated images (left) and contours of the reduced two-particle quasi-energy (right). Depicted is a two-particle scattering process during which two particles initially in the band-minimum (red balls) scatter into higher quasi-energy states (blue balls). This process conserves the reduced two-particle quasi-energy or equivalently can be viewed as one particle scattering into the lower shifted dispersion (grey ball).}
    \label{fig:floquet_disp}
\end{figure}

The relevance of these considerations become even more apparent for  two-particle scattering. As a second example, we consider scattering of two particles occupying an energy band (on a lattice) that has a bounded dispersion, e.g. each particle experiences the band structure of the form $\epsilon_k^0/(\hbar \omega)=-0.4  \cos(k) $ depending on the dimensionless wavevector $k$, shown in Fig.~\ref{fig:floquet_disp} with its periodic repetitions. From a naive interpretation of the single-particle spectrum, in which one ignores the periodically repeated spectra,  one would say that two particles at the bottom of the band $k=0$ with quasi-energy $\epsilon=-0.4 \hbar \omega$ would be forbidden from scattering by energy conservation. However, the total two-particle energy is $\epsilon_2=-0.8 \hbar \omega$ which would have to be mapped to a quasi-energy $0.2 \hbar \omega$ to lie in the range $-\hbar\omega \to \hbar\omega/2$. Thus, scattering is in fact allowed to a number of states for which $\cos(k_1) + \cos(k_2)=-0.5$. One such possible scattering event is depicted in the figure where the two initial particles in the band minimum (red) scatter to a final state (blue).
One might also consider the reduced two-particle dispersion of which contours are shown in Fig.~\ref{fig:floquet_disp}. Clearly, scattering from the centre to a region around the corners of the BZ now becomes possible, so two particles in the band minimum are not stable anymore. Those are processes that become allowed only within the Floquet description and will be called inelastic. If the (two-particle) quasi-energies are defined to be continuous with respect to the variable $k$ those correspond to a term $m \ne 0 $.  For example, using the viewpoint of the repeated zone for the quasienergy, the above scattering process involves the transition of one particle  (depicted grey in Fig.\ref{fig:floquet_disp}) to a copy of the Floquet band shifted down by $\hbar\omega$ (i.e. a transition with $m=1$).

Based on these examples, and the applications below, we provide a
general definition of ``inelastic scattering'' of Floquet-Bloch
waves. We consider the Floquet-Bloch spectrum for a single particle,
and define the quasi-energy for $m=0$, $\epsilon^{0}_{k,\tau}$, to be a
{\it continuous} function of the wavevector $k$ (which will be a
vector in dimensions $d>1$). The index $\tau$ accounts for any other
discrete quantum numbers -- e.g. band, or spin indices -- which
characterize the Floquet-Bloch state. The full set of Floquet modes is
obtained via $\epsilon^{m}_{k,\tau} = \epsilon^{0}_{k,\tau} + m
\hbar\omega$. In any scattering event, the particle (or particles)
must start and finish in states labelled by these indices
$(k$,$\tau$,$m$) (at long times before and after the collision).  We
define all those scattering events which involve a change of either
the discrete label $\tau$ or of the Floquet index $m$ (or both) to be
``inelastic''. This definition of inelastic scattering accounts both
for collisions in which the band index (of one or both) particles
changes, and for collisions that would not have occurred under a naive
interpretation of the energy bands where the repeated copies spaced by
$\hbar\omega$ are ignored.

\subsection{Toy model\label{subsec:toy_model}}
To illustrate the preceding discussion of the scattering properties of Floquet states, we consider a toy model for scattering in the presence of Raman dressing of internal states. We consider a single particle (or relative particle co-ordinate) with two internal states, and the Hamiltonian $H(t) = H_0(t) + V$ with
\begin{equation}
H_0(t)= \frac{p^{2}}{2 M} \mathbbold{1} + \begin{pmatrix}
                                      0 & \Omega e^{-i \omega t} \\
                                      \Omega e^{i \omega t} & -\hbar \omega
                                     \end{pmatrix}
\end{equation}
and 
\begin{equation}
 V= \delta(x) \begin{pmatrix}
               g_0 +g_1 & g_c \\
               g_c &  g_0 -g_1
              \end{pmatrix} \, .
\end{equation}
The Floquet modes are given as
\begin{equation}
 \Phi^{m}_{k,\tau} =\frac{1}{\sqrt{2 L}}e^{i m \omega t} e^{i k x} \left(\begin{array}{c} 1\\ \tau e^{i \omega t}\end{array} \right)
\label{eq:toyvector}
\end{equation}
with quasi-energies $\epsilon^{m}_{k,\tau}= \frac{\hbar^{2} k^{2}}{2 M} +\tau \Omega + m \hbar \omega$, where $L$ is the system size and labels are the quasi-momemtum $k$, the band index $\tau = \pm 1$ and the Floquet mode number $m$. Note that this actually corresponds to two shifted copies of a single parabolic dispersion which is shown in Fig.~\ref{fig:floquet_disp2} and for which the implications within the Floquet framework have been discussed above.

We begin the discussion of the scattering by a treatment within the Floquet Fermi golden rule.
The rate of scattering from initial state $\Psi_{\rm i}$ to final state $\Psi_{\rm f}$ is given by Eq.~(\ref{eq:ffgr}), 
with $\Phi^{m}_{{\rm (i/f)}}$ the Floquet modes associated to $\Psi^{m}_{{\rm (i/f)}}$ with quasi-energies $\epsilon^{m}_{{\rm i/f}}=\epsilon_{{\rm (i/f)}} + m \hbar \omega$ as described in section~\ref{sec:floquet}. 
The argument of the quasi-energy conserving $\delta$-function reads 
\begin{equation}\label{eq:toy_energy_conservation}
 \frac{\hbar^{2} k_{{\rm f}}^{2}}{2 M} =   \frac{\hbar^{2} k_{{\rm i}}^{2}}{2 M} + (\tau_{{\rm i}}-\tau_{{\rm f}}) \Omega - m \hbar \omega
\end{equation}
which shows that a transition to a final state with $m < 0$ ($m > 0$) corresponds to an absorption (emission) of energy $m \hbar \omega$ from the driving field. This may be converted  into kinetic energy or into a change of the bands, or both. Transitions with $m=0$ can still convert the difference between the band energies into kinetic and vice-versa, but the total energy of the states remain the same.

We take the initial and final state to be $\Phi_{{\rm i}}=\Phi^{0}_{k_{\rm i},\tau_{\rm i}}$ and $\Phi^{m}_{{\rm f}}=\Phi^{m}_{k_{\rm f},\tau_{\rm f}}$ and the matrix element is computed as 
\begin{equation}
\begin{split}
\brakett{\Phi^{m}_{{\rm f}}}{V}{\Phi^{0}_{{\rm i}}}  &=1/(2L)\, \delta_{m,0} \left[(g_0+g_1) + (g_0-g_1)\tau_{\rm i} \tau_{\rm f}  \right]\\ 
                                         & +1/(2L)\, \delta_{m,1} \,\tau_{\rm i} g_c  + \delta_{m,-1} \,\tau_{\rm f}  g_c 
\end{split}
\end{equation}

As the Floquet modes (\ref{eq:toyvector}) contain two frequency components at $m$ and $(m+1)$, single-particle scattering within FFGR allows at most the absorption of a single quantum $\hbar \omega$. Moreover, clearly for $g_c=0$ no inelastic processes ($\Delta m =0$) take place and particles can only scatter elastically between the two bands. However, for $g_c \ne 0$ particles can absorb energy during scattering ($\Delta m \neq 0$).  

The elastic scattering rate is given by
\begin{equation}
 \gamma_{k_{\rm i},\tau\rightarrow k_{\rm f},\tau} =\frac{1}{L} \frac{2 \pi}{\hbar}  \left|g_{0}\right|^{2} \rho_{\tau}(\epsilon_{\rm i}).
\end{equation}
For the inelastic rates we distinguish between those processes which only convert kinetic energy into band energy and vice versa which for our basis choice correspond to no change $m$ and those that change $m$. The inelastic rate for band changing collisions with no energy absorption is 
\begin{equation}
 \gamma_{k_{\rm i},\tau_{\rm i}\rightarrow k_{\rm f},-\tau_{\rm i}} =\frac{1}{L} \frac{2 \pi}{\hbar}  \left|g_{1}\right|^{2} \rho_{-\tau_{\rm i}}(\epsilon_{\rm i}) \, ,
\end{equation}
assuming that a band-changing transition is energetically allowed by Equation~(\ref{eq:toy_energy_conservation}) with $m=0$,
and the inelastic scattering rate with a change in total energy is given by
\begin{equation}
 \gamma_{k_{\rm i},\tau_{\rm i}\rightarrow k_{\rm f},\tau_{\rm f}} =\frac{1}{L} \frac{\pi}{2 \hbar}  \left|g_c\right|^{2} \rho_{\tau_{\rm f}}(\epsilon_{\rm i} \pm \hbar \omega)
\end{equation}
with the density of states $\rho_{\tau}(\epsilon)$ defined for the single particle dispersion $\epsilon^0_{k,\tau}$ per unit length.

In 1D the corresponding cross sections read 
$\sigma_{\rm el}= \frac{2 M^{2} g_0^{2}}{\hbar^{4}} \frac{ 1}{k^{2}}$ for elastic collisions for which $k=k_{\rm i}=k_{\rm f}$  and $\sigma^{m=0}_{\rm inel}= \frac{2 M^{2} g_1^{2}}{\hbar^{4}} \frac{ 1}{k_{\rm i} k_{\rm f}}$ for band changing energy-conserving collisions for which $k_{\rm i}$ and $k_{\rm f}$ satisfy Equation~(\ref{eq:toy_energy_conservation}) with $m=0$.
The inelastic scattering cross section for absorption/emission of $\hbar \omega$ is $\sigma^{m \ne 0}_{\rm inel}=\frac{2 M^{2} g_c^{2}}{4\hbar^{4}} \frac{ 1}{k_{\rm i} k_{\rm f}}$ where $k_{\rm i}$ and $k_{\rm f}$ satisfy Equation~(\ref{eq:toy_energy_conservation}) with $m=\pm 1$.
The divergences in the cross sections at low energies (small $k_{\rm i}$) stem from two factors. Firstly from the division by the incoming flux which accounts for the $1/k_{\rm i}$ factor present in all cross sections, and secondly from the final density of states which is proportional to $1/k_{\rm f}$ in 1D.  

Assuming a regime in which band-changing collisions are not allowed in collisions without absorption of energy, i.e. initial particles in the lower band and $\epsilon_{\rm i}<2 \Omega$, and further $\epsilon_{\rm i}< \hbar \omega$ and $\hbar \omega > 2 \Omega$ such that emission of energy during scattering is not possible and band-changing collisions are allowed with the absorption of energy, the ratio of the total cross sections is given by
\begin{equation}\label{eq:toy_cross_section_ratio}
 \frac{\sigma^{1D}_{\rm inel}}{\sigma^{1D}_{\rm el} } = \frac{g_c^{2}}{4 g_0^{2}} \left[\frac{1}{\sqrt{1+ \hbar \omega/\epsilon^{\rm kin}_{\rm i}}} + \frac{1}{\sqrt{1+ (\hbar \omega-2 \Omega)/\epsilon^{\rm kin}_{\rm i}}} \right]
\end{equation}
with $\epsilon^{\rm kin}_{\rm i} = \hbar^{2} k_{\rm i}^{2}/(2 M)$.
These expressions suggest that to achieve strong elastic scattering as compared to inelastic scattering it is advantageous to work at small $\epsilon^{\rm kin}_{\rm i}$. However, at very low energies the Born-Approximation becomes invalid, certainly breaking down when $\sigma_{\rm el} \gtrsim 1$, thus for $\epsilon^{\rm kin} \lesssim \frac{M g_0^{2}}{\hbar^{2}}$. Using this value in Equation~(\ref{eq:toy_cross_section_ratio}) we obtain a natural lower limit for the ratio of inelastic to elastic cross sections. 

The corresponding expression in 2D is
\begin{equation}
 \frac{\sigma^{2D}_{\rm inel}}{\sigma^{2D}_{\rm el} } = \frac{g_c^{2}}{4 g_0^{2}} \times 2 \, ,
\end{equation}
where the factor of 2 is due to the fact that the inelastic scattering cross section has two contributions from the band-changing and the band-conserving scattering processes. Thus, in 2D there is no energy dependence, and the relative size of inelastic and elastic scattering is simply controlled by the ratio of the relevant interaction parameters. 
In 3D one finds
\begin{equation}
 \frac{\sigma^{3D}_{\rm inel}}{\sigma^{3D}_{\rm el} } = \frac{g_c^{2}}{4 g_0^{2}} \left[\sqrt{1+ \hbar \omega/\epsilon^{\rm kin}_{\rm i}} + \sqrt{1+ (\hbar \omega-2 \Omega)/\epsilon^{\rm kin}_{\rm i}} \right] 
\end{equation}
which shows that inelastic scattering becomes increasingly important compared to elastic scattering at low kinetic energies. Unless $g_c/g_0$ is very small, this could lead to experimental difficulties in achieving stable strongly correlated phases of dressed-state particles at low energies in 3D settings. This model provides a simple example of how the suppression of inelastic compared to elastic scattering for low-energy particles may be favoured by the confinement of free motion to low dimensions.

We emphasize that the inelastic scattering rate and cross sections only depend on $\omega$ via the final density of states. In particular, in the limit $\omega \rightarrow \infty$, the rate vanishes in 1D, but is constant in 2D and divergent in 3D. Thus, the dynamics in higher than 2 dimensions are not in any way described by the average Hamiltonian even for $\omega \rightarrow \infty$, but rather inelastic scattering with an infinitely high energy transfer occurs in this case. (In view of the contact interaction, the matrix element remains nonzero for arbitrarily high momentum transfer, hence arbitrarily large final state energy.)

We now consider the effects of terms that are higher order in the scattering potential. From the higher order terms in the Dyson series equation~(\ref{eq:Dyson_Series}) one can see that to order $V^{n}$ transitions with an energy absorption/emission of $n \hbar \omega$ are allowed for this specific model if $g_c \ne 0$.  
However, in this case it is more transparent to perform a unitary transformation to the eigenstates of the non-interacting Hamiltonian. 
\begin{equation}
\begin{split}
 \tilde{H}&=U^{\dagger}H U - i \hbar U^{\dagger}\partial_t U \\
         &=\frac{p^{2}}{2 M} \mathbbold{1} + \begin{pmatrix}
                                      \Omega & 0 \\
                                      0 & -\Omega
                                     \end{pmatrix}
                                     +\tilde{V}(t)
                                     \end{split}
\end{equation}
with
\begin{equation}
 \tilde{V}(t) = \begin{pmatrix}
                  g_0 & g_1 \\
                  g_1  & g_0 
                 \end{pmatrix} 
                +  g_c \begin{pmatrix}
                  \cos\omega t & i \sin \omega t \\
                  - i \sin \omega t  & \cos \omega t 
                  \end{pmatrix} \, .
\end{equation}
In this representation $\tilde{V}(t)$ contains two frequency components at $\pm \omega$ and at order $\tilde{V}^{n}$ allows the absorption of $n \hbar \omega$ of energy. This derivation has the additional advantage that it provides a natural explanation for the inelastic scattering in this model. Whenever the unitary transformation that diagonalises the time-periodic non-interacting Hamiltonian $H_0(t)$ commutes with the interaction $U^{\dagger}V U=V$ (or, more generally, leaves it time-independent) no inelastic scattering can occur. This exactly corresponds to the case in which $V^{\rm I}(t)$, equation~(\ref{eq:V_interaction_picture}), only picks up the trivial phase dependence due to the difference in quasi-energies.

This toy model is special in two aspects. Each Floquet mode contains only two frequency components, because the Hamiltonian contains only rotating-wave terms.
Moreover, in the internal state basis each internal state component has a single oscillation frequency. As a consequence, inelastic scattering only occurs if these internal states are coupled by the interaction, i.e. if $g_c \ne 0$. If one adds counter-rotating terms to the Hamiltonian,  the Floquet modes do in fact contain all frequency components and inelastic scattering is possible even for $g_c =0$. 

From the discussion of this toy model we draw the following conclusions.
Firstly, the scattering properties of a time-periodic Hamiltonian are not encapsulated by some effective time-independent Hamiltonian: of the infinite set of momentum states that have the same quasi-energy, and therefore could be coupled by scattering, we have found that the rate of coupling depends both on the detailed time-dependence of the Floquet modes and on the structure of the interaction. In particular, the scattering properties of a time-periodic Hamiltonian with frequency $\omega$ cannot be described by an effective ``time-averaged'' Hamiltonian even in the limit $\omega \rightarrow \infty$. Secondly,  we have shown how the Floquet Fermi golden rule may be used to compute transition rates to lowest order in the interaction potential, and that higher order corrections captured by the full Dyson Series can modify the picture emerging from FFGR but do not change the qualitative scattering properties.

%==========================================================================
\section{\label{sec:app}Applications}
%==========================================================================

Following these preliminary considerations, and the development of the formalism of scattering theory for particles in time periodic Hamiltonians, we now turn to discuss applications to situations of physical interest. We shall consider the two-particle scattering processes in cases where the one-particle states are Bloch waves arising from some ``dressed'' states. We consider two cases that are representative of physical implementations that have recently been studied in experiments: the use of Raman coupling of internal states to generate gauge fields in the continuum; and the use of periodic modulation of site energies to form vector potentials on optical lattices.
Our interest will be in the sizes of ``inelastic" two-body scattering processes (which have deleterious effects of heating) as compared to the remaining elastic processes (which are required for the formation of strongly correlated phases). Although our approach may be applied to fermions or bosons, or to two distinguishable particles, and may be extended to any general interaction potentials, for simplicity we focus on the case of bosons with contact interactions.

%==========================================================================
\subsection{\label{subsec:raman}Continuum/Raman coupling model}

We consider a model for the creation of artificial vector potentials in the continuum by dressing of two internal states, $\sigma= \pm$, similar to the experimental implementations\cite{Spielman_Theory,Spielman_Experiment1,Spielman_Experiment2,Spielman_Experiment3,Chinese_group_on_Potassium,Peter_Engels_Washington1,Peter_Engels_Washington2,Cheng_Chin_subbands_1,Cheng_Chin_subbands_2}. However, the states may be either internal (spin) states of the atomic species coupled by optical transitions or subbands of a suppressed additional spatial dimension coupled by a suitable time-periodic potential perturbation \cite{Cheng_Chin_subbands_1,Cheng_Chin_subbands_2}. For simplicity we will mainly treat the system in 1 dimension assuming tight confinement in the other two directions. We will briefly comment on the extension to a 2D model by adding free motion in a second dimension.

%==========================================================================
The system is described by the Hamiltonian
\begin{equation}
\begin{aligned}
 H(t) &= \int dx \, \Psi_{\sigma'}^{\dagger}(x) \left[\frac{p^{2}}{2M} \mathbbold{1}_{\sigma'\sigma} + \mathbb{V}_{\sigma'\sigma}(x,t)\right] \Psi_{\sigma}(x),
\end{aligned}
\end{equation}
where $\Psi^{\dagger}_{\sigma}(x)$ is a creation operator for bosons in internal state $\sigma$. The coupling matrix $\mathbb{V}$ describes the internal dynamics of the atoms interacting with the laser field. It is given by
\begin{equation}\label{eq:raman_couplingmatrix}
 \mathbb{V}(x,t) = \begin{pmatrix} 
               \Delta/2    &\hbar \Omega/2 e^{-i \omega t +2 i k_r x} \\
               \hbar \Omega/2 e^{i  \omega t -2 i k_r x}         &-\Delta/2
              \end{pmatrix},
\end{equation}
with an energy splitting $\Delta$ between internal states and the coupling of strength $\hbar \Omega$ between internal states due to the laser fields taken to be of the rotating wave form. As described in \cite{Spielman_Theory} such a system may arise as the effective two-level description of Raman-coupled spin-states in which case the splitting and coupling strength are also to be understood as effective quantities for the two photon transitions involved.

The  interaction Hamiltonian is
\begin{equation}
\begin{aligned}
 H_{\text{int}} =\frac{1}{2} \int dx \, &\sum_{\sigma} g_{\sigma} \Psi_{\sigma}^{\dagger}(x)\Psi_{\sigma}^{\dagger}(x) \Psi_{\sigma}(x) \Psi_{\sigma}(x)\\
                 + g_2 &\sum_{\sigma} \Psi_{-\sigma}^{\dagger}(x)\Psi_{\sigma}^{\dagger}(x) \Psi_{-\sigma}(x) \Psi_{\sigma}(x)\\
                   + g_c &\sum_{\sigma}\Psi_{\sigma}^{\dagger}(x)\Psi_{\sigma}^{\dagger}(x) \Psi_{-\sigma}(x) \Psi_{-\sigma}(x)
\end{aligned}
\end{equation}
describing general contact interactions with species-dependent strength $g_{\sigma}$, inter-species coupling with strength $g_2$ and species-changing coupling with strength $g_c$.
The relative sizes of these coupling depend on the physical origin of the two internal states. For two (hyperfine) spin states, the $g_c$ term does not conserve the spin projection and is therefore is not present if spin-rotation symmetry is preserved\footnote{We note that for strong dipolar interactions, such processes are allowed for spin states, with the internal angular momentum taken up by relative orbital angular momentum of the two particles.}. However, if the two internal states are two states of position motion -- for example two vibrational subbands\cite{Cheng_Chin_subbands_1,Cheng_Chin_subbands_2}  -- then $g_c$ is proportional to the usual contact interaction modified by a geometric factor describing the wave function overlap between bands. In the specific case of Ref.\cite{Cheng_Chin_subbands_1,Cheng_Chin_subbands_2} the two internal states are the $s$- and $p$-bands of an optical lattice, and the couplings are $g_{\sigma} \propto g \int dx \, \left \vert w_{\sigma}(x)\right \vert^4$, $g_2 \propto g \int dx \,\left \vert w_s(x)\right \vert^2\left \vert w_p(x)\right \vert^2$ and $g_c \propto g \int dx \, w^{*}_s(x)w^{*}_s(x) w_p(x)w_p(x)$ where $g$ is the appropriate 1 dimensional contact interaction strength and $w_{\sigma}$ the Wannier orbital of the band $\sigma=s(p)$. Thus the couplings ($g_\sigma$, $g_2$ and $g_c$) are all non-zero and of comparable magnitude.
\subsubsection{Single-particle states}
Our discussion of the single particle states follows the one given in \cite{Gauge_fields_Review} with the main exception that the explicit time-dependence of the states is kept within the Floquet theory description.

The non-interacting Hamiltonian $H_0$ couples only two components and can be expressed with respect to the operators $\phi^{\dagger}_{1}(k) = \phi^{\dagger}_{1,k+ k_r}$ and $\phi^{\dagger}_{2}(k) = \phi^{\dagger}_{2,k- k_r} e^{i \omega t}$, where $\phi^{\dagger}_{\sigma,k}$ creates an internal state $\sigma$ particle in a plane-wave state. Then, $H_0$ reduces to a sum over independent $2\times2$ blocks of the form
\begin{equation}
 H_0(k)=\begin{pmatrix}
             \frac{\hbar^2(k+k_r)^2}{2M}-\hbar \delta/2      &   \hbar \Omega/2 \\
              \hbar \Omega/2                                  &   \frac{\hbar^2(k-k_r)^2}{2M} + \hbar \delta/2 \\
          \end{pmatrix}  
\end{equation}
where $\hbar \delta/2=\hbar \omega/2 - \Delta/2$ and we dropped an overall constant energy shift $\hbar \omega/2$. Note that this implies that the eigenstates will be mixtures of different (internal) states at different momenta where the composition will depend on the quasi-momentum $k$.
We choose the recoil energy $E_r=\hbar^2 k_r^2/2M$ as the unit of energy and $k_r$ as the unit of momentum defining dimensionless parameters $\tilde{\Omega}=\hbar \Omega/ E_r$, $\tilde{\delta}=\hbar \delta/ E_r$ and $\tilde{k}=k/k_r$. The Hamiltonian becomes
\begin{equation}
 H_0(k)=E_r \begin{pmatrix}
             (\tilde{k}+1)^2 -  \tilde{\delta}/2&  \tilde{\Omega}/2 \\
               \tilde{\Omega}/2             & (\tilde{k}-1)^2 + \tilde{\delta} /2\\
          \end{pmatrix}
\end{equation}
and the Floquet modes are 
\begin{equation}
 \Phi^{m}_{k,\tau}= N_{k,\tau} \begin{pmatrix}  \tilde{\delta} + 4\tilde{k} +\tau \sqrt{\tilde{\Omega}^2 +\left(4 \tilde{k}-\tilde{\delta} \ \right)^2}  \\
                        \tilde{\Omega} 
                       \end{pmatrix}e^{i m \omega t}
%              = \begin{pmatrix}
%                   a_{\tau}(k)\\
%                   b_{\tau}(k) 
%                \end{pmatrix}
%                 \, .
\end{equation}
with the normalisation factor 
\begin{equation}
N_{k,\tau}= 1/\sqrt{L} \left[ \left(\tilde{\delta} + 4\tilde{k} +\tau \sqrt{\tilde{\Omega}^2 +\left(4 \tilde{k}-\tilde{\delta}  \right)^2}\right)^{2}+ \tilde{\Omega}^{2} \right]^{-1/2} 
\end{equation}
for a system of size $L$. Defining the components of the $\tau=\pm$ eigenvector with respect to $\phi^{\dagger}_{1}(k)$, $\phi^{\dagger}_{2}(k)$ as $a_{\tau}(k)$ and $b_{\tau}(k)$ respectively these states read in the original basis as
\begin{equation}
\Phi^m_{k,\tau} (x)= \begin{pmatrix} 
		      a_{\tau}(k) e^{i k_r  x} \\
		      b_{\tau}(k) e^{-i k_r x}e^{i \omega  t }
		      \end{pmatrix}
		      e^{i(k x+ m \omega t)}
 \, .
\end{equation}
Note at this point that the coupling in the rotating wave approximation leads to a wave function in which each component in the internal state basis has a single oscillation frequency, i.e. $ m \, \omega$ and $(m+1) \, \omega$ for the components of  $\Phi^m_{k,\tau}$. As a consequence, inelastic scattering processes can only occur if these internal states are coupled by the interaction as has been mentioned in the discussion of the toy model before.

%%%%%%%%%%%%%%%%%%%%%%%%%%%%%%%%%%%%%%%%%%
    \begin{figure}
    \subfloat%[First sub-figure\label{subfig-1:dummy}]
    {%
      \includegraphics[width=0.22\textwidth]{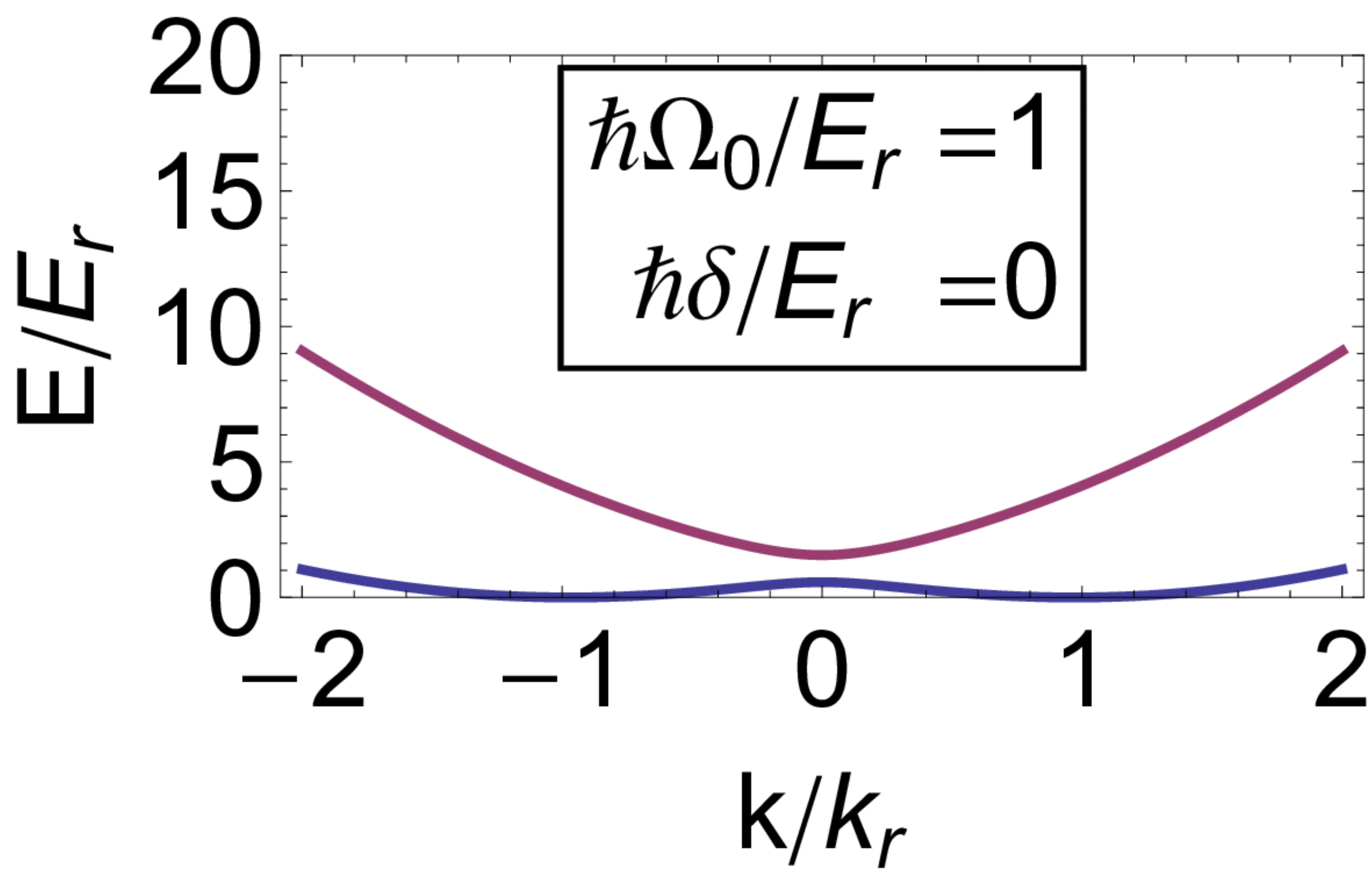}
    }
    \hfill
    \subfloat%[First sub-figure\label{subfig-2:dummy}]
    {%
      \includegraphics[width=0.22\textwidth]{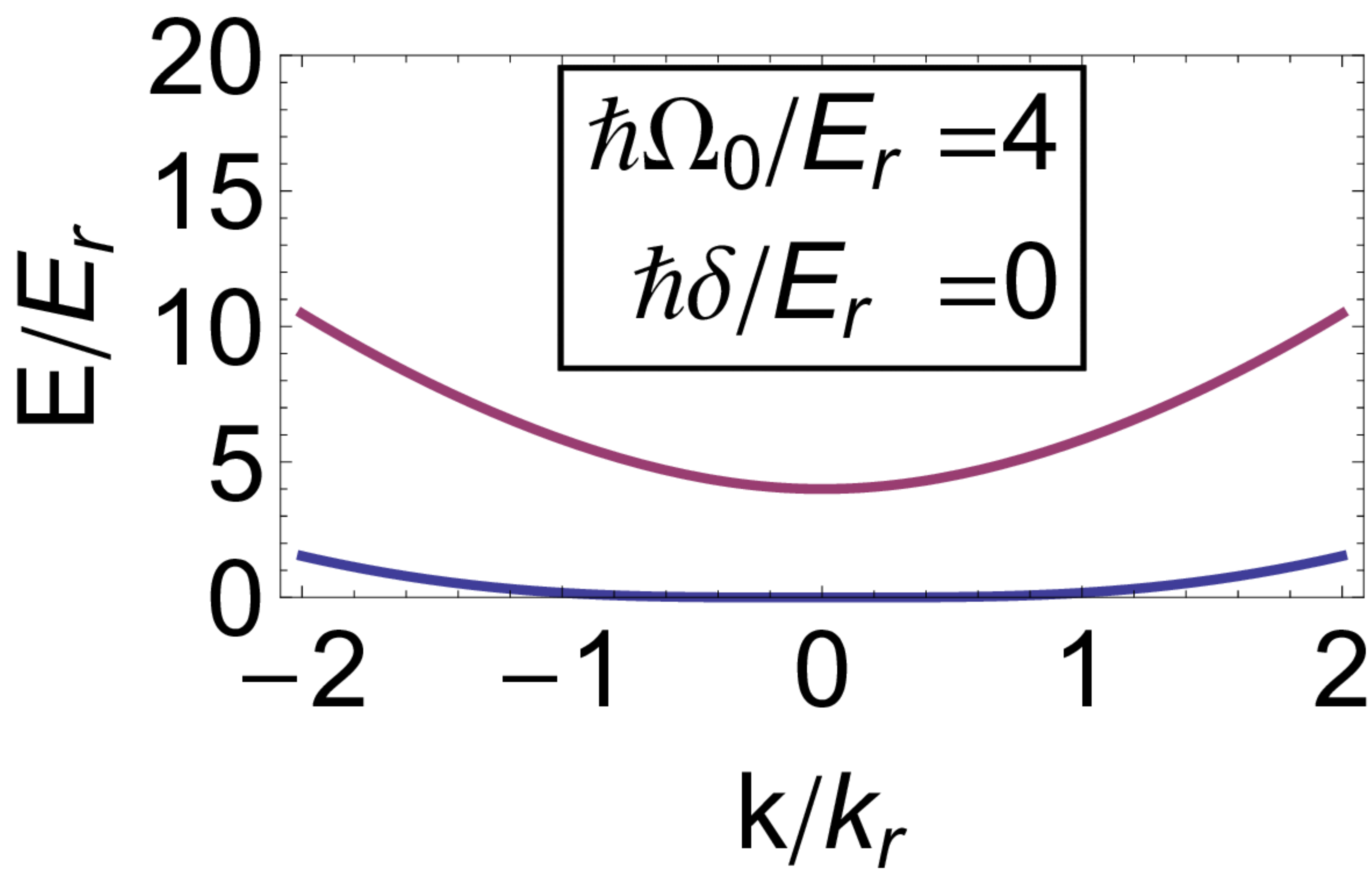}
    }
    \\
    \subfloat%[First sub-figure\label{subfig-1:dummy}]
    {%
      \includegraphics[width=0.22\textwidth]{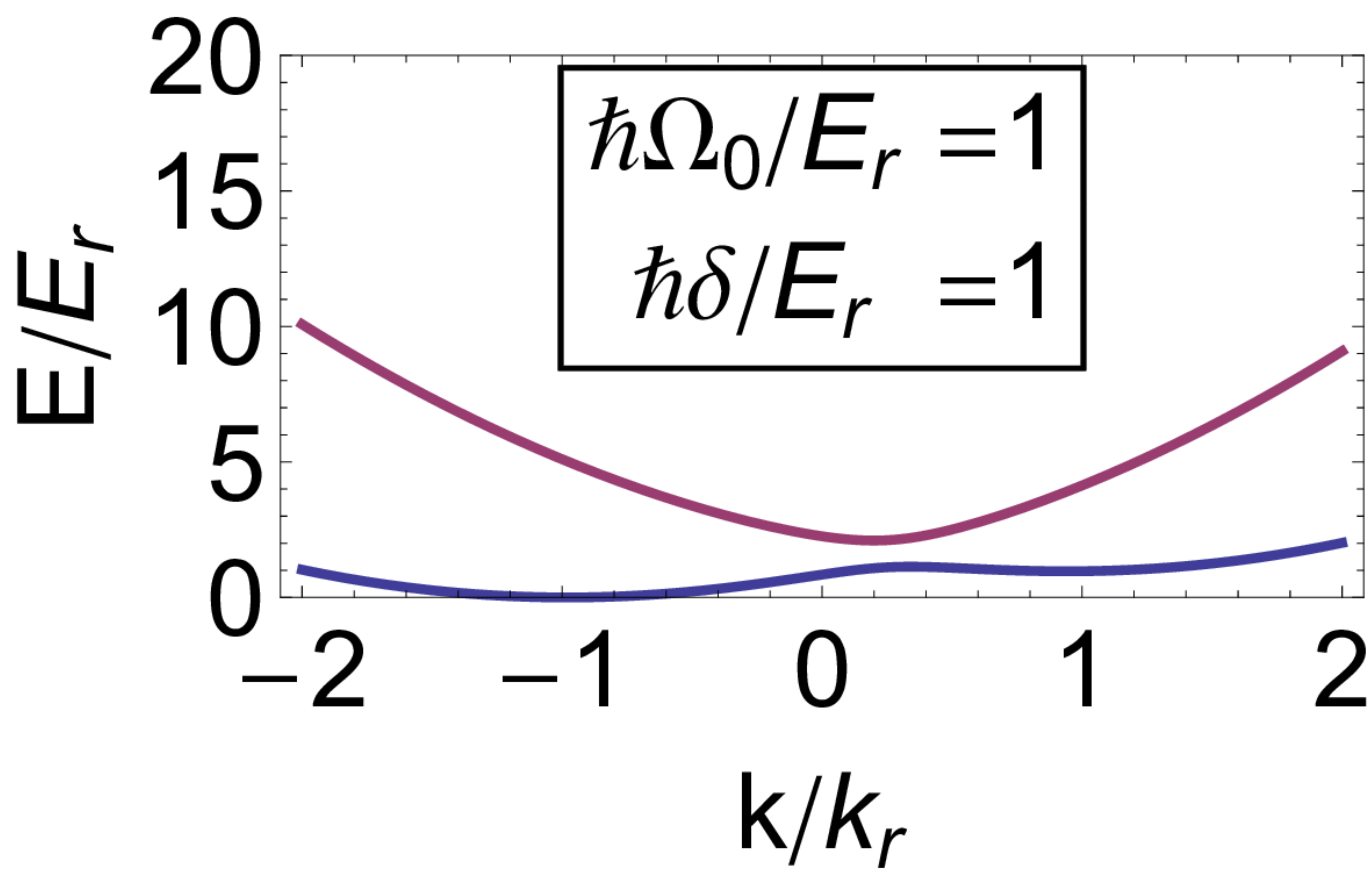}
    }
    \hfill
    \subfloat%[First sub-figure\label{subfig-2:dummy}]
    {%
      \includegraphics[width=0.22\textwidth]{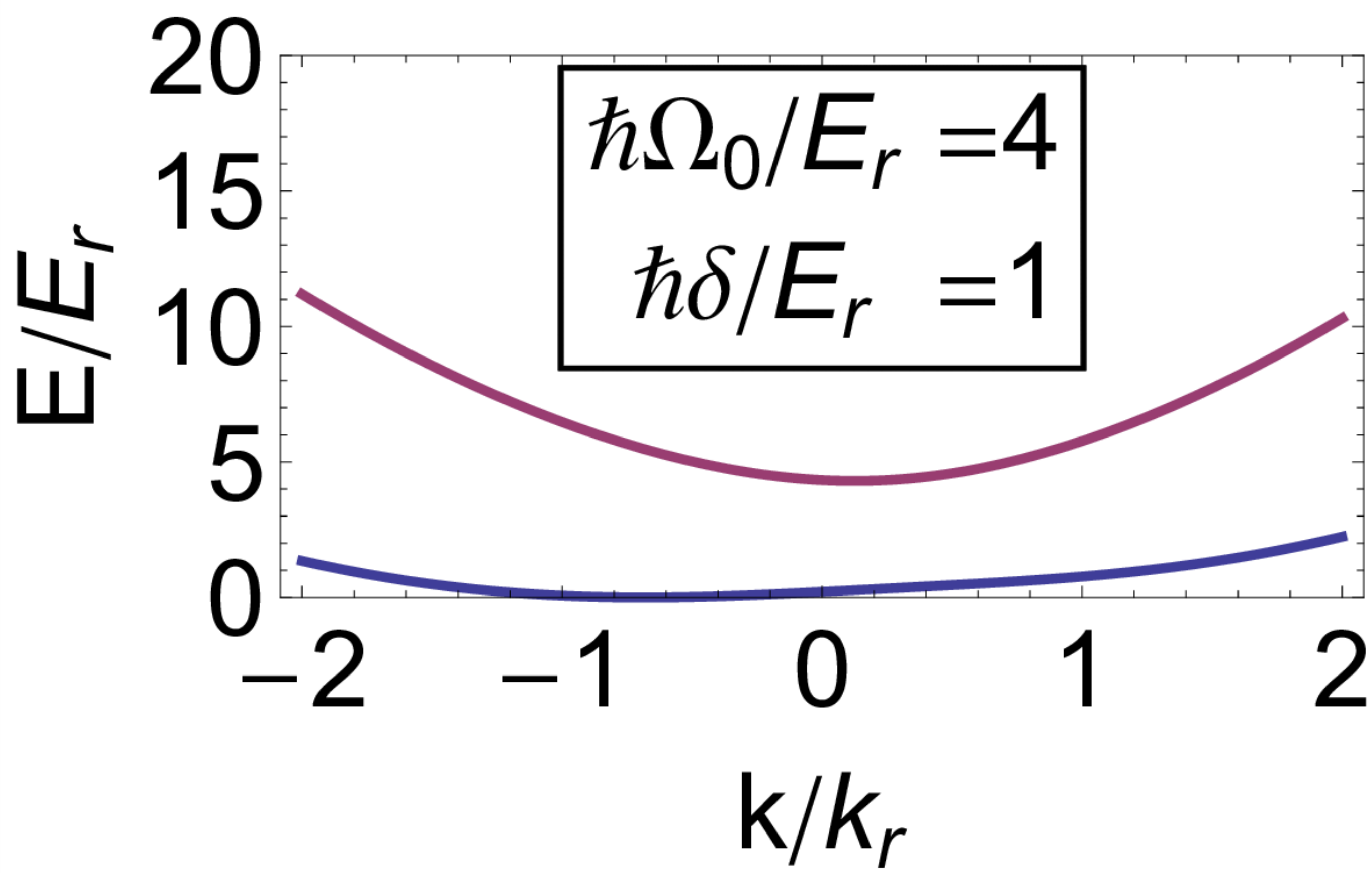}
    }
    \caption{(Color online) Dispersion $E/E_r$ as a function of $k/k_r$, see Equation~(\ref{eq:raman_quasienergies}), including an energy offset to have zero minimum. The top row shows the dispersion for $\hbar \delta/E_r=0$ and $\hbar \Omega/E_r=1$ to the left and $\hbar \Omega/E_r=4$ to the right and the bottom row for the same parameters in the case of $\hbar \delta/E_r=1$. }
    \label{fig:raman_dispersion}
  \end{figure}
  %%%%%%%%%%%%%%%%%%%%%%%%%%%%%%%%%%%%%%%%
The corresponding quasi-energies are
\begin{equation}\label{eq:raman_quasienergies}
 \epsilon^m_{k,\tau}=E_r \left[ \tilde{k}^2 +  \tau \sqrt{\tilde{\Omega}^2 +\left(4 \tilde{k} - \tilde{\delta} \right)^2}  \right]+ m \hbar \omega .
\end{equation}
The dispersion for different characteristic values of the parameters is shown in Fig.~\ref{fig:raman_dispersion}. Firstly, for no detuning $\hbar \delta/E_r=0$ both bands are symmetric around $k/k_r=0$. The character of the lower band changes as a function of $\hbar \Omega/E_r$ as discussed in \cite{Spielman_Experiment3}. For $\hbar \Omega/E_r <4$ (left column of the Fig.~\ref{fig:raman_dispersion}) it has three distinct extrema of which the one at $k/k_r=0$ is a maximum and two global degenerate minima at $k/k_r=\pm |k_0|$. For $\hbar \Omega/E_r \ge 4$ (right column of Fig.~\ref{fig:raman_dispersion}) it only has a single global minimum at $k/k_r=0$. The gap between upper and lower band is in both cases given by $\hbar \Omega$. For non-zero detuning, both bands become skewed lifting the symmetry under $k \rightarrow -k$ and the degeneracy between the minima of the lower band present for $\hbar \Omega/E_r <4$. 
The shift of the minimum of the dispersion to non-zero $k$ can be interpreted as the effect of a non-zero electromagnetic vector potential\cite{dalibardreview}.% As this shift depends on the couplings and detunings of the laser fields, this can be designed in such a way as to give rise to a non-vanishing magnetic field configuration for the adiabatic motion of particles in the lower band \cite{Spielman_Theory}. 

Note that the non-interacting Hamiltonian $H_0$ is invariant under two continuous symmetry operations: modified spatial translations generated by $\mathcal{O}_{\hat{p}}= \mathbbold{1} \hat{p} + k_r \sigma_z$, and modified temporal translations generated by $\mathcal{O}_{\hat{E}}= \mathbbold{1} i \hbar \partial_t -\hbar \omega \sigma_z/2$ where $\sigma_z= \mbox{diag}(1,-1)$ denotes the third Pauli matrix. The corresponding finite symmetry operations are translations multiplied by a state dependent phase factor, $\mbox{diag}(e^{i k_r a},e^{-i k_r a}) T_{x\rightarrow x+a}$ and $\mbox{diag}(e^{-i \omega a/2},e^{i \omega a/2}) T_{t \rightarrow t+a}$. These symmetries imply the conservation of both the momentum $k$ and the energy $E$ and the single-particle eigenstates. The single-particle states can thus be characterised by their momentum $k$ and energy $E$, both of which can take unbounded values: there is no BZ for momentum or energy, owing to the existence of the continuous symmetry. It is for this reason that no BZ structure (in energy of momentum) appears in  Fig.~\ref{fig:raman_dispersion}.

\subsubsection{Two-body scattering}

We now study whether, through their mutual interaction, two particles
that both start in plane-waves states in the lower band can undergo
scattering into the higher band or scattering into higher
quasi-momentum states in the same band via the absorption of energy
quanta $\hbar \omega$ from the time-dependent fields.

In discussing the two-body scattering, it is interesting to consider
the interplay of the interparticle interaction $H_{\text{int}}$ and
the above finite symmetry operations.  One finds that these symmetries
commute with the $g_{\sigma}$ and $g_2$ terms, but that both
symmetries are broken by the $g_c$ coupling term. However there
remains a discrete symmetry, namely, when including the $g_c$
interaction term the full Hamiltonian is still invariant under
discrete spatial and temporal translations by $x \rightarrow x+
\pi/k_r$ and $t \rightarrow t+ 2\pi/\omega$. Thus, while in the
non-interacting model both momentum $k$ and energy $E$ are strictly
conserved, in the presence of the $g_c$ interaction term scattering
processes that change the momentum by integer multiples of $2 k_r$ and
the energy by multiples of $ \hbar \omega$ are allowed; or, put
differently, for $g_c\neq 0$ only quasi-momentum and quasi-energy
remain conserved quantities.

We compute the two-body scattering rate
using the FFGR (\ref{eq:ffgr}). The initial state consists of two particles in the lower band  $\Phi^m_{k,-}$ with their momentum centred at the minimum $k_0$ of the single-particle dispersion, i.e. $k_1=k_0+k$ and $k_2=k_0-k$ respectively, i.e.
\begin{equation}
 |\psi_{\rm i}\rangle =\Psi^{\dagger}_{k_1,-} \Psi^{\dagger}_{k_2,-}\vac ,
\end{equation}
and the final state with two particles in any of the bands with momentum $q_1$ and $q_2$
\begin{equation}
 |\psi_{\rm f}\rangle =|\psi_{q_1,\tau_1;q_2;\tau_2}\rangle=\Psi^{\dagger}_{q_1,\tau_1} \Psi^{\dagger}_{q_2,\tau_2}\vac .
\end{equation}
The general two-particle Floquet mode can be written as a four-component spinor in the basis of (internal) states ($\ket{1} \ket{1}$, $\ket{1} \ket{2}$, $\ket{2} \ket{1}$ and $\ket{2} \ket{2}$) as
\begin{eqnarray}
\nonumber
 \Phi^{m}_{\tau_1,k_1;\tau_2,k_2} & =&  P\begin{pmatrix}
                           a_{\tau_1}(k_1) a_{\tau_2}(k_2) e^{i k_r (x+y)}\\
                            a_{\tau_1}(k_1) b_{\tau_2}(k_2)  e^{-i k_r (y-x)} e^{i \omega t}\\
                            b_{\tau_1}(k_1) a_{\tau_2}(k_2) e^{-i k_r (x-y)}  e^{i \omega t} \\
                            b_{\tau_1}(k_1) b_{\tau_2}(k_2) e^{-i k_r (x+y)} e^{2 i \omega t}
                          \end{pmatrix}  \\
 & & \times e^{ i k_1 x + i k_2 y} e^{i m \omega t}
\label{eq:raman_2_particle_state}
\end{eqnarray}
where $P$ denotes symmetrisation of the wave function under exchange of single-particle quantum numbers as is appropriate for the bosonic particles considered.

As we are interested in inelastic processes with the absorption of a non-zero number of photons, the relevant matrix element is $\brakett{\Phi^{m}_{\rm f}}{ H_{\text{int}}}{\Phi^{0}_{\rm i}}$ for non-zero $m$. Therefore, the usual scalar product $\braket{\Phi^{m}_{\rm f}}{ H_{\text{int}}}{\Phi^{0}_{\rm i}}$ contains an overall oscillating factor of $\exp[-i m \omega t]$ stemming from the last factor in equation~(\ref{eq:raman_2_particle_state}). This factor can only be cancelled to yield a non-zero time-average if different components of the spinors are coupled by $H_{\text{int}}$.
Thus, the only relevant coupling for inelastic scattering is the one given by $g_c$ coupling the states $\ket{1} \ket{1}$ to $\ket{2} \ket{2}$. For this process the energy of exactly two two-photon transitions, i.e. $m=2$ in equation~(\ref{eq:ffgr}), is absorbed; simultaneously,  the centre of mass momentum changes by $4 k_r$, owing to the fact that the running waves providing the Raman coupling cause a momentum transfer of $2 k_r$ for each two-photon absorption.

%%%%%%%%%%%%%%%%%%%%%%%%%%%%%%%%%%%%%%%%%%%%%%%%%%%%%
\begin{figure} 
\centering 
\subfloat{\subfigimg[hsep=0.07\textwidth,width=0.45\textwidth]{\bf (a)}{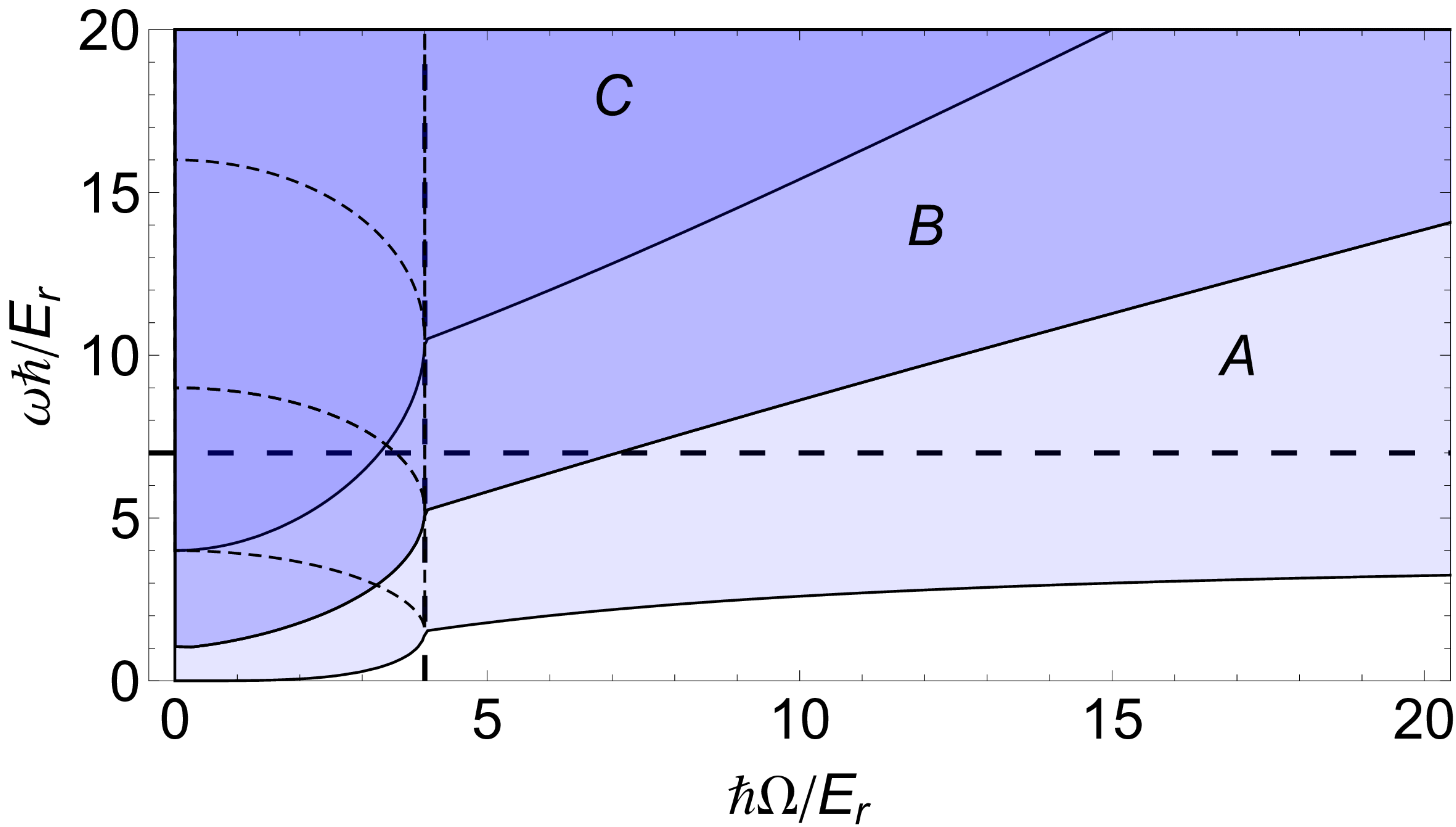}}
\hfill
\subfloat{\subfigimg[hsep=0.07\textwidth,width=0.45\textwidth]{\bf (b)}{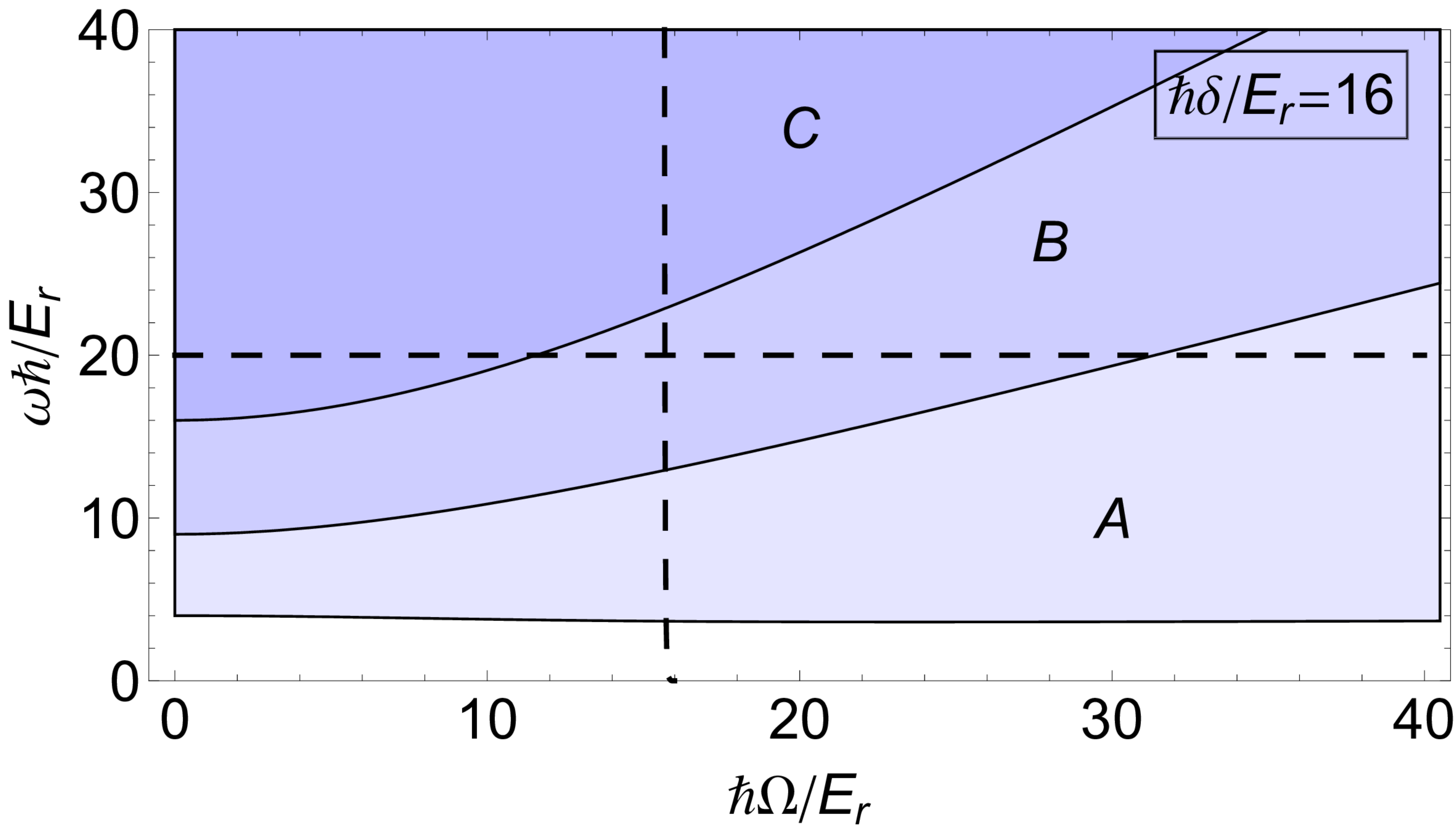}} 
\caption{(Color online) Stability diagram for  (a) $\hbar \delta/E_r=0$  and (b)  $\hbar \delta/E_r=16$  with initial state of two particles with quasi-momentum $k=k_0$ in lower band. For no detuning and $\hbar \Omega/E_r<4$ both cases of $k=+ |k_0|$ (dashed lines) and $k=-|k_0|$ (full lines) are shown, with detuning the minimum is unique. Shaded regions correspond to parameter regimes in which inelastic scattering is allowed. The bottom region (A) in light blue corresponds to inelastic scattering where both particles remain in the lower bands, i.e. the $(-,-)$-final state, in the middle region (B) particles can scatter either into the $(-,-)$ or the $(+,-)$ -final state and in the top region (C) scattering into all states $(-,-)$,$(+,-)$ and $(+,+)$ is allowed. The thick dashed lines of constant $\omega \hbar /E_r$ and of constant $\Omega \hbar /E_r$ correspond to the cuts along which the scattering rate is shown in Fig.~\ref{fig:raman_scatteringrate} and Fig.~\ref{fig:raman_scatteringrate_2d}.
\label{fig:raman_stability}}
\end{figure}
%%%%%%%%%%%%%%%%%%%%%%%%%%%%%%%%%%%%%%%%%
In the following we focus on the results for the case of two low-energy particles, $k\to 0$. Exploiting the quasi-energy conservation in the FFGR and the constraint on the final centre of mass momentum those parameter regions in which such inelastic processes are allowed energetically can be derived and are shown in Figure~\ref{fig:raman_stability}. Generally, as $\Omega$ increases the gap to the higher band increases as well as making transitions from the initial state in $(-,-)$ to those with at least one excited particle in a higher band [$(+,-)$, $(-,+)$ or $(+,+)$] impossible as only an energy of $2 \hbar \omega$ is available. 
Conversely as $\omega$ is increased there is a threshold above which particles can be excited into higher bands. 
For no detuning $\hbar \delta/E_r=0 $ the dispersion is symmetric and has two degenerate minima for $\hbar \Omega/E_r<4$ at  $k=\pm |k_0| \ne 0$. Therefore, there are two distinct initial states with $k=\pm |k_0|$ and stability regions for both cases are shown. The thin dashed lines that split off and go up for $\hbar \Omega/E_r<4$ corresponds to $k=+ |k_0|$ and the bold that go down to $k=- |k_0|$. This is readily explained by the fact that for $k=- |k_0|$ particles starting in the left minimum of the dispersion get scattered close to the right minimum when increasing their crystal momentum by $2k_r$ and therefore have a lower energy and threshold $\omega$. This breaking of the symmetry $k\rightarrow -k$ that is apparent in the dispersion is due to the fact the the coupling matrix in equation~(\ref{eq:raman_couplingmatrix}) contains a running wave term $\exp[-i \omega t +2 i k_r x]$ which explicitly sets a direction in space.
With detuning the degeneracy is lifted and the minimum is unique for all parameter values.

\begin{figure}
    \subfloat%[First sub-figure\label{subfig-1:dummy}]
    {%
      \includegraphics[width=0.22\textwidth]{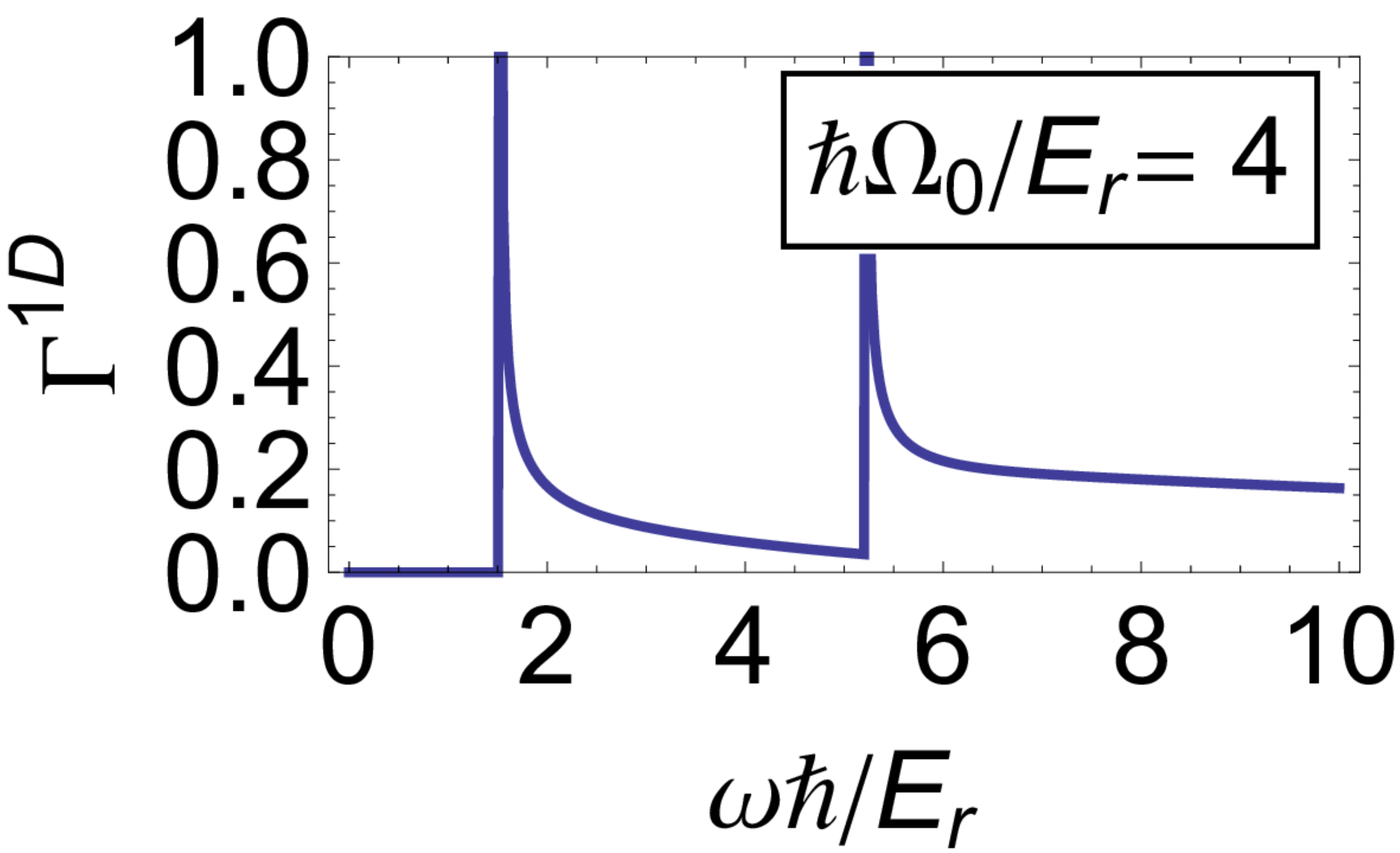}
    }
    \hfill
    \subfloat%[First sub-figure\label{subfig-2:dummy}]
    {%
      \includegraphics[width=0.22\textwidth]{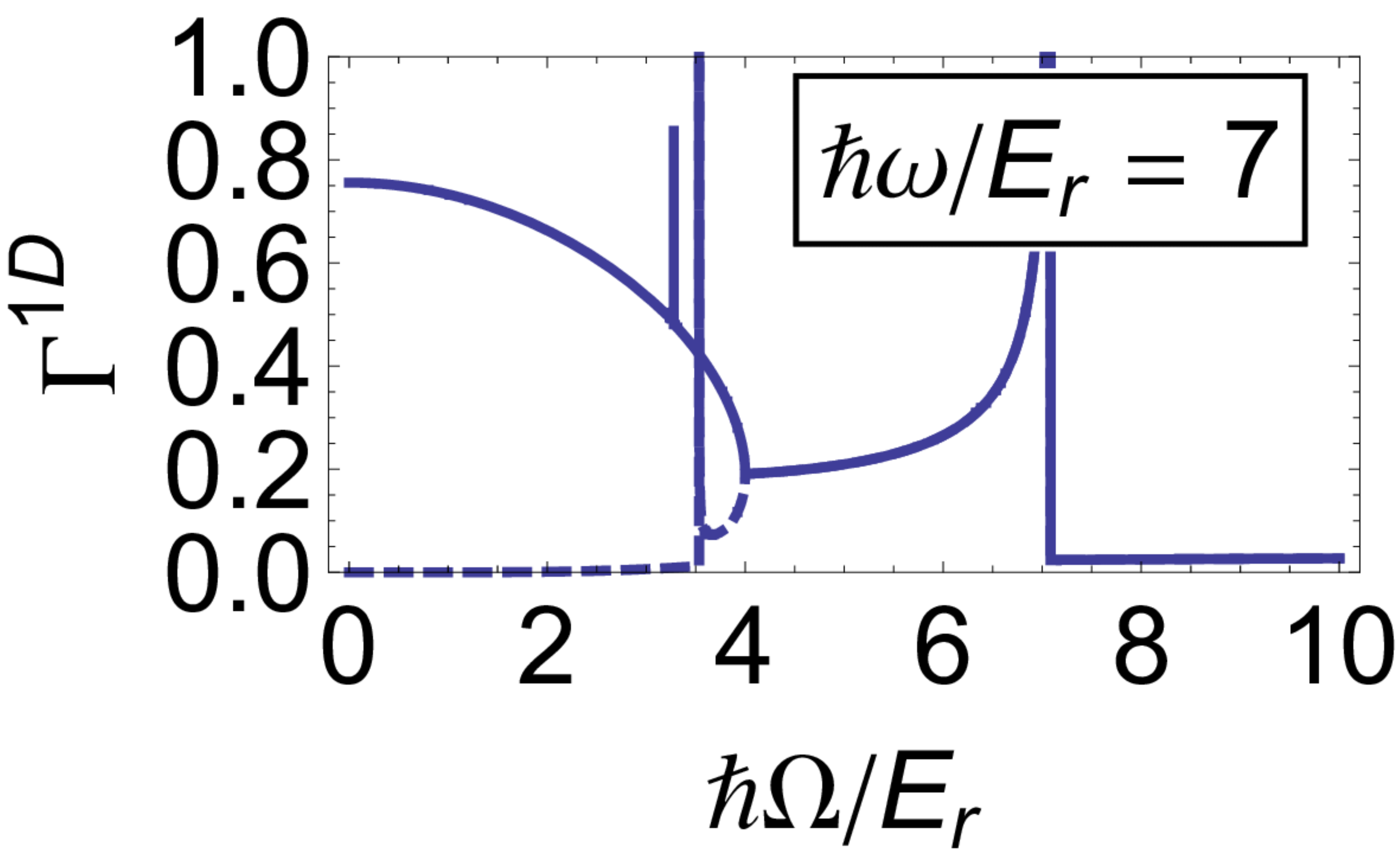}
    }
    \hfill
   \subfloat%[First sub-figure\label{subfig-2:dummy}]
    {%
      \includegraphics[width=0.22\textwidth]{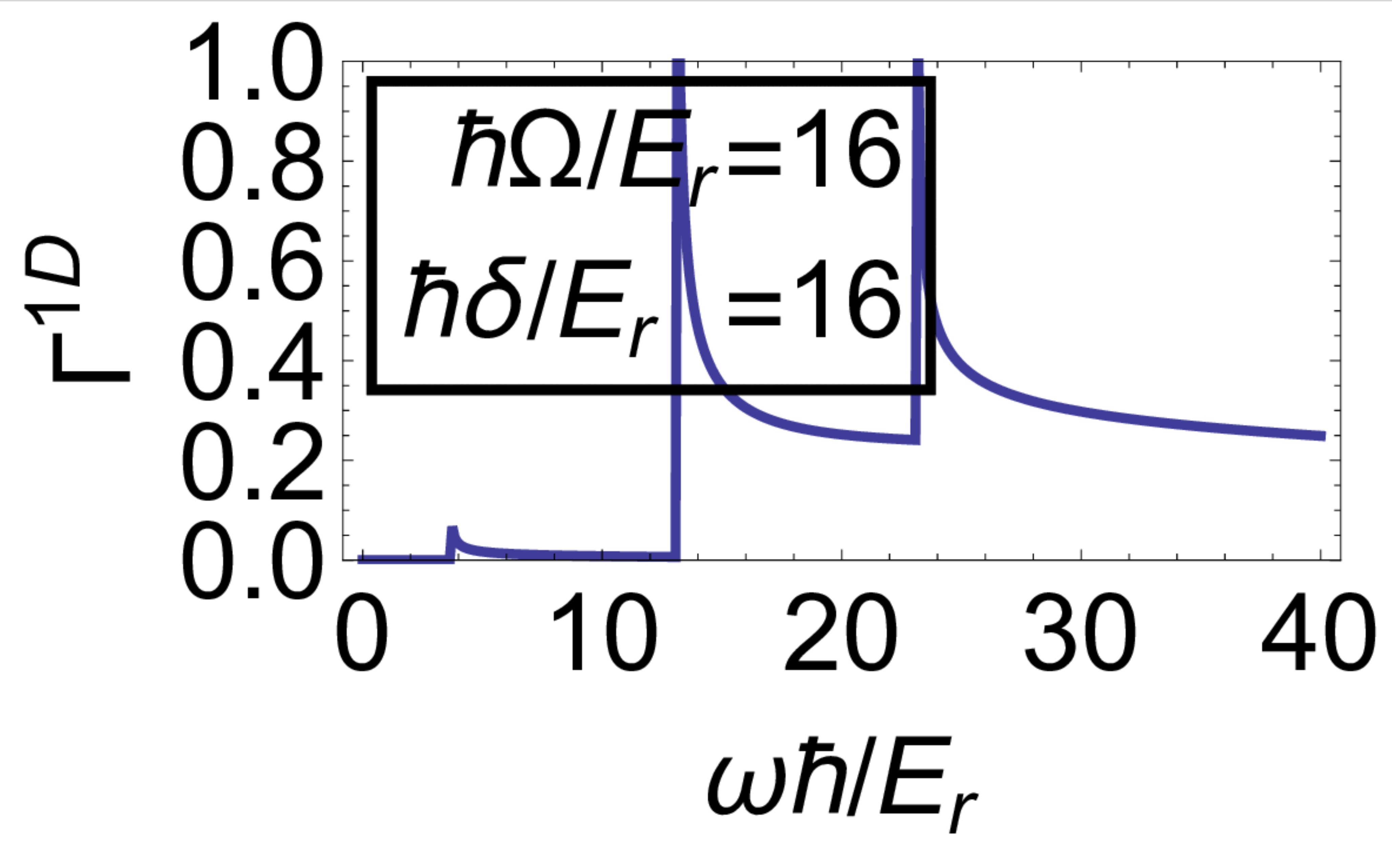}
    }
    \hfill
       \subfloat%[First sub-figure\label{subfig-2:dummy}]
    {%
      \includegraphics[width=0.22\textwidth]{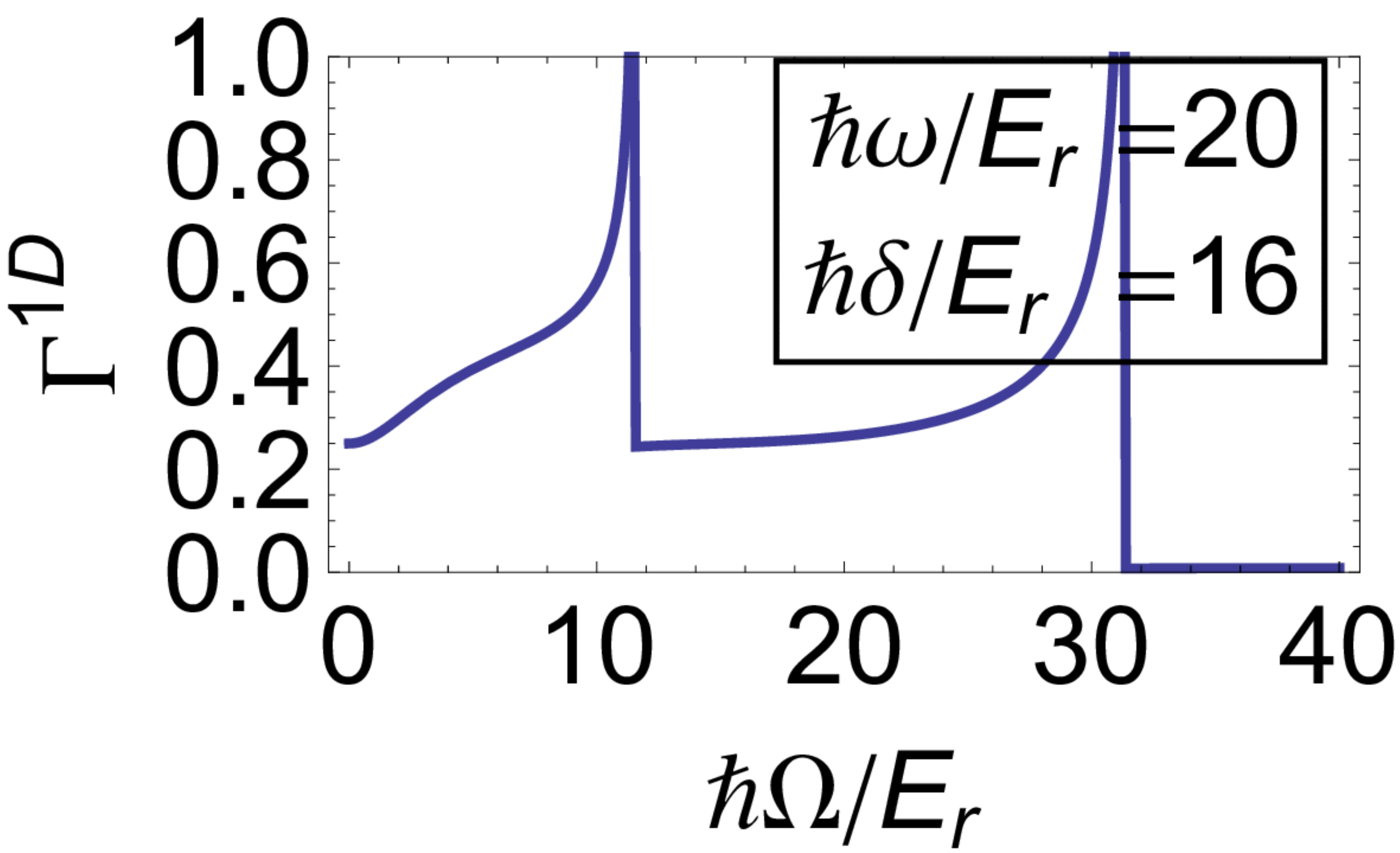}
    }
    \caption{(Color online) Dimensionless scattering rate $\Gamma^{\rm 1D}$, see Equation~(\ref{eq:raman_scatteringrate}) for an initial state with particles in the lower band with momentum $k=\pm |k_0|$ getting inelastically scattered. In the top row for no detuning $\hbar \delta/E_r=0$ as a function of $\omega \hbar /E_r$ for fixed $\hbar \Omega /E_r = 4$ to the left and as a function of $\hbar \Omega /E_r$ for fixed $\omega \hbar /E_r=7$ to the right (dashed $k=+|k_0|$ , full $k=-|k_0|$) and at the bottom for $\hbar \delta/E_r=16$ as a function of $\omega \hbar /E_r$ for fixed $\hbar \Omega /E_r = 16$ to the left and as a function of $\hbar \Omega /E_r$ for fixed $\omega \hbar /E_r=20$ to the right as indicated in Fig.~\ref{fig:raman_stability}. The rate shows divergences at the opening/closing of scattering channels corresponding to the borders in Fig.~\ref{fig:raman_stability} at which the density of states of the final states diverges.}
    \label{fig:raman_scatteringrate}
  \end{figure}

From these considerations the allowed final states can be parametrised as $\Phi^{m}_{\rm f}(q)=\Phi^{m}_{\tau_1,k_0+2k_r+q;\tau_2,k_0 +2k_r-q}$. To compute the total scattering rate one integrates over these final states
\begin{equation}\label{eq:raman_scatteringrate}
 \begin{aligned}
  \frac{dn}{dt} &= \frac{2 \pi} {\hbar} \sum_{\tau_1\tau_2;m\ne0}\frac{L}{2\pi} \int dq\, \left\vert \brakett{\Phi^{m}_{\rm f}(q)}{ H_{\text{int}}}{\Phi^{0}_{\rm i}} \right \vert^{2} \\
                &\quad \times \delta(\epsilon_{\rm i}-\epsilon_{\rm f}(q)-m \hbar \omega) \\
                &= \frac{1} {\hbar L} \frac{k_r}{E_r} \sum_{\tau_1\tau_2;m\ne0}L^2 \int d\epsilon_{\rm f} \frac{d(q/k_r)}{d(\epsilon_{\rm f}/E_r)}  \\
                & \quad \left\vert \brakett{{\Phi^{m}_{\rm f}}(q)}{ H_{\text{int}}}{\Phi^{0}_{\rm i}} \right \vert^{2} \delta(\epsilon_{\rm i}-\epsilon_{\rm f}-m \hbar \omega)\\
                &=\frac{g_c^{2}} {\hbar L} \frac{k_r}{E_r} \Gamma^{\rm 1D}
 \end{aligned}
\end{equation}
which defines the intensive dimensionless scattering rate $\Gamma^{\rm 1D}$ for inelastic processes. Due to the dependence on the 1D-density of states $\Gamma^{\rm 1D}$ will diverge at the borders of the stability regions in Fig.~\ref{fig:raman_stability} whenever a scattering channel opens or closes and the density of states of the final state diverges. A plot of $\Gamma^{\rm 1D}$ for characteristic parameter values is shown in Fig.~\ref{fig:raman_scatteringrate} with these divergences clearly visible. Away from those points the dimensionless rate is $\Gamma^{\rm 1D} \approx 0.2-0.4$.
We remark again that for coupled spin-states $g_c=0$ and no inelastic scattering occurs, whereas for subbands $g_c \ne 0$ generally.
%%%%%%%%%%%%%%%%%%%%%%%%%%%%%%%%%%%%%%%%%%
  %%%%%%%%%%%%%%%%%%%%%%%%%%%%%%%%%%%%%%%%%%%%%%%%%%%%%%%  

  The elastic scattering rates are comparatively easier to compute. For simplicity we focus on the case of spin-independent coupling strengths $g_{\sigma}=g_2=g$, no detuning $\tilde{\delta}=0$ and consider the limit of $k\rightarrow 0$ for which the leading behaviour can be given explicitly. A more detailed discussion of the elastic scattering properties can be found in \cite{Spielman_Experiment2}.
  
  As mentioned before the non species changing interaction terms ($g_{\sigma}$ and $g_2$) conserve the total momentum and only the relative momentum can be changed during scattering. Moreover, in first order $g_c$ does not contribute to the elastic scattering rate as it always changes both the energy and the total momentum of the colliding particles. Neglecting higher order effects of $g_c$ and in the limit of $k\rightarrow 0$ the particles behave like spinless bosons with a modified dispersion relation interacting via a contact interaction and all differences that occur in their elastic scattering is entirely due to density of states effects.
 
 The elastic scattering rates within FGR are given by
  \begin{equation}\label{eq:raman_el_scatteringrate}
    \begin{aligned}
   \Gamma_{el}&= \frac{1}{(\tilde{\Omega}/4)^2-1} \frac{1}{k/k_r} && \text{for }\tilde{\Omega} < 4\\
   \Gamma_{el}&= \frac{2}{(k/k_r)^3}&&\text{for }\tilde{\Omega} = 4\\
   \Gamma_{el}&= \frac{1}{1-4/\tilde{\Omega}} \frac{1}{k/k_r} &&\text{for }\tilde{\Omega} > 4
  \end{aligned}
  \end{equation}
  where in the case of $\tilde{\Omega} = 4$ the dispersion is quartic $\epsilon(k) \propto k^4$ and consequently the divergence is $1/k^3$ instead of the ususal $1/k$ for a parabolic dispersion. We emphasize again that these rates are the same as for undressed particles with the modified dispersion interacting via a contact interaction. In the limit of $k\rightarrow 0$ the dressing of particles only changes the dispersion and the density of states, not the interactions themselves.
  
  To relate both the inelastic and elastic scattering rates to the corresponding cross sections, the rates have to be divided by the incident flux. 
For simplicity we again focus on the case of $\tilde{\delta}=0$ and the limit of $k\rightarrow 0$ for which the incoming flux for our initial state is
 \begin{equation}\label{eq:raman_flux}
 \begin{aligned}
   J_{in}&= \frac{2\hbar k}{M L}   \left[1-\left(\frac{\tilde{\Omega}}{4} \right)^2\right]  && \text{for }\tilde{\Omega} < 4\\
   J_{in}&= \frac{2\hbar k}{M L}   \left(\frac{k}{2 k_r} \right)^2 &&\text{for }\tilde{\Omega} = 4\\
   J_{in}&= \frac{2\hbar k}{M L}   \left[1-\frac{4}{\tilde{\Omega}} \right] &&\text{for }\tilde{\Omega} > 4
  \end{aligned}
 \end{equation}
For $\tilde{\Omega}\ne 4$ these factors together with Equation~(\ref{eq:raman_el_scatteringrate}) and Equation~(\ref{eq:raman_scatteringrate}) give a divergence of $1/k^2$ and $1/k$ for the elastic and inelastic scattering cross section respectively in the same way as discussed in the toy model above. The case of $\tilde{\Omega} = 4$ is special as the dispersion then becomes quartic $\epsilon(k) \propto k^4$. The corresponding $1/k^6$ and $1/k^3$ behaviour of the elastic and inelastic scattering cross sections is entirely due to the dispersion and density of states effects and is not related to the dressing of the states. In both cases the divergence at low $k$ signals a failure of the Born-Approximation.

Following the discussion of the toy model, in 1D the elastic rate should dominate over the inelastic rate at low $k$ and at lower overall interaction strengths when the ratio of $g_c/g$ is kept fixed.

  \subsubsection{Extension to two dimensions}

The above model has motion only along one dimension, as relevant for
the motion along tubes with transverse confinement frequencies large
compared to $\omega$. In systems with weak confinement in the
transverse directions there are additional inelastic scattering
channels. Here we consider the case of a two-dimensional system as is
required for generate non-vanishing effective magnetic field. For now,
we ignore any spatial dependence of the laser fields along the second
direction, which we denote $y$. The setting is then a two-dimensional
system tightly confined in the $z$-direction, with the Raman-lasers
running along the $x$-direction and free motion in $y$.

The discussion straightforwardly generalises to this case. We define
\begin{equation}                                                        
\epsilon_{\rm f}(q,k_y)=\epsilon_{\rm f}(q,k_y=0) + E_y= \epsilon_q+E_y
 \end{equation}
where the additional energy is given by $E_y=2 E_r  (k_y/k_r)^2$ and $k_y$ is the relative momentum in the $y$-direction of a two-particle state. Note that inelastic scattering processes remain gapped in this case even for particles remaining in the same band as the absorption of photons is always coupled to a change in the centre of mass momentum in this model which changes the energy.

\begin{figure}
    \subfloat%[First sub-figure\label{subfig-1:dummy}]
    {%
      \includegraphics[width=0.22\textwidth]{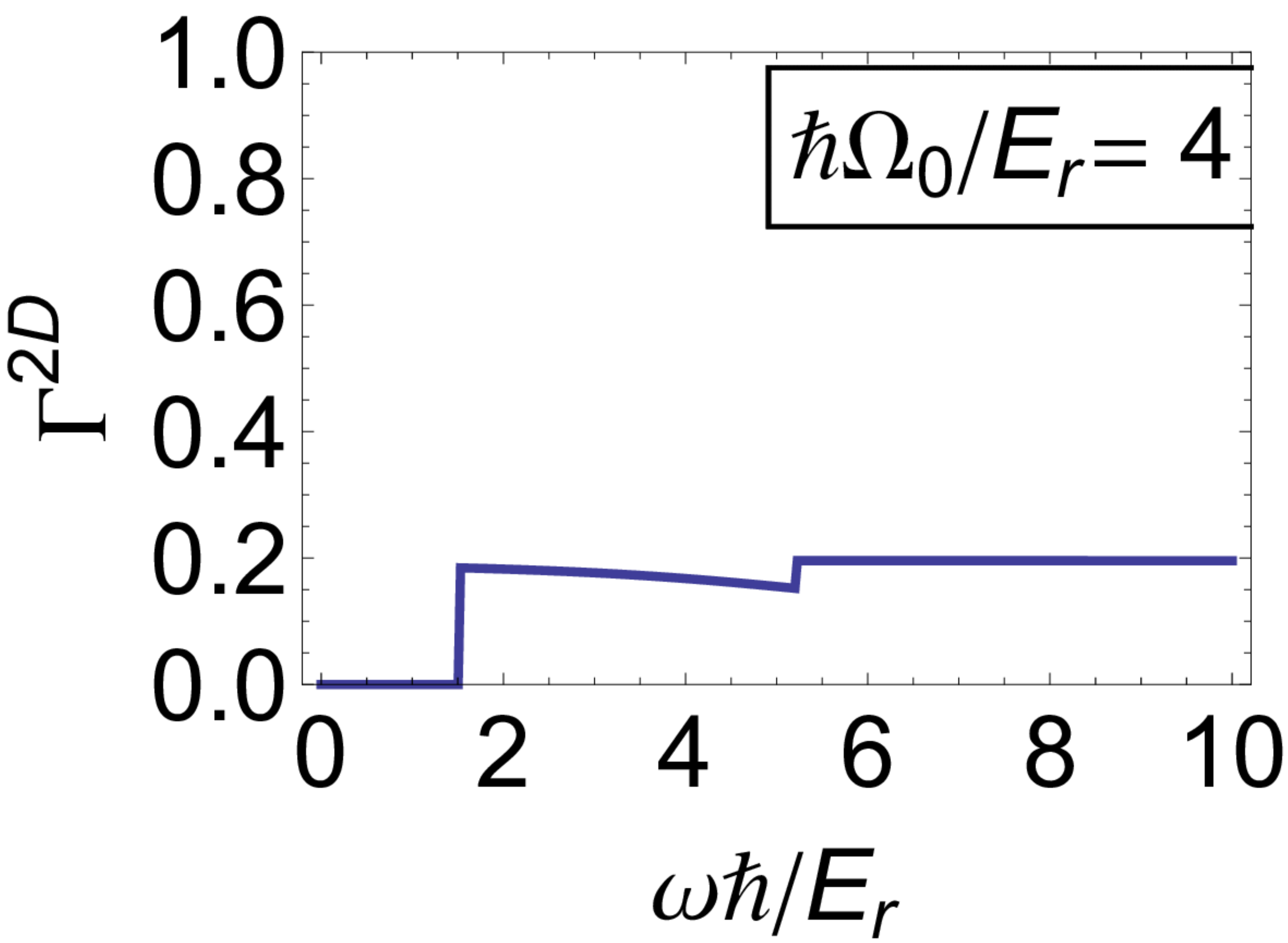}
    }
    \hfill
    \subfloat%[First sub-figure\label{subfig-2:dummy}]
    {%
      \includegraphics[width=0.22\textwidth]{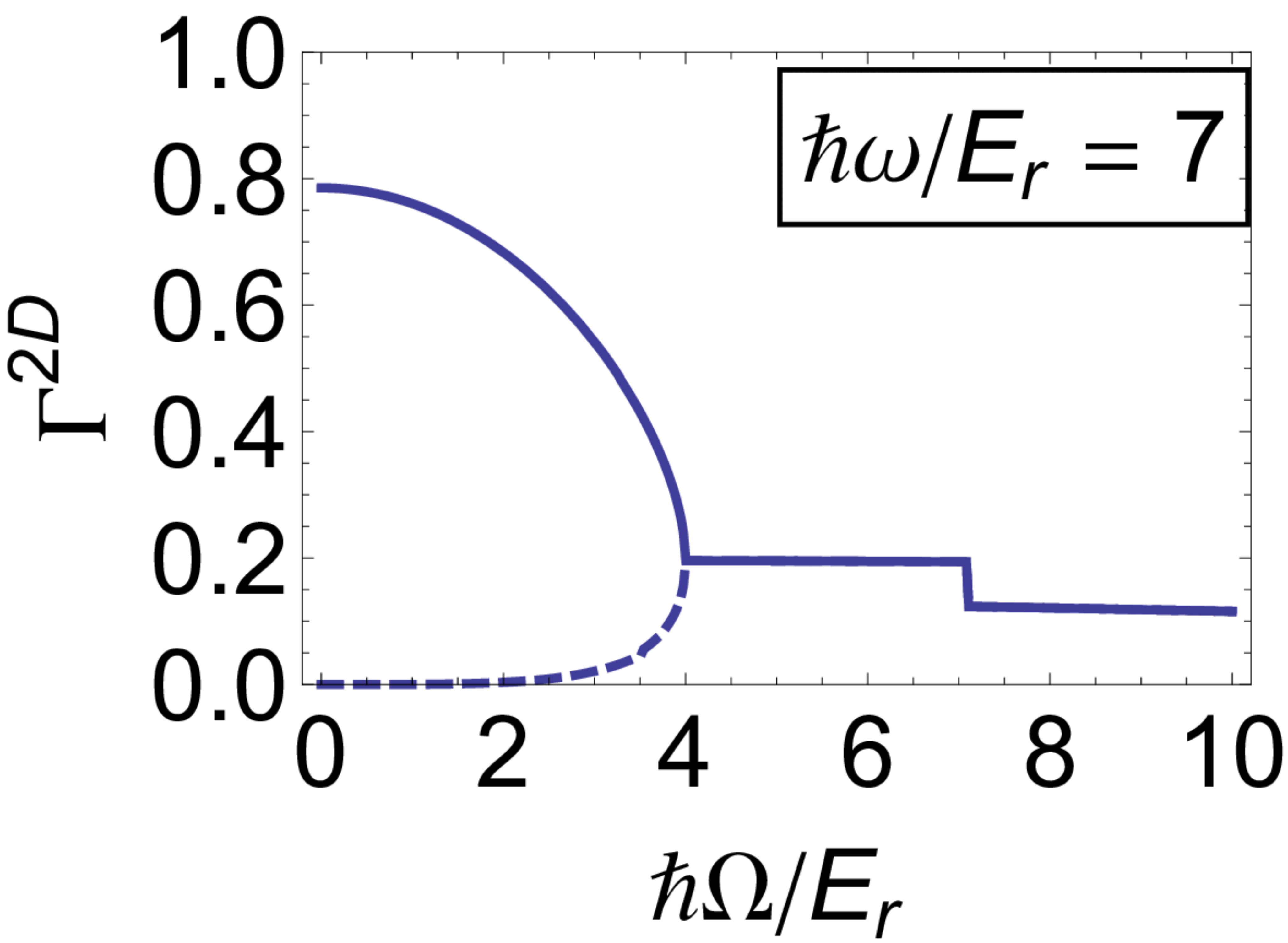}
    }
    \hfill
   \subfloat%[First sub-figure\label{subfig-2:dummy}]
    {%
      \includegraphics[width=0.22\textwidth]{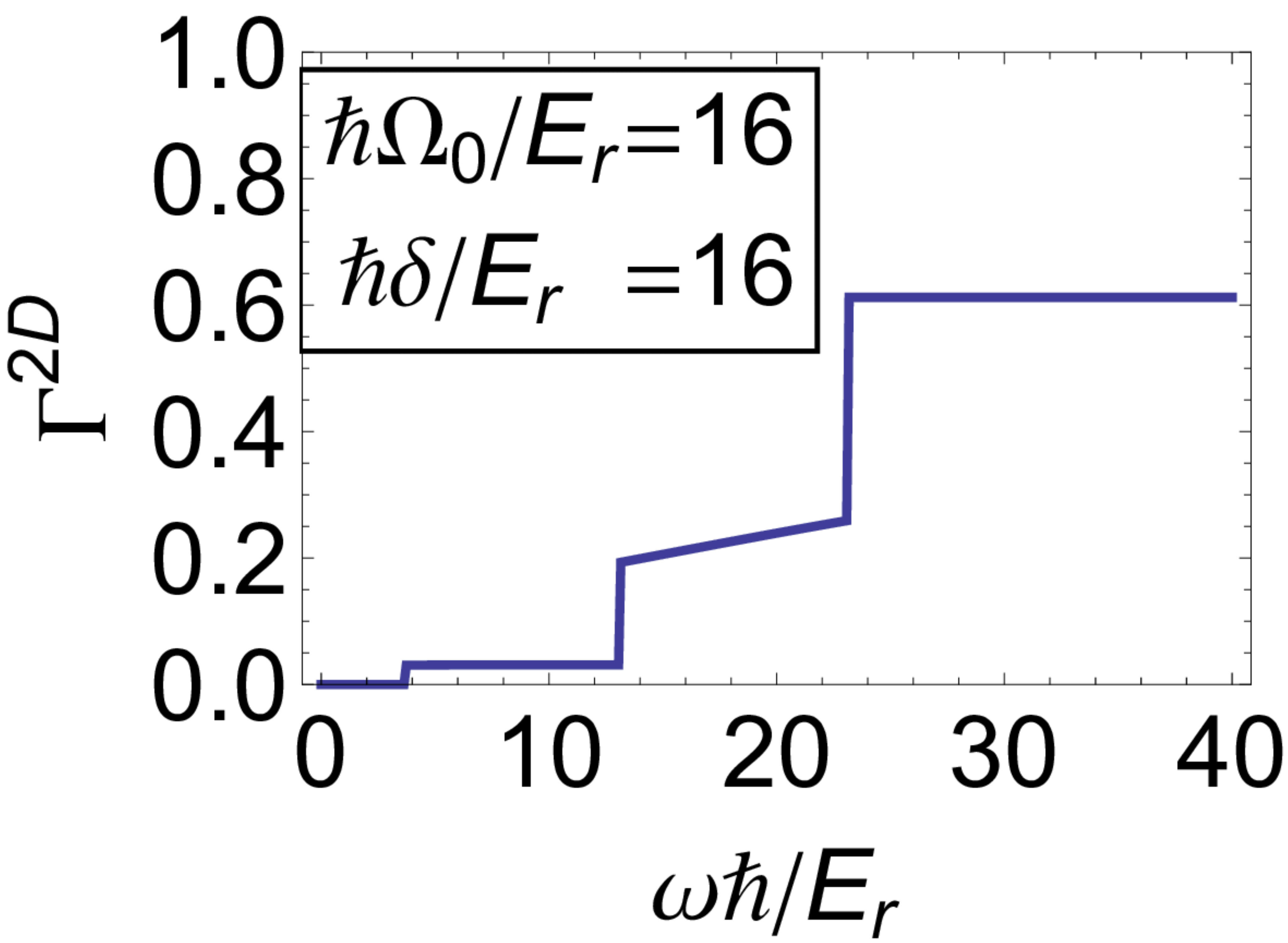}
    }
    \hfill
       \subfloat%[First sub-figure\label{subfig-2:dummy}]
    {%
      \includegraphics[width=0.22\textwidth]{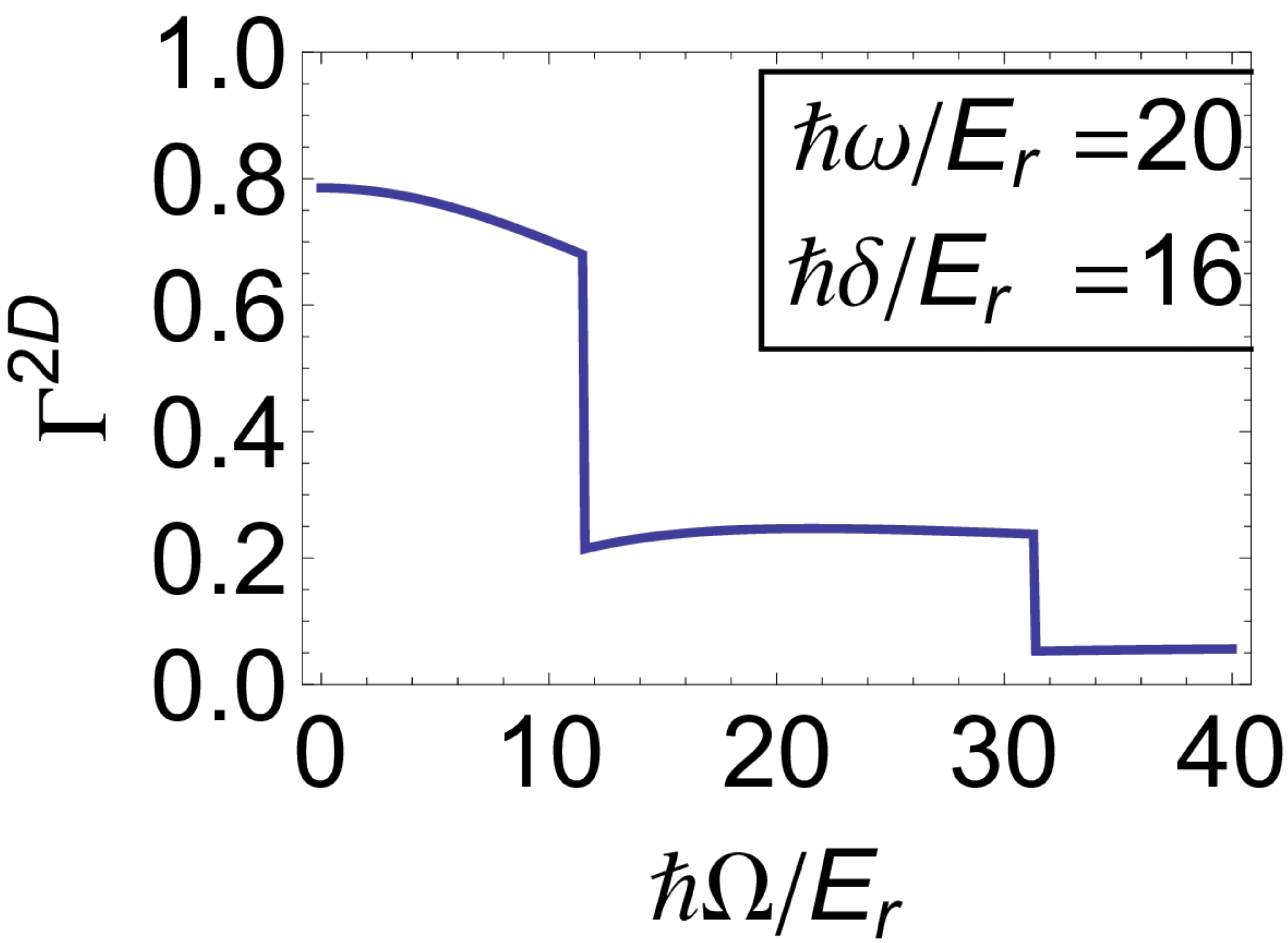}
    }
    \caption{(Color online) Dimensionless scattering rate $\Gamma^{\rm 2D}$, see Equation~(\ref{eq:raman_scatteringrate_2d}) for an initial state with particles in the lower band with momentum $k=\pm |k_0|$ and relative momentum $k_y=0$ getting inelastically scattered for the extension to a 2D setting with free motion in a transverse direction. In the top row for no detuning $\hbar \delta/E_r=0$ as a function of $\omega \hbar /E_r$ for fixed $\hbar \Omega /E_r = 4$ to the left and as a function of $\hbar \Omega /E_r$ for fixed $\omega \hbar /E_r=7$ to the right (dashed $k=+|k_0|$ , full $k=-|k_0|$) and at the bottom for $\hbar \delta/E_r=16$ as a function of $\omega \hbar /E_r$ for fixed $\hbar \Omega /E_r = 16$ to the left and as a function of $\hbar \Omega /E_r$ for fixed $\omega \hbar /E_r=20$ to the right as indicated in Fig.~\ref{fig:raman_stability}. The rate shows jumps at the opening/closing of scattering channels corresponding to the borders in Fig.~\ref{fig:raman_stability}.}
    \label{fig:raman_scatteringrate_2d}
  \end{figure}

Taking an initial state with no relative momentum $k_y^i=0$ and a final state with relative momentum $k^{\rm f}_y=q_y$  we define the matrix element of the interaction Hamiltonian as
\begin{equation}
 I^m_{q,q_y}=\brakett{\Phi^{m}_{\rm f}(q,q_y)}{ H_{\text{int}}}{\Phi^{0}_{\rm i}}
\end{equation}
and get for the inelastic scattering rate
\begin{equation}\label{eq:raman_scatteringrate_2d}
 \begin{aligned}
  \frac{dn}{dt} &= \frac{2 \pi} {\hbar} \sum_{\tau_1\tau_2;m\ne0}\left(\frac{L}{2\pi}\right)^2 \int dq_y  \\
  &\qquad   \int dq \left\vert I^m_{q,qy} \right \vert^{2}  \delta(\epsilon_{\rm i}-\epsilon_{\rm f}(q,q_y)-m \hbar \omega) \\
  &=\frac{1} {2 \pi \hbar L^2} \frac{k_r^2}{E_r^2} \sum_{\tau_1\tau_2;m\ne0}L^4 \int dE_y \frac{d (q_y/k_r)}{d(E_y/E_r)} \\
  &  \qquad \int d\epsilon_q \frac{d(q/k_r)}{d(\epsilon_q/E_r)} \left\vert I^m_{q,qy} \right \vert^{2}  \delta(\epsilon_{\rm i}-\epsilon_q-E_y-m \hbar \omega)\\
                &=\frac{g_c^{2}} {2 \pi \hbar L^2} \frac{k_r^2}{E_r} \Gamma^{\rm 2D}
 \end{aligned}
\end{equation}
where as before $\Gamma^{\rm 2D}$ is a dimensionless intensive rate constant for inelastic scattering processes. Note that $g_c$ is now defined differently, while it was an effective quantity for a 1-dimensional system before it is now the corresponding quantity for a 2D confined system. This two-dimensional rate is not expected to diverge at the opening or closing of scattering channels anymore, but rather to exhibit jumps which is confirmed in Fig.~\ref{fig:raman_scatteringrate_2d}.
Note that
the scattering rate does not vanish in the limit $\omega \rightarrow \infty$. The situation is the same as in the toy model discussed in section~\ref{subsec:toy_model}, where the rate vanishes in 1D simply due to the decreasing density of states, whereas in 2D with a constant density of states this is no longer the case. In the 3D case (not shown), the density of states increases as $\sqrt{\omega}$ for large drive frequency, again leading to large scattering for  $\omega \rightarrow \infty$. 
In terms of the relation of inelastic to elastic scattering, the expectation is that in 2D elastic and inelastic scattering should scale in the same way as functions of $k$ for low momenta as the density of states is $k$ independent, whereas in 3D the inelastic rate should dominate at low $k$ because of the suppression of elastic scattering due to the vanishing density of states.
%%%%%%%%%%%%%%%%%%%%%%%%%%%%%%%%%%%%
  %%%%%%%%%%%%%%%%%%%

%==========================================================================
\subsection{\label{subsec:lattice}Modulated lattice}
%==========================================================================
We now turn to  consider a model of a lattice with modulated on-site energies. This is a simplified version of the modulation protocols used in \cite{MagneticFields_1,MagneticFields_2,munichuniform} used to create artificial magnetic fields in optical lattices. Despite its simplifications it should still capture the novel scattering properties which become important due to the periodic driving. 
Our discussion describes generic features of two-particle scattering in the second class of proposals for the periodic driving of optical lattices mentioned in the section~\ref{sec:intro}, the modulation of on-site energies. Our work is complementary to that of Ref.  \cite{Stability_Floquet_condensate} which considered many-particle systems subjected to another form of periodic drive, of a shaken lattices.

\begin{figure}
      \includegraphics[width=0.45\textwidth]{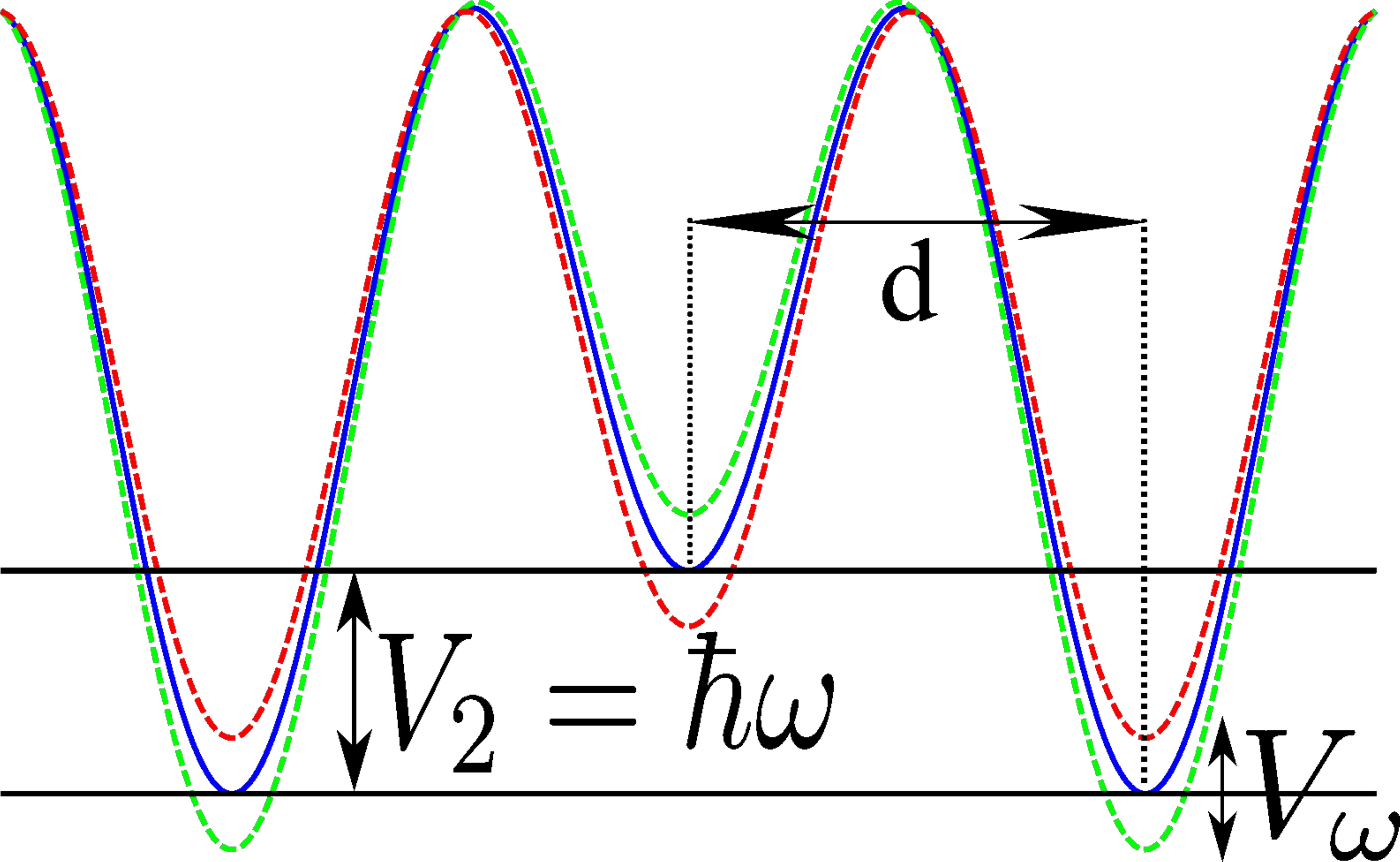}
    \caption{(Color online) Sketch of the one-dimensional time-periodically driven lattice potential given in Eq.~(\ref{eq:lattice_H0}). The lattice is staggered with an energy offset $V_2$ between neighbouring sites which suppresses tunneling along the lattice. Tunnelling is then restored by resonantly modulating the site-energies with a modulation strength $V_{\omega}$ at frequency $\hbar \omega = V_2$}
    \label{fig:modulated_lattice_sketch}
\end{figure}

Our model consists of a one-dimensional  superlattice with time-periodic modulation of on-site energies sketched in Fig.~(\ref{fig:modulated_lattice_sketch}).
The superlattice causes a staggered energy offset between sites, and the site-modulation resonantly restores the suppressed tunnelling along the lattice. We shall assume that the resulting bandwidth $\Delta_w$ is small compared to the modulation frequency, $\Delta_w \ll \hbar\omega$.
However, we shall allow for the possibility that $\hbar\omega$ is close to the interband transition energy $\Delta_g$, allowing inelastic scattering into this higher band. We therefore retain two bands of the original one-dimension lattice (i.e. there are two
Wannier states per local minimum of the potential).
After a discussion of the one-dimensional model we comment on the inclusion of an additional free direction of motion.
Consider bosons described by a field-operator $\Psi(x)$ loaded into such an optical superlattice. The resulting Hamiltonian is then given by $H=H_0(t) + H_{\text{int}}$,
%%%%%%%%%
\begin{align}
  \begin{split}
 H_0(t)&= \int dx \, \Psi^{\dagger}(x) \Big[\frac{- \hbar^{2}}{2M} \frac{d^{2}}{dx^{2}} +V_1 \sin^{2}(k x) \\
       & \quad \quad \quad \quad \quad \quad \quad  + V_2 \cos^{2}(k x/2)  \Big] \Psi(x)  \\
  & +\int dx \, \Psi^{\dagger}(x) \left[V_{\omega} \cos^{2}(k x/2 +\omega t/2) \right] \Psi(x) ,
  \end{split} \label{eq:lattice_H0}\\
 H_{\text{int}} &=\frac{g}{2} \int{dx \, \Psi^{\dagger}(x) \Psi^{\dagger}(x) \Psi(x) \Psi(x)} \label{eq:Lattice_Hint},
\end{align}
%%%%%%%%%
where the first and second line of equation~(\ref{eq:lattice_H0}) describe the kinetic energy and superlattice potential with strength $V_1$ and $V_2$ created by standing laser fields and the third line gives the modulation of on-site energies with strength $V_{\omega}$ and modulation frequency $\omega$ which can be created by two running-wave beams as described in \cite{MagneticFields_2}. Energies and lattice depth will be measured in terms of the recoil energy $E_r = \frac{\hbar^{2} k_r^{2}}{2 m}$ which we define with respect to the unstaggered lattice, i.e. $k_r=2 k$, and we assume the lattice to be deep $V_1>E_r$ in order to justify the tight-binding description made in the following.

The last part, equation~(\ref{eq:Lattice_Hint}), gives the usual 1D contact interaction between atoms of strength $g$. Assuming a tight-confinement by a harmonic potential in the transverse radial direction, it is given by $g= \frac{4 \hbar^2 a_s}{a_{\perp}^2 M}$, where $a_s$ is the 3D s-wave-scattering length of the true interaction potential and $a_{\perp}=\sqrt{2\hbar/(M \omega_{\perp})}$ the radial confinement length of the harmonic trap with frequency $w_{\perp}$ \cite{1D_effscattering}.

We begin by mapping the Hamiltonian $H_0(t)$ onto a tight-binding Hamiltonian with two orbitals per lattice site. To this end the bosonic field-operator $\Psi(x)$ is expanded in terms of the Wannier functions of the two lowest bands of the Hamiltonian with $V_2=V_{\omega}=0$, i.e. in the Wannier functions of the simple optical lattice without the superlattice potential. Writing
\begin{equation}
 \Psi(x) = \sum_j w_1(x-x_j) a_j + w_2(x-x_j) b_j\,,
\end{equation}
where $a$  ($b$) are field operators for Wannier states in the first (second) band,
one obtains the tight-binding model as
\begin{equation}\label{eq:H0_lattice}
\begin{aligned}
H_0(t) &= \sum_{ij} \left( -t_{ij}^{(1)} a_i^{\dagger} a_j -  t_{ij}^{(2)} b_i^{\dagger} b_j + h.c.\right) \\
  &+ \sum_j 1/2\left[1+(-1)^{j} \right][V_2 + V_{\omega} \cos(wt)]a^{\dagger}_j a_j  \\
  &+\sum_j 1/2\left[1+(-1)^{j} \right][V_2 + V_{\omega} \cos(wt)+ \Delta_g]b^{\dagger}_j b_j ,
\end{aligned}
\end{equation}
where,
\begin{equation}
 t_{ij}^{(n)} = \int dx \, w_n^{*}(x-x_i) \left[\frac{- \hbar^{2}}{2M} \frac{d^{2}}{dx^{2}} +V_1 \sin^{2}(k x) \right] w_n(x-x_j) 
 \end{equation}
and $\Delta_g$ is the energy gap between the first ($a$) and second ($b$) band.
The superlattice potential $V_2 \cos^{2}(k x/2)$ is seen to lead to a staggering in the tight-binding model which suppresses tunnelling along the lattice due to the energy difference $V_2$ between neighbouring sites.
Tunnelling can then be restored by modulating the lattice on resonance $\hbar \omega = V_2$ whereby the necessary energy is provided by absorption and emission of photons. 
The parameters need to satisfy $ V_2 >t_{ij}^{(n)}$ such that in the staggered lattice tunnelling is suppressed. Moreover, to obtain clearly separated bands the gap $\Delta_g$ should be bigger than the band-width of the Bloch bands. Finally, to avoid resonant excitation from the lowest to the highest band $\hbar \omega$ should be smaller than the gap. Additionally, the time-dependent modulation $V_{\omega}$ should not be too strong, as will become apparent in the derivation below. In the high frequency regime the relevant quantity to measure the effect of the modulation is $\kappa = V_{\omega}/(\hbar \omega)$ which should be of order 1, whereas for the low frequency regime $V_{\omega}$ should be comparable to $V_2$ and smaller than $V_1$.
This leads to a hierarchy of energy-scales $\Delta_g , V_1 > V_2=\hbar \omega \simeq V_{\omega} >t_{ij}^{(n)} $. 

In the following we will treat the inter-particle interaction, equation~(\ref{eq:Lattice_Hint}), as a perturbation of the non-interacting Hamiltonian, thus, our study is limited to the weakly interacting regime. In particular, we will be precluded from accessing the Hubbard regime in which instead of the Bloch-waves used below, the Wannier states would be more suitable. The discussion of corresponding processes in the strongly interacting regime that would lead to heating and interband-transitions is beyond the scope of the present work.
\subsubsection{Single-particle states}
We proceed to obtain the single-particle spectrum of the non-interacting Hamiltonian $H_0$. As it is translationally invariant with respect to translations by 2 lattice sites $l\rightarrow l+2$ it is best analysed in momentum space.
Defining the Fourier-Transform of the operators as
\begin{align}
 a_k&= \frac{1}{\sqrt{N}} \sum_l a_l e^{-i k l d} , \\
 a^{\dagger}_k&= \frac{1}{\sqrt{N}} \sum_l a^{\dagger}_l e^{i k l d} ,
\end{align}
where $d$ is the lattice spacing which is set to 1 in the following, one obtains the Hamiltonian in momentum space
\begin{equation}\label{eq:modulated_lattice_H0_k}
\begin{aligned}
 H_0(t) &= \sum_{k} \epsilon_k^{(1)} a^{\dagger}_k a_k +(\epsilon_k^{(2)}+\Delta_g) b^{\dagger}_k b_k \\
        & + \sum_k 1/2 [V_2 + V_{\omega} \cos(\omega t)](a^{\dagger}_{k+\pi} a_k + b^{\dagger}_{k+\pi} b_k)
\end{aligned}
\end{equation}
with
\begin{equation}
 \epsilon_k^{(n)} = \sum_l t_l^{(n)} \cos(l k ) ,
\end{equation}
and $l=|i-j|$.
The staggering of the lattice induces a coupling between the two momentum components at $k$ and $k+\pi$.

\begin{figure}
      \includegraphics[width=0.48\textwidth]{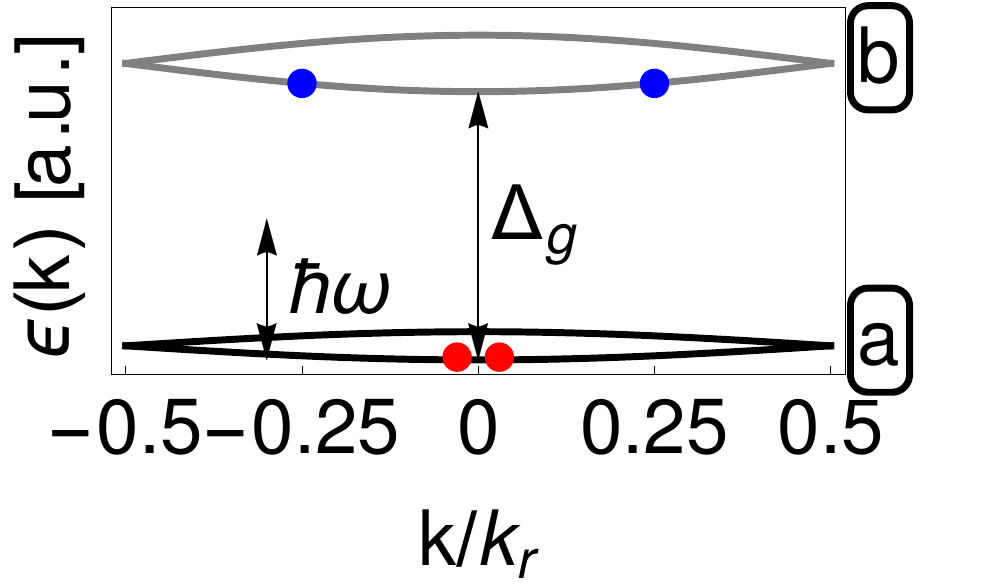}
    \caption{(Color online) Quasienergies of the resonantly modulated lattice, Eq.~(\ref{eq:lattice_dispa}) and (\ref{eq:lattice_dispb}), as a function of the quasi-momentum $k$ in arbitrary energy units. The two lowest bands of the original lattice ($a$) and ($b$) both split into two subbands $\tau=\pm$ which are degenerate at the Brillouin zone boundaries. Depicted is a typical situation in which the energy of the periodic modulation $\hbar \omega$ is larger than the bandwidth of the lowest band and smaller than the bandgap $\Delta_g$. The balls show our initial state with two particles in the lowest band and a possible final state with two particles in the upper band after scattering. For this plot a nearest neighbour tight-binding dispersion $\epsilon^{(n)}(k) = {\rm t}^{(n)} \cos(k) $ is assumed with parameters ${\rm t}^{a}=1.1$, ${\rm t}^{b}=2.3$ and $\hbar \omega =4.8$, $\Delta_g=10$ and $\kappa =1$}
    \label{fig:lattice_dispersion}
  \end{figure}

We proceed to diagonalise this Hamiltonian by the use of a rotating wave like approximation for the case of resonant modulation $\hbar\omega = V_2$. Details of the derivation are given in appendix~\ref{app:floquet_modes}.
The Floquet modes turn out to be
\begin{equation}\label{eq:lattice_eigenstates}
\begin{split}
  \Phi^{(n),m}_{\tau,k}(t) &= 1/2 \left[ \left(f(t) + \tau e^{i \omega t}  \bar{f}(t)\right) c^{\dagger}_{k,n} \right.\\
   & \left. \quad \quad + \left(f(t) -\tau e^{i \omega t}  \bar{f}(t)\right) c^{\dagger}_{k+\pi,n} \right] e^{i m \omega t}\vac ,
   \end{split}
\end{equation}
where we denote by $c^{\dagger}_{k,n}$ the creation operator for a state in band $n=a$ or $b$, the states are characterised by an additional subband index $\tau= \pm$ and the time-periodic function $f(t)=  \exp[i \kappa \sin(\omega t)]$ with $\kappa=V_{\omega}/(\hbar \omega)$ was defined. The corresponding quasi-energies are
\begin{align}
\epsilon^{a,0}_{\tau,k} &= \tau \epsilon^{a}_k J_{-1}(\kappa) , \label{eq:lattice_dispa}\\
\epsilon^{b,0}_{\tau,k} &= \tau \epsilon^{b}_k J_{-1}(\kappa) + \Delta_g  . \label{eq:lattice_dispb}
\end{align}
where $J_{-1}$ denotes the Bessel function of the first kind.
Note that the quasi-energies are not reduced to a Floquet BZ here, but rather defined to keep the association with the original lowest first ($a$) and second ($b$) bands that are gapped in energy by $\Delta_g$ in the static Hamiltonian. The resulting bandstructure is depicted in Fig.~\ref{fig:lattice_dispersion}.
%%%%%%%%%%%%%%%%%%%%%%%%%
%%%%%%%%%%%%%%%%%%%%%%%%%

The modulation of the lattice now shows its effect in two ways. Firstly, the tunnelling is restored with a modified strength of $\epsilon^{n}_k J_{-1}(\kappa)$. Secondly, the population of momentum components oscillates in time between  $k$ and $k+\pi$ with equal amplitudes as the energy gap of $V_2$ is bridged by the energy of the modulation $\hbar \omega=V_2$.

Following the definitions in section~\ref{subsec:FFGR} scattering processes in which particles change the band from $a$ to $b$ and those for which particles stay within a band, but scatter into higher energy single-particle states in the same band will be called inelastic. The first process leads to loss of particles from the lowest band, whereas the second process may lead to heating within the band.
\subsubsection{Two-particle scattering}
To consider the effects of the collisions of the atoms given by the interaction Hamiltonian~(\ref{eq:Lattice_Hint}) we will treat them to first order within Fermi's golden rule. This will preclude the discussion of strongly correlated many-body phases, but is still sufficient to see the relevant 2-particle physics and their relevance to heating.

We apply FFGR (\ref{eq:ffgr}) for an  initial state consisting of two particles in the lowest band with the same subband index $\tau_{\rm i}=\pm$ with crystal-momentum $+k$ and $-k$ respectively, i.e.
\begin{equation}
 \Psi_{\rm i}=\Psi^{a \dagger}_{\tau_{\rm i},k} \Psi^{a \dagger}_{\tau_{\rm i},-k}\vac 
\end{equation}
and a final state containing two particles in the upper band in subbands $\tau_1$, $\tau_2$ with momenta $q_1$ and $q_2$
\begin{equation}
 \Psi_{\rm f}=\Psi^{b \dagger}_{\tau_1,q_1} \Psi^{b \dagger}_{\tau_2,q_2}\vac .
\end{equation}
This is the only relevant inelastic scattering process allowed within FFGR for the case in which $\hbar \omega > 4  \epsilon^{a}_k J_{-1}(\kappa)$ which forbids the absorption of a quantum of energy $\hbar \omega$ within the lowest band. Because of the resonance condition $\hbar \omega = V_2$, this corresponds to strong suppression of tunnelling for which $V_2$ was assumed to be large compared to the bandwidth. Such a situation with the initial and an allowed final state is shown in Fig~\ref{fig:lattice_dispersion}.

\begin{figure*} 
\centering 
\subfloat{\subfigimg[hsep=0.07\textwidth,width=0.45\textwidth]{\bf (a)}{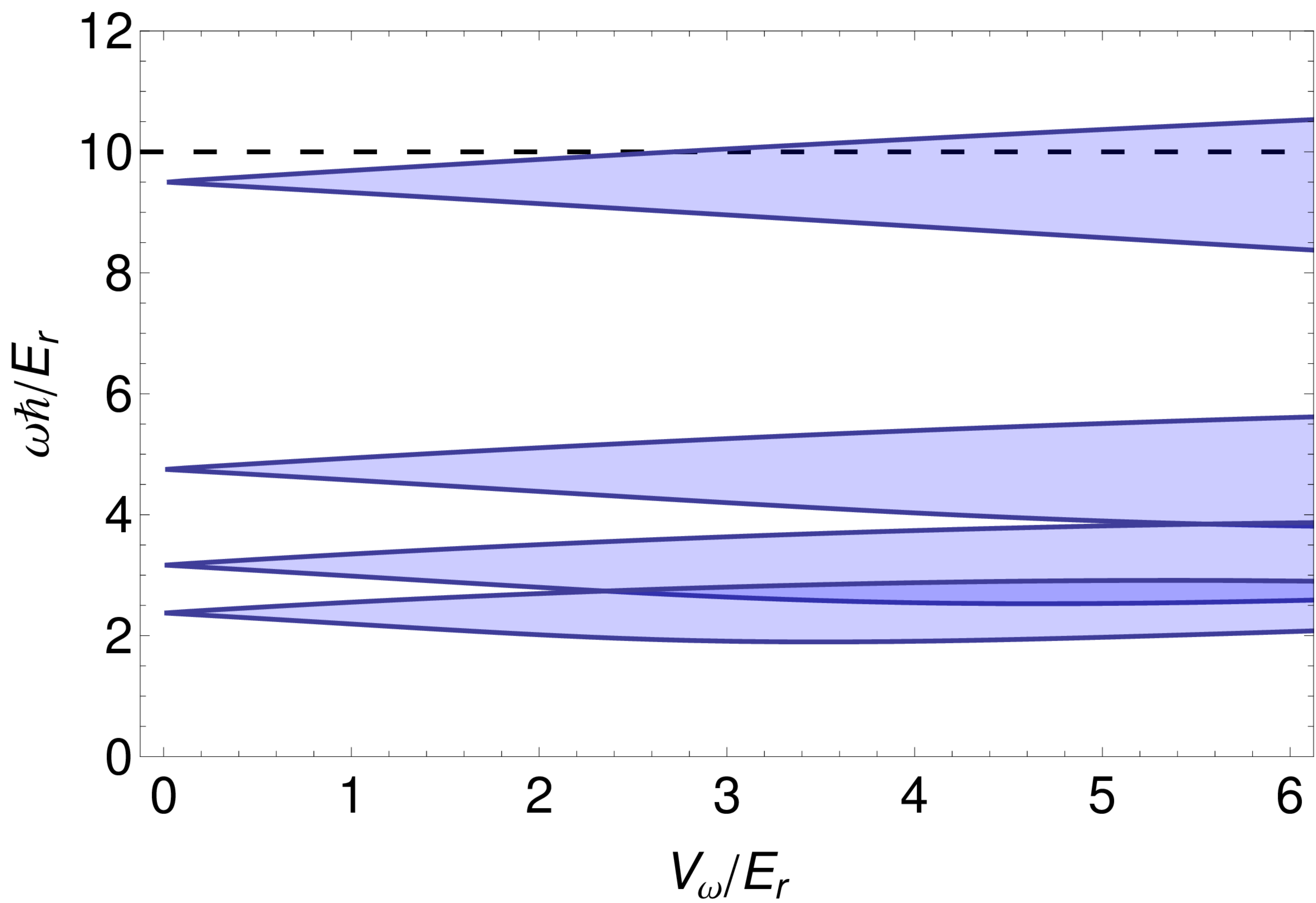}}
\hfill
\subfloat{\subfigimg[hsep=0.07\textwidth,width=0.45\textwidth]{\bf (b)}{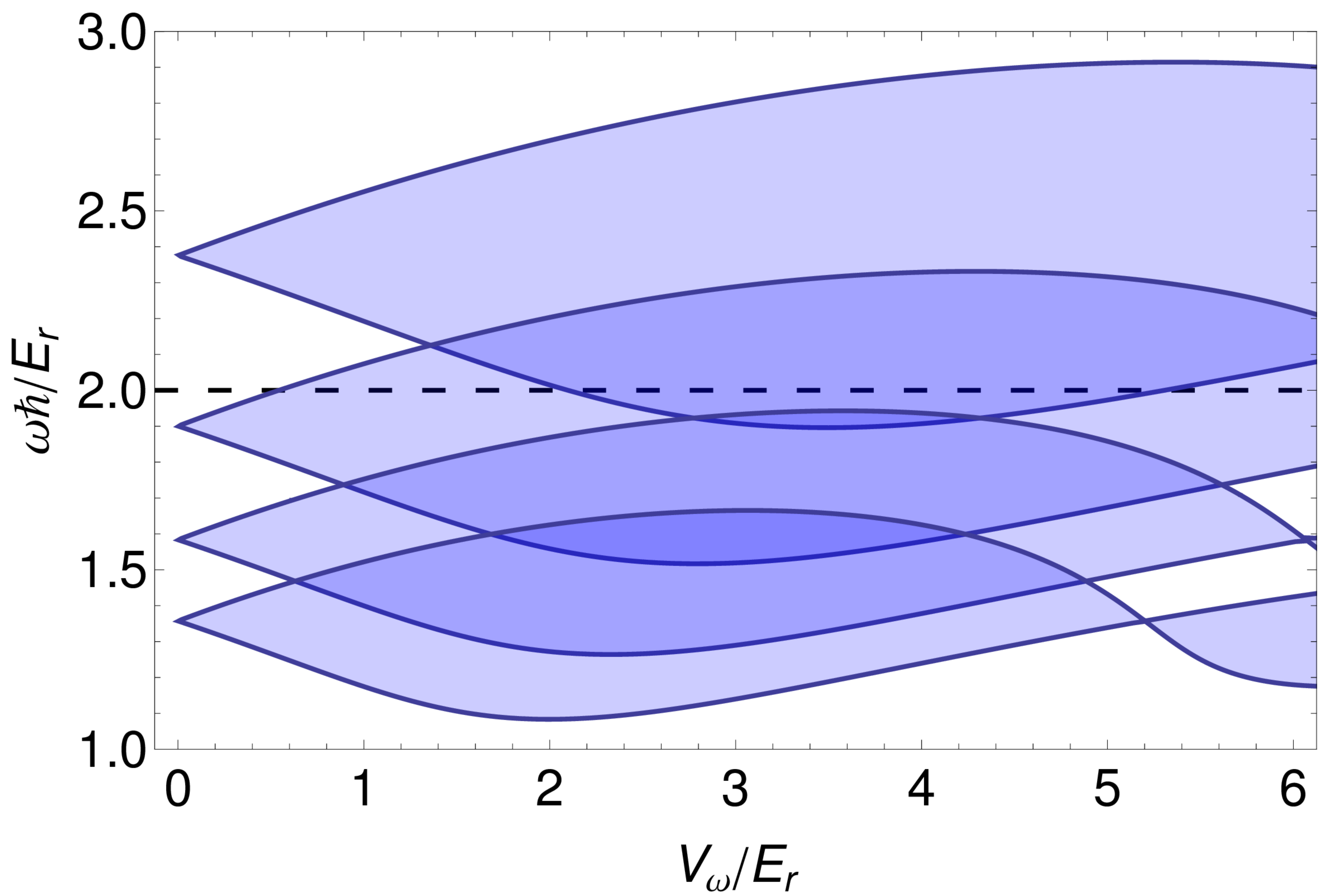}} 
 \caption{(Color online) Stability diagram of the lowest band in a time-periodically modulated one-dimensional lattice of depth $V_1=6 E_r$ for which the band gap is $4.75 E_r$. Shaded regions correspond to energetically allowed scattering from the ground state into the first excited band. Different lobes correspond to different orders of the instability $m$ starting with $m=1$ at the top and increasing downwards. In (a) only the first 4 such lobes are shown for clarity, in (b) the region $1\le\omega \hbar/E_r \le3$ with the $m=4,5,6,7$ lobes which overlap the $m=5$ lobe are shown. The dashed lines of constant $\omega \hbar /E_r$ correspond to the cuts along which the scattering rate is shown in Fig.~\ref{fig:Lattice_scatteringrate}.}
    \label{fig:Lattice_stability}
\end{figure*}

Due to the structure of the single-particle states that contain two momentum components at $k$ and $k+\pi$ scattering is allowed into states with momenta $q_1=q$, $q_2=-q$ and $q_1=q+\pi$, $q_2=-q$. Within the reduced BZ the second case $(\tau_1,q+\pi)$ actually corresponds to $(-\tau_1,q)$.
As we consider the case in which the bands are well separated, i.e. $4 (J_{1,{\rm eff}}+J_{2,{\rm eff}})< \Delta_g$, transitions from the lower to the upper band require the absorption of a non-zero number $m_0$ of photons.
Specifically, the conservation of quasi-energy in the FFGR then picks the representative state $\Phi_{\rm f}^{m}$ with $m=-m_0$ and energy conservation reduces to
\begin{equation}\label{eq:lattice_energyconservation}
  2 \epsilon^{a}_k J_{-1}(\kappa)=
                        \epsilon^{b}_q J_{-1}(\kappa) \left[ \tau_1\pm \tau_2 \right] + (2 \Delta_g-m_0 \hbar \omega)
\end{equation}
where the $+$ ($-$) sign corresponds to the cases $q_1+q_2=0$ ($q_1+q_2=\pi$) described above. 
This equation fixes the momentum $q_{\rm f}$ of the final state depending on the band gap $\Delta_g$, the amplitude of the driving $V_{\omega}$ and the driving frequency $\omega$.
Assuming for the moment that an arbitrary number of photons may be absorbed within FFGR, equation~(\ref{eq:lattice_energyconservation}) then allows us to derive a stability diagram purely based on quasi-energy conservation in which scattering is allowed or forbidden, see Fig.~\ref{fig:Lattice_stability}. Depending on the modulation strength $\kappa=V_{\omega}/{(\hbar \omega)}$ and frequency $\omega \hbar /E_r$ there are stable regions in which no energy absorption takes place, regions where a single transition with a unique $m$ and regions where multiple transitions with different photon numbers $m$ are allowed.
%%%%%%%%%%%%%%%%%%%%%%%
To compute the matrix element appearing in FFGR
\begin{equation}\label{eq:FFGRmatrixelement}
  I^{m}_{\tau_{\rm i},\tau_1,\tau_2;k,q} =\brakett{\Phi^{-m}_{\tau_1,q;\tau_2,-q}}{H_{\text{int}}/g}{ \Phi^{0}_{\tau_{\rm i},k;\tau_{\rm i},-k}}
\end{equation}
we can make use of the expansion in the basis of Bloch functions $\psi_{n}(x,k) $ of band $n$, similarly to the steps performed in \cite{Stability_Floquet_condensate}. Thus, the field operator $\Psi(x)$ in equation~(\ref{eq:Lattice_Hint}) is expanded as
\begin{equation}
 \Psi_{n}(x) = \sum_k \phi_{n}(k) \psi^{k}_{n}(x) ,
\end{equation}
and the interaction Hamiltonian is rewritten as
\begin{equation}
\begin{aligned}
\frac{ H_{\text{int}}}{g}&= \frac{1}{2}\sum_{\{n_i\}} \int_0^L dx \, \Psi_{n_1}^{\dagger}(x)\Psi_{n_2}^{\dagger}(x) \Psi_{n_3}(x) \Psi_{n_4}(x) \\
                &=\sum_{\{n_i,k_i\}} W_{n_1n_2n_3n_4}^{k_1k_2k_3k_4} \, \phi^{\dagger}_{n_1}(k_1) \phi^{\dagger}_{n_2}(k_2) \phi_{n_3}(k_3) \phi_{n_4}(k_4),
\end{aligned}
\end{equation}
where we defined the matrix elements of the interaction between Bloch waves
\begin{equation} \label{eq:Bloch_functions_matrix_elements}
 W_{n_1n_2n_3n_4}^{k_1k_2k_3k_4} =\frac{1}{2} \int_0^L dx \, \bar{\psi}^{k_1}_{n_1}(x) \bar{\psi}^{k_2}_{n_2}(x) \psi^{k_3}_{n_3}(x) \psi^{k_4}_{n_4}(x) .
\end{equation}
The explicit expressions for coupling matrix elements $I^{m}_{\tau_{\rm i},\tau_1,\tau_2;k,q}$, equation~(\ref{eq:FFGRmatrixelement}), are given in the appendix~\ref{app:matrixelements}.
If one assumes that the matrix element between Bloch functions $W_{bbaa}^{k_1k_2k_3k_4}$ defined in Eq.~(\ref{eq:Bloch_functions_matrix_elements}) is completely momentum-independent, then the matrix element $ I^{m}_{\tau_{\rm i},\tau_1,\tau_2;k,q}$ in equation~(\ref{eq:FFGRmatrixelement}) vanishes for $m \ne 0$ and there is no inelastic scattering. This occurs if the Bloch waves themselves become momentum-independent, e.g. in the case of an infinitely deep lattice. Thus, for sufficiently deep lattices there is no inelastic scattering and no coupling to the higher-band within the FFGR.
Generically, the interaction matrix elements $W_{bbaa}^{k_1k_2k_3k_4}$ do depend on momentum and $ I^{m}_{\tau_{\rm i},\tau_1,\tau_2;k,q}$ is non-vanishing for any $m$. This implies the possibility of absorption of arbitrary integer numbers of energy $\hbar \omega$ during scattering and thus justifies the assumption made in deriving the stability diagram.

\begin{figure}
    \subfloat%[First sub-figure\label{subfig-1:dummy}]
    {%
      \includegraphics[width=0.22\textwidth]{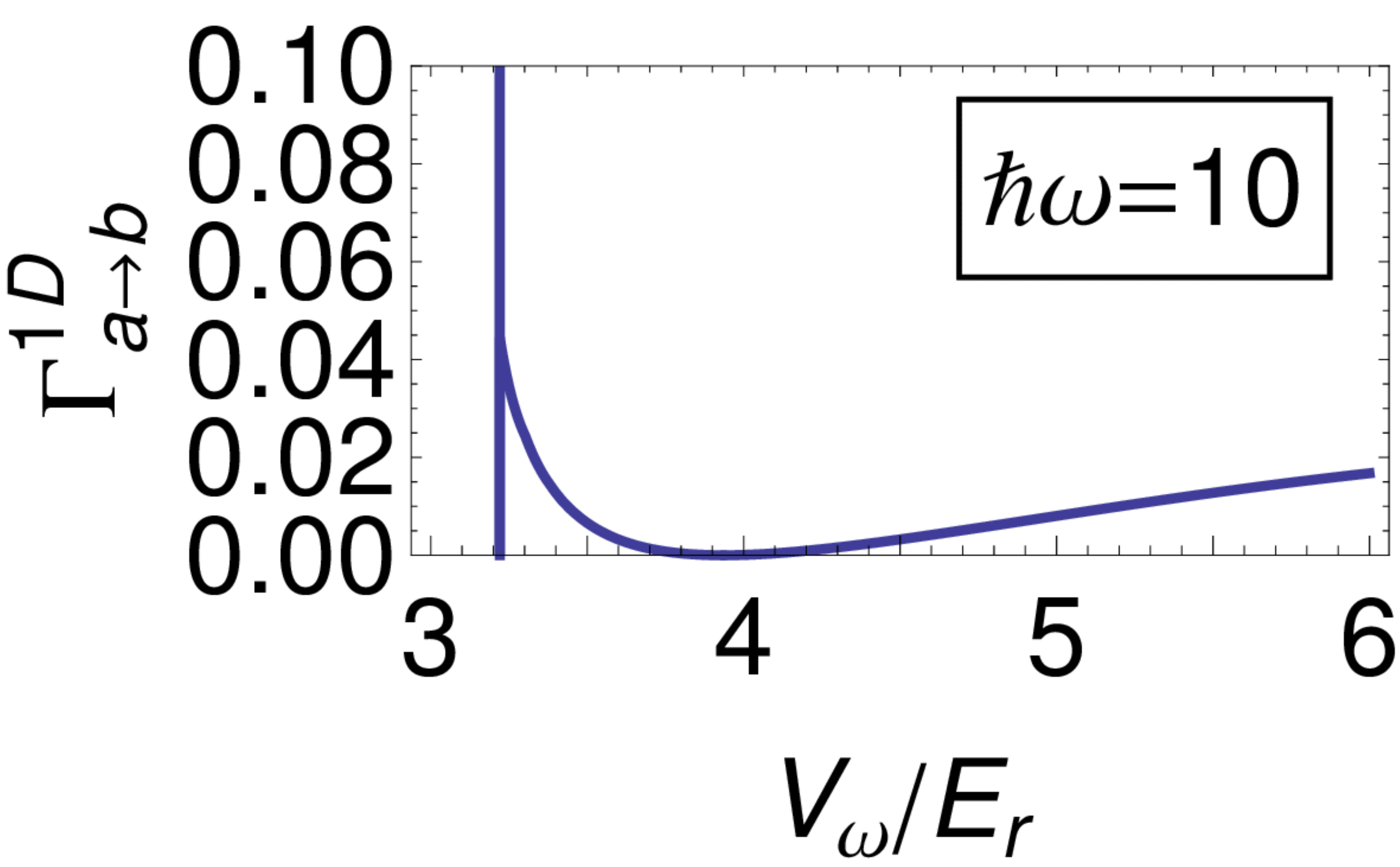}
    }
    \hfill
    \subfloat%[First sub-figure\label{subfig-2:dummy}]
    {%
      \includegraphics[width=0.22\textwidth]{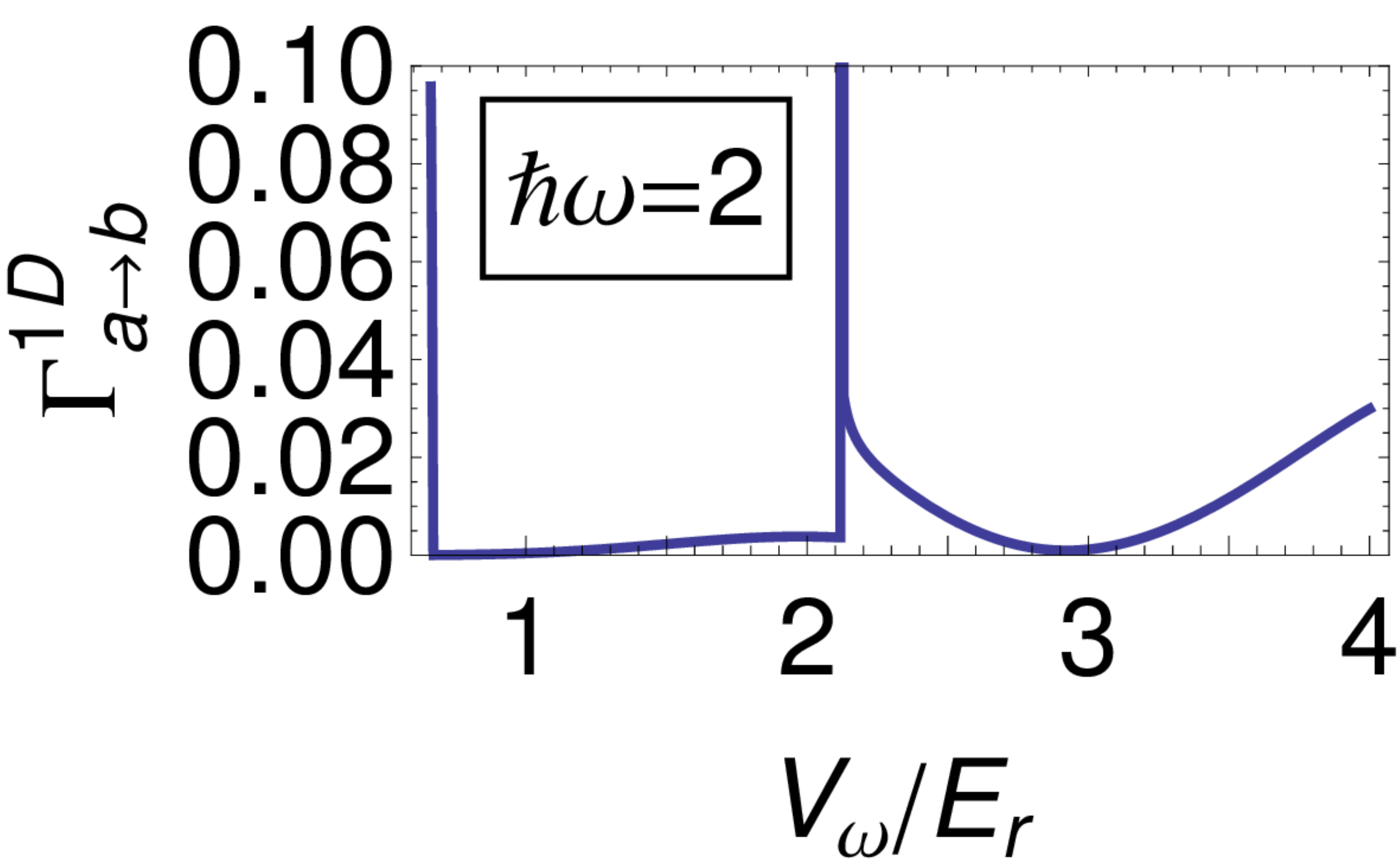}
    }
    \caption{(Color online) Dimensionless scattering rate $\Gamma^{\rm 1D}$, see Equation~(\ref{eq:lattice_scatteringrate}) along the cuts $\hbar \omega /E_r = const.$ as indicated in Fig.~\ref{fig:Lattice_stability} for particles in the first band ($a$) with momentum $k=0$ scattering into the second band ($b$). The left corresponds to $\hbar \omega /E_r = 10$ for which $m=1$ is the only scattering channel whereas the right corresponds to $\hbar \omega /E_r = 2$ for which $m=5$ transitions are allowed for $ 0.6\le V_{\omega}/E_r\le6.9$ and both $m=4,5$ for $ 2.1\le V_{\omega}/E_r\le5.3$.}
    \label{fig:Lattice_scatteringrate}
  \end{figure}

To obtain the total inelastic scattering rate one integrates over all allowed final states
\begin{equation}\label{eq:lattice_scatteringrate}
\begin{aligned}
 \frac{dn}{dt} &= \frac{L} {\hbar} \sum_{\substack{\tau_1\tau_2 \\ m\ne0}}   \int dq \, g^2 \left \vert I^{m}_{\tau_{\rm i},\tau_1,\tau_2;k,q} \right \vert^{2} \delta(\epsilon_{\rm i}-\epsilon_{\rm f}(q)+m \hbar \omega)\\
               &= \frac{ g^2 L } {\hbar} \sum_{\substack{\tau_1\tau_2 \\ m\ne0}}  \int d \epsilon_{\rm f} \frac{dq}{d\epsilon_{\rm f}} \left \vert I^{m}_{\tau_{\rm i},\tau_1,\tau_2;k,q} \right \vert^{2}   \delta(\epsilon_{\rm i}-\epsilon_{\rm f}+m \hbar \omega)\\
               &= \frac{ g^2 } {\hbar E_r} \frac{1}{L d} \sum_{\substack{\tau_1\tau_2 \\ m\ne0}} L^2 \int d \epsilon_{\rm f} \frac{d(qd)}{d(\epsilon_{\rm f}/E_r)} \\
               & \quad \times \left \vert I^{m}_{\tau_{\rm i},\tau_1,\tau_2;k,q} \right \vert^{2}   \delta(\epsilon_{\rm i}-\epsilon_{\rm f}+m \hbar \omega)\\
               &= \frac{ g^2 } {\hbar E_r} \frac{1}{L d} \Gamma^{\rm 1D} \,
\end{aligned}
\end{equation}
which defines the intensive dimensionless scattering rate $\Gamma^{\rm 1D}$ for scattering into the higher band. This equation may be compared to (19) in \cite{Stability_Floquet_condensate} which is identical apart from the different definition of $I^{m}_{\tau_{\rm i},\tau_1,\tau_2;k,q}$ which contains the dependence on the driving protocol and the relevant physics. $\Gamma^{\rm 1D}$ depends on the lattice via the band structure $\epsilon_k$ and the gap $\Delta_g$, the modulation strength and frequency which determine both the effective band structure $ \epsilon_k J_{-1}(V_{\omega}/w)$ and the eigenstates via their dependence on $f(t)=  \exp[i \kappa \sin(\omega t)] = \sum_n J_n(\kappa) \exp[i n \omega t]$. Therefore, the scattering rate will show a complicated behaviour, possibly with zeros inherited from the Bessel functions. Moreover, the rate will diverge at the thresholds for scattering, i.e. at the envelop functions of the shaded areas in Fig~(\ref{fig:Lattice_stability}), where the momentum of the final state is at the edges of the BZ and the dispersion is flat yielding a diverging 1-D density of states. These expectations are confirmed in Fig~(\ref{fig:Lattice_scatteringrate}) which shows the scattering rate $\Gamma^{\rm 1D}$ for transitions of particles in the lower band ($a$) into the higher band ($b$) along the cuts indicated in Fig~(\ref{fig:Lattice_stability}).

%%%%%%%%%%%%%
  %%%%%%%%%%%%%%%%%%%%
As an order of magnitude estimate for the decay of particles starting in the lower band, consider a gas of density $N/L \approx 1/d$, $a_s = \SI{5}{\nano\metre}$, $d_{\perp} =\SI{100}{\nano\metre}$, $m = \SI{100}{u}$ and take $\Gamma^{\rm 1D} \approx 0.05$ to obtain
$N/(dN/dt) \approx \SI{40}{\milli\s}$. From this estimate, experiments in the unstable region would be seriously affected by the scattering into higher bands and a single-band approximation would not be valid.
We can conclude that experiments on modulated lattices should take care to work in regions of parameter space where transitions are not allowed to avoid rapid scattering into higher bands. From Fig.~\ref{fig:Lattice_stability} this corresponds to avoiding single ($m=1$) and multi-photon ($m>1$) resonances in which the gap $2 \Delta_g$ to lift two particles into the higher band is bridged by a number of $m$ photons. Multi-photon scattering processes may also be reduced by keeping the modulation amplitude $V_{\omega}$ small compared to $\hbar \omega$. However, for this specific model there is parameter space available to avoid any resonant scattering into the higher bands while still keeping within the limits of the approximations made. An example for such suitable parameter values would be given by working at $\hbar \omega / E_r \approx 4$ which allows modulation strengths $\kappa=V_{\omega}/{(\hbar \omega)}$ sufficiently high to explore both the maximum and the first zero of $J_{-1}(\kappa)$, thus, completely tuning the effective dispersion of the resulting bands.

\subsubsection{Extension to weakly-confined system}
We now discuss the inclusion of an additional free degree of motion. Such a model is relevant for experiments in which the confinement in the transverse direction is relatively weak. In this case particles may absorb energy during collisions from the driving fields and may scatter into states with fast motion in the transverse direction, which may either lead to heating or to loss from the experimentally relevant region. This additional direction of motion would correspond to the $z$-direction in the case of 2D optical lattices used to simulate magnetic fields (in the $xy$-plane).

We assume that the motion in the $z$-direction is free, so the previous discussion generalises straightforwardly by including the additional energy $E_z=2 E_r (k_z/k_r)^2 $ and integrating over the plane-wave states of the transverse direction. The scattering into the higher band still requires a minimal energy and the additional degree of freedom does not change the stability regions. However, as the energy in the transverse direction is unbounded, arbitrarily high energy may be absorbed from the driving fields, which corresponds to the presence of non-zero terms for all $m$ in FFGR higher than the minimal $m$ required to scatter into the higher band.

Additionally, inelastic scattering within the same band now becomes possible which was forbidden by the smallness of the bandwidth compared to the modulation energy before, as any amount of energy can be absorbed in the transverse direction irrespectively of how small or high the driving frequency is. For these processes any scattering with $m \ne 0$ corresponds to inelastic scattering following the definitions made at the end of section~\ref{subsec:FFGR}. Therefore, the system is always suscpetible to inelastic scattering if motion in the transverse direction is free. 

The inelastic scattering rate is now given by
\begin{equation}\label{eq:lattice_scatteringrate_2d}
\begin{aligned}
 \frac{dn_{a,\tilde{a}}}{dt} &= \frac{2 \pi} {\hbar} \sum_{\substack{\tau_1\tau_2 \\ m\ne0}}\left(\frac{L}{2\pi}\right)^2 \int dq_z \int dq \, g^2 \left \vert I^{m;a\rightarrow \tilde{a}}_{\tau_{\rm i},\tau_1,\tau_2;k,q} \right \vert^{2} \\
               &  \quad \times \delta(\epsilon^a_{\rm i}-\epsilon^{\tilde{a}}_{\rm f}(q,q_z)+m \hbar \omega)\\
               &= \frac{ g^2 } {2 \pi \hbar E_r^2} \frac{1}{L^2 d^2} \sum_{\substack{\tau_1\tau_2 \\ m\ne0}}L^4 \int d E_z \frac{d(q_zd)}{d(E_z/E_r)}\int d \epsilon_{q}   \\
               & \quad  \frac{d(qd)}{d(\epsilon_q/E_r)}   \left \vert I^{m;a\rightarrow \tilde{a}}_{\tau_{\rm i},\tau_1,\tau_2;k,q} \right \vert^{2} \delta(\epsilon^a_{\rm i}-\epsilon^{\tilde{a}}_q -E_z+m \hbar \omega)\\
               &= \frac{ g^2 } {2 \pi \hbar E_r} \frac{1}{L^2 d^2} \Gamma^{\rm 2D}_{a\rightarrow \tilde{a}} \,
\end{aligned}
\end{equation}
where we split the final state energy into the part due to the motion in the lattice and the free part via $\epsilon_q(q)=\epsilon_{\rm f}(q,k_z)-E_z(k_z)$ and defined the generalised matrix element  $I^{m;a\rightarrow \tilde{a}}_{\tau_{\rm i},\tau_1,\tau_2;k,q}$ for transitions with two particles initially in band $a$ to a final state with two particles in band $\tilde{a}$. 

\begin{figure}
    \subfloat%[First sub-figure\label{subfig-1:dummy}]
    {%
      \includegraphics[width=0.22\textwidth]{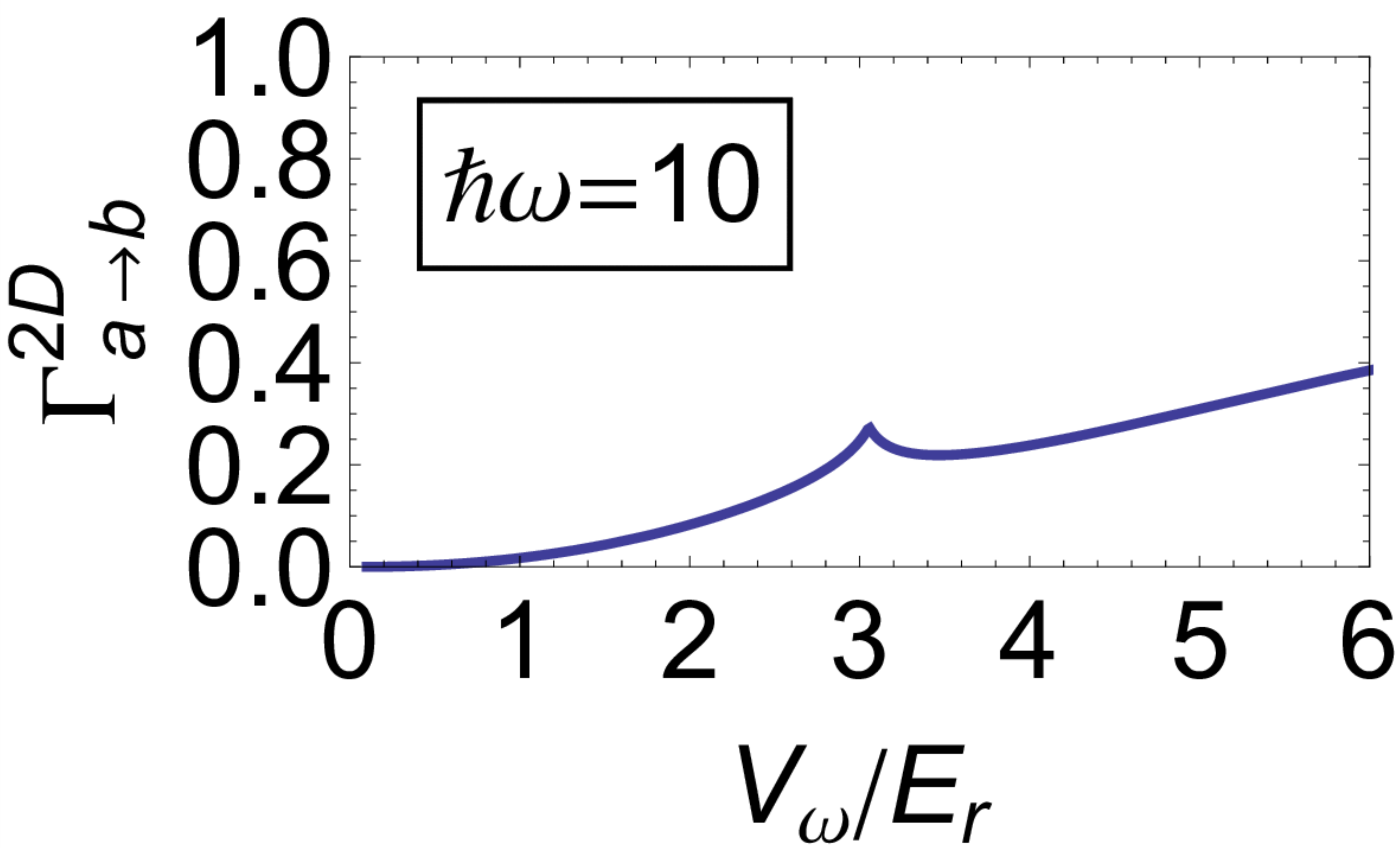}
    }
    \hfill
    \subfloat%[First sub-figure\label{subfig-2:dummy}]
    {%
      \includegraphics[width=0.22\textwidth]{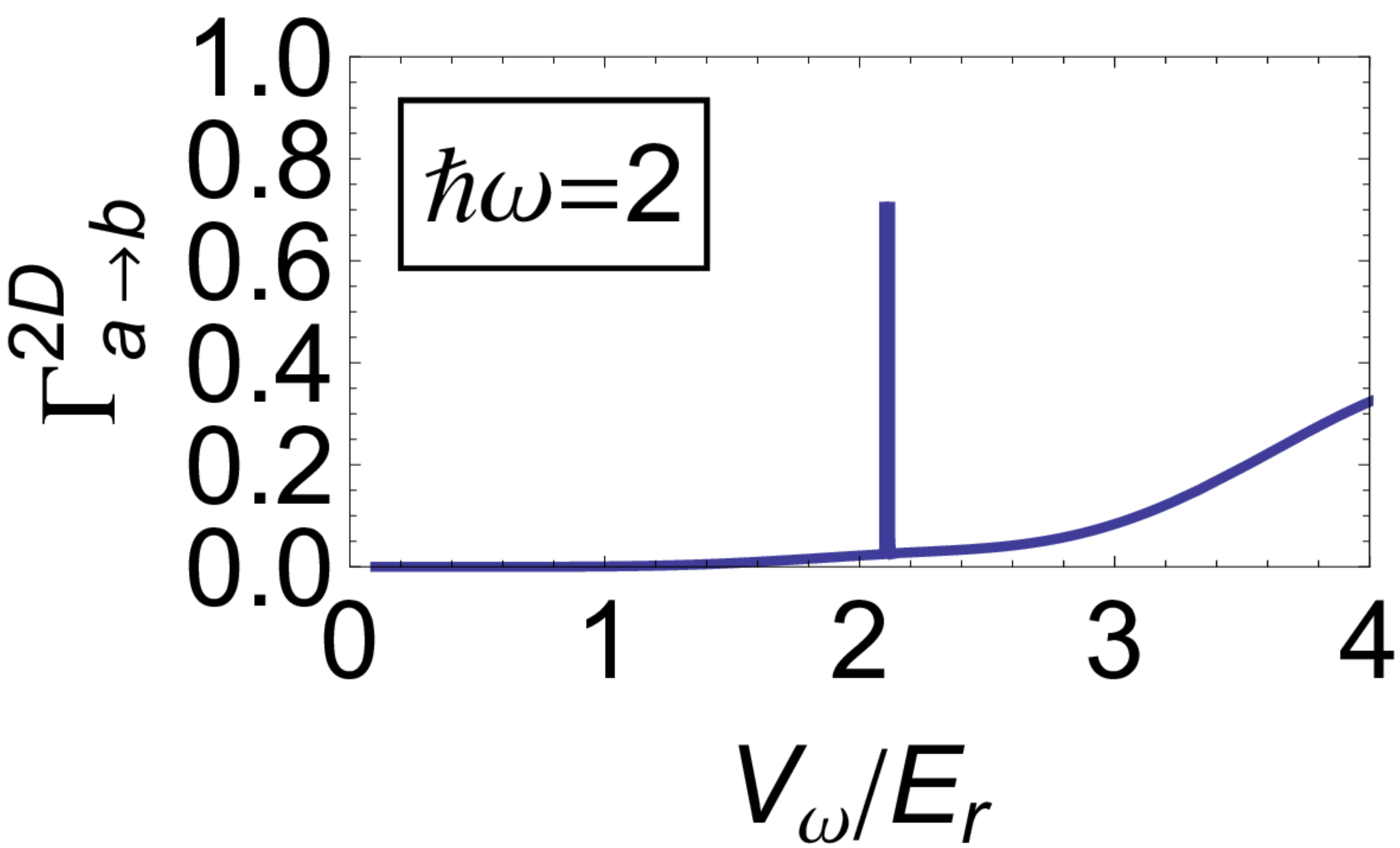}
    }
    \caption{(Color online) Dimensionless scattering rate $\Gamma^{\rm 2D}_{a\rightarrow b}$ for the extension to a weakly-confined system with a free transverse degree of motion, see Equation~(\ref{eq:lattice_scatteringrate_2d}) along the cuts $\hbar \omega /E_r = const.$ as indicated in Fig.~\ref{fig:Lattice_stability} for particles in the first band with quasi-momentum $k=0$ and relative momentum $k_y=0$ scattering into the second band, integrated over the final states with crystal momentum $q$ and relative momentum $q_y$. The left corresponds to $\hbar \omega /E_r = 10$ for which $m\ge1$ are the available inelastic scattering channel whereas the right corresponds to $\hbar \omega /E_r = 2$ for which $m\ge5$ transitions are allowed for $ 0.6\le V_{\omega}/E_r\le6.9$ and $m\ge4$ for $ 2.1\le V_{\omega}/E_r\le5.3$.}
    \label{fig:Lattice_scatteringrate_2d_1}
  \end{figure}

\begin{figure}
    \subfloat%[First sub-figure\label{subfig-1:dummy}]
    {%
      \includegraphics[width=0.22\textwidth]{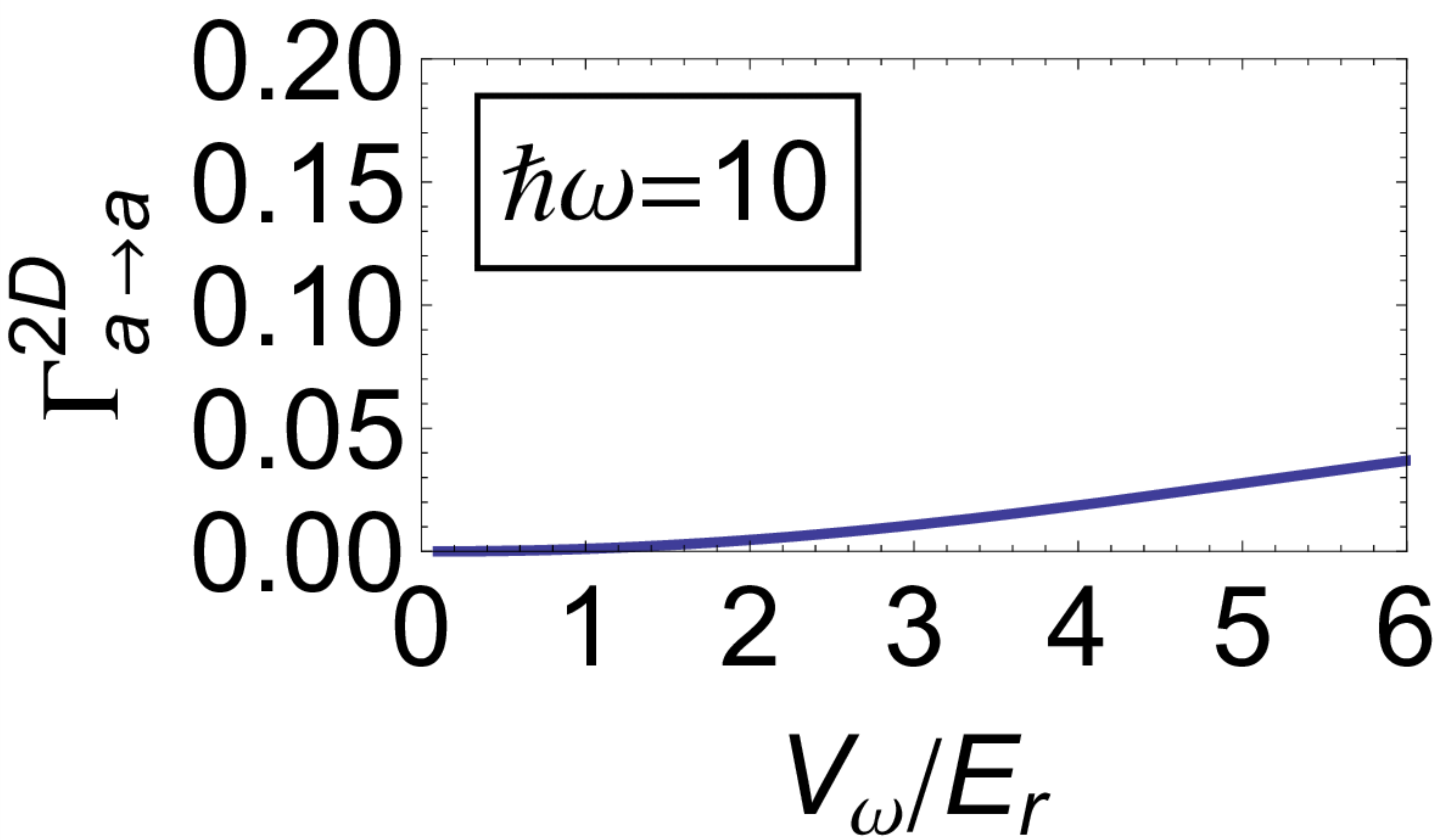}
    }
    \hfill
    \subfloat%[First sub-figure\label{subfig-2:dummy}]
    {%
      \includegraphics[width=0.22\textwidth]{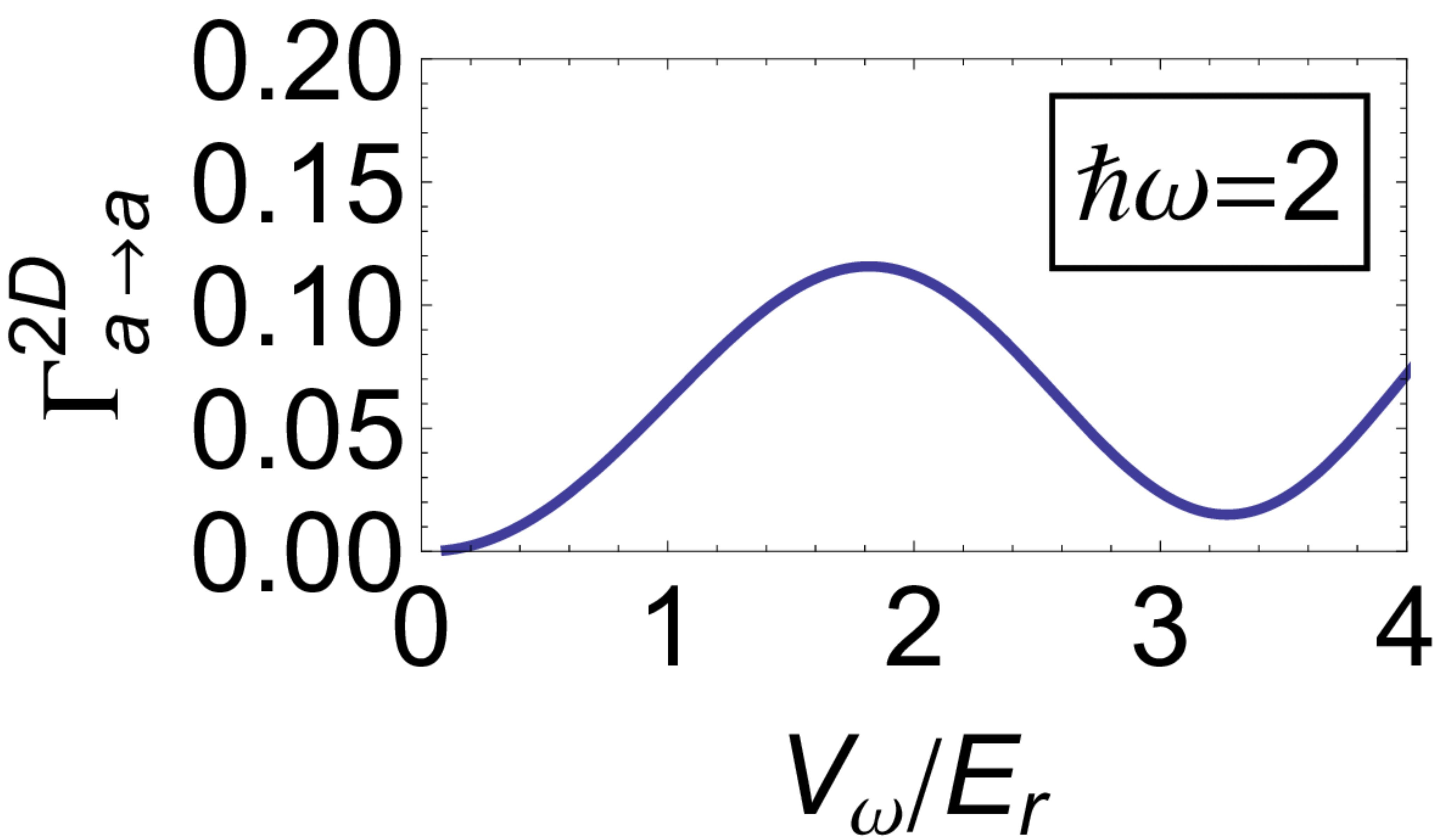}
    }
    \caption{(Color online) Dimensionless scattering rate $\Gamma^{\rm 2D}_{a\rightarrow a}$ for the extension to a weakly-confined system with a free transverse degree of motion, see Equation~(\ref{eq:lattice_scatteringrate_2d}) along the cuts $\hbar \omega /E_r = const.$ as indicated in Fig.~\ref{fig:Lattice_stability} for particles in the first band with quasi-momentum $k=0$ and relative momentum $k_y=0$ scattering inelastically and remaining in the first band, integrated over the final states with crystal momentum $q$ and relative momentum $q_y$. The left corresponds to $\hbar \omega /E_r = 10$ and the right corresponds to $\hbar \omega /E_r = 2$. In both cases all processes with $m\ge1$ are inelastic and allowed as the particles remain in the same band and the energy in $y$ direction is not gapped.}
    \label{fig:Lattice_scatteringrate_2d_2}
  \end{figure}

The dimensionless rate constants $\Gamma^{\rm 2D}_{a\rightarrow b}$ and $\Gamma^{\rm 2D}_{a\rightarrow a}$ are shown in Fig.~\ref{fig:Lattice_scatteringrate_2d_1} and Fig.~\ref{fig:Lattice_scatteringrate_2d_2} respectively. For the scattering into the higher band $\Gamma^{\rm 2D}_{a\rightarrow b}$ the rates are of the same order as in the 1D-case. The inelastic rate $\Gamma^{\rm 2D}_{a\rightarrow a}$ for particles remaining in the lower band depends strongly on $\hbar \omega$. For high oscillation frequencies most of the energy must be absorbed in the transverse direction for which the density of states decreases as $1/\sqrt{E_z}$ and consequently the total rate remains small.
 %%%%%%%%%%%%%%%%%%%%%%%%%%%%5 
%%%%%%%%%%%%%%%
%%%%%%%%%%%%%%%55
%%%%%%%%%%%%%%%%%%%%%%%%%%%%

To relate these considerations to recent experiments in Ref.~\cite{munichchern} we provide a rough estimate of the relevant inelastic scattering processes. The experiment simulates the Hofstadter model in a 2 dimensional driven optical lattice. The flux per cell is $\pi/2$, so the lowest Wannier band splits into four Hofstadter subbands. Firstly, based on the experimental parameters, we conclude that the restriction to the lowest Wannier band is justified as scattering into the higher Wannier bands should be forbidden by quasi-energy conservation or very highly suppressed. Thus, the dominant process should be the absorption of energy within the same Wannier band and into weakly confined transverse directions, i.e. inelastic scattering between the 4 Hofstadter subbands.

The results of Ref.~\onlinecite{munichchern} indeed show repopulation dynamics in which particles from the lowest Hofstadter subband are transferred to the higher subbands. The rate of transfer into the highest subband is observed to be approximately $\gamma_{\rm exp} \approx\SI{10}{Hz}$ per particle. In our model, the collision of two particles in the lowest subband  can lead to both being transferred to the highest subband, leading to a rate $\gamma_{\rm model} = 2 \frac{dn_{a\rightarrow a}}{dt} N$ per particle. To connect to the experimental 2D setup, we extend our 1D model to 2D by assuming that particles collide and remain in the lowest Wannier band of the optical lattice. Based on the optical lattice depth of $V_y=\SI{10}{E_r}$ this leads to $g^{2D}= 2 g^{3D}/d$. Further we assume confinement in the transverse direction, i.e. $L_z= \sqrt{\pi} a_{z}$ with $a_z$ the oscillator length in the transverse direction. The rate in our model then is $\gamma_{\rm model} =  (g^2_{2D}/(d L_z h E_r) \Gamma^{2D}_{a\rightarrow a} \rho$ where we introduced the two dimensional particle density $\rho =N/(L_x L_y)$. For the experimental parameters of $\rho d^2 \approx 20$ and with $\Gamma^{2D}_{a\rightarrow a}=0.25$ this yields $\gamma_{\rm model} \approx\SI{9}{Hz}$.

Inelastic scattering within the same subband can be reduced by either working at larger $\hbar \omega/E_r$ as this then requires a large amount of energy to be absorbed in the $z$-direction or by working in sufficiently deep lattices in which the inelastic processes of the type discussed become strongly suppressed. Such inelastic scattering processes can also be eliminated by adding an additional optical confining potential in the transverse direction which depending on the parameters of the experiment might be required to avoid losses and heating.

%==========================================================================
\section{\label{sec:summary}Summary}
%==========================================================================

We have studied the scattering processes of Floquet-Bloch waves in
periodically driven systems in the weakly interacting regime, as relevant to recent experiments
creating artificial gauge fields for gases of cold atoms. 
An extension of this work to the strongly interacting regime would be 
desirable to explore the stability and heating processes in the Mott 
and other strongly correlated phases, but is beyond the scope of the present work.
We have
described a formalism that allows the computation of elastic and
inelastic two-body scattering rates of particles in Floquet-Bloch
states, and have illustrated the consequences for model systems that
are representative of experimental situations: where energy can be
absorbed through transitions into other Floquet-Bloch bands or to motion in weakly-confined directions.  Notably we
have shown that, in general, the scattering cannot be understood in
terms of some effective time-independent Hamiltonian even for rapid
modulation.  Our results provide a framework by which the relative
sizes of elastic and inelastic two-body scattering processes can be
determined.  As experiments move towards the realization of
strongly correlated phases of matter in artificial gauge fields, it
will be crucial to determine the parameter regions in which the
elastic interactions which are responsible for the emergence of the
interesting physics remain dominant compared to the inelastic
processes which can limit the experimentally achievable temperatures
through particle loss or heating.

After completion of this work, we learned of a related study by Choudhury and Mueller \cite{Stability_Floquet_condensate_2} in which instabilites through excitation of transverse motion were also explored.

\begin{acknowledgments}
We thank two anonymous referees for their constructive and helpful comments. We are grateful to Monika Aidelsburger, Michael Lohse and Immanuel Bloch for discussions of experimental parameters. 
This work was supported by EPSRC Grant EP/K030094/1.
\end{acknowledgments}

\appendix
\section{Derivation of the Floquet-modes for the modulated lattice}\label{app:floquet_modes}
In this appendix we derive the single-particle Floquet-modes of the modulated lattice in a rotating-wave like approximation.

We start from the non-interacting Hamiltonian given in momentum space in Eq.~(\ref{eq:modulated_lattice_H0_k})
\begin{equation}
\begin{aligned}
 H_0(t) &= \sum_{k} \epsilon_k^{(1)} a^{\dagger}_k a_k +(\epsilon_k^{(2)}+\Delta_g) b^{\dagger}_k b_k \\
        & + \sum_k 1/2 [V_2 + V_{\omega} \cos(\omega t)](a^{\dagger}_{k+\pi} a_k + b^{\dagger}_{k+\pi} b_k).
\end{aligned}
\end{equation}

We abbreviate $V_c(t)=1/2 [V_2 + V_{\omega} \cos(\omega t)]$ and write the Hamiltonian in the basis of coupled momentum states at $k$ and $k+\pi$ as
\begin{equation}
\begin{aligned}
 H_0(t) &= \begin{pmatrix}
             a^{\dagger}_k & a^{\dagger}_{k+\pi} 
          \end{pmatrix}                 
          \begin{pmatrix}
           \epsilon^{a}_k  &V_c(t) \\
          V_c(t) & - \epsilon^{a}_k
          \end{pmatrix}
	  \begin{pmatrix}
             a_k \\
             a_{k+\pi} 
          \end{pmatrix}   \\
          & + \begin{pmatrix}
             b^{\dagger}_k & b^{\dagger}_{k+\pi} 
          \end{pmatrix}                 
          \begin{pmatrix}
           \epsilon^{b}_k +\Delta_g &V_c(t) \\
          V_c(t)& - \epsilon^{b}_k + \Delta_g
          \end{pmatrix}
	  \begin{pmatrix}
             b_k \\
             b_{k+\pi} 
          \end{pmatrix} ,
\end{aligned}
\end{equation}
where $k$ is now restricted to lie in the reduced BZ $- \pi/2< k \le\pi/2$ corresponding the supercell of two inequivalent sites in the real space lattice.

Without the modulation term $V_{\omega}$ this model can be easily solved exactly. The result will be a ground-state of bosons with two momentum components at $k=0$ and $k=\pi$ with a relative occupation that depends on the strength of $V_2$. In the limit of strong staggering $V_2 \gg \epsilon^{(n)}_k$, both components are equally occupied, corresponding in real space to occupation of the lower-energy sites, and the excited states correspond to occupation of the higher-energy sites gapped by an energy difference of $V_2$ and both subbands are flat as a function of $k$. The undriven bandstructure is given by
$E_{\pm}(k) = \pm \sqrt {\epsilon(k)^{2}+ (V_2/2)^{2}}$ with $\epsilon(k) $ corresponding to either the $a$ or $b$ band. Assuming a nearest neighbour tight-binding dispersion $\epsilon(k)=-{\rm t} \cos(k)$ the bandstructure displays a gap of $V_2$ for $k=\pm \pi/2$ between the $\pm$ subbands and the bandwidth of the subbands is suppressed by a factor of $t/V_2$ compared to the case of vanishing $V_2$.

As tunnelling should be strongly suppressed, i.e. $V_2 > \epsilon^{(n)}_k$, we first change basis to the eigenstates for $\epsilon^{(n)}_k = 0 $, $\alpha_k^{\pm}=1/\sqrt{2} (a_k \pm a_{k+\pi}) $, and treat the resulting off-diagonal terms as a small coupling. With this definition one obtains
\begin{equation}
\begin{aligned}
 H_0(t) &= \begin{pmatrix}
             \alpha^{\dagger}_{k+} & \alpha^{\dagger}_{k-} 
          \end{pmatrix}                 
          \begin{pmatrix}
          V_c(t) & \epsilon^{a}_k \\
            \epsilon^{a}_k& - V_c(t)
          \end{pmatrix}
	  \begin{pmatrix}
             \alpha_{k+} \\
             \alpha_{k-} 
          \end{pmatrix}   \\
          & + \begin{pmatrix}
             \beta^{\dagger}_{k+} & \beta^{\dagger}_{k-} 
          \end{pmatrix}                 
          \begin{pmatrix}
          \Delta_g+V_c(t)  & \epsilon^{b}_k \\
           \epsilon^{b}_k &  \Delta_g -V_c(t)
          \end{pmatrix}
	  \begin{pmatrix}
             \beta_{k+} \\
             \beta_{k-} 
          \end{pmatrix} .
\end{aligned}
\end{equation}
We now perform a unitary transformation to eliminate the diagonal terms via
\begin{equation}
U_c(t)= \begin{pmatrix}
          \exp[-i \kappa \sin(\omega t)/2 ]& 0 \\
           0    & \exp[i(\omega t + \kappa \sin(\omega t)/2)  ]
        \end{pmatrix}
\end{equation}
where we defined $\kappa=V_{\omega}/(\hbar \omega)$ and use the resonance condition $V_2= \hbar \omega$ to get 
\begin{equation}
\begin{aligned}
 H_0(t) &= \begin{pmatrix}
             \tilde{\alpha}^{\dagger}_{k+} & \tilde\alpha^{\dagger}_{k-} 
          \end{pmatrix}                 
          \begin{pmatrix}
           0 & \epsilon^{a}_k(t) \\
            \bar\epsilon^{a}_k(t)& 0 
          \end{pmatrix}
	  \begin{pmatrix}
             \tilde\alpha_{k+} \\
             \tilde\alpha_{k-} 
          \end{pmatrix}   \\
          & + \begin{pmatrix}
             \tilde\beta^{\dagger}_{k+} & \tilde\beta^{\dagger}_{k-} 
          \end{pmatrix}                 
          \begin{pmatrix}
           \Delta_g& \epsilon^{b}_k(t) \\
           \bar\epsilon^{b}_k(t) &\Delta_g
          \end{pmatrix}
	  \begin{pmatrix}
             \tilde\beta_{k+} \\
             \tilde\beta_{k-} 
          \end{pmatrix} ,
\end{aligned}
\end{equation}
where $\epsilon^{(n)}_k(t) = \epsilon^{(n)}_k \exp[i(\omega t+ \kappa \sin(\omega t))]$ and an overall constant energy shift of $\hbar \omega/2$ was dropped.
The exponential is expanded in terms of Bessel functions as
\begin{equation}
 \epsilon^{(n)}_k(t)=\epsilon^{(n)}_k e^{i \omega t} \sum_n J_n(\kappa) e^{i n \omega t} ,
\end{equation}
and this contains a term constant in time and oscillating terms. In particular, for a nearest neighbour tight-binding dispersion $\epsilon^{(n)}_k =-{\rm t}^{(n)} \cos(k)$ the time-dependent dispersion simplifies to $\epsilon^{(n)}_k(t)=-{\rm t}^{(n)} \cos(k) e^{i \omega t} \sum_n J_n(\kappa) e^{i n \omega t} = -{\rm t}^{(n)}(t) \cos(k)$ and the modulation is seen to lead to a time-dependent hopping strength  ${\rm t}^{(n)}(t)$ . Finally, we perform a rotating wave-approximation and only keep the constant term $ \epsilon^{(n)}_k(t) \approx \epsilon^{(n)}_k J_{-1}(\kappa)$. Thus, one obtains the eigenstates as
\begin{align}
 \Psi^{a}_{\tau,k} \vac&= 1/\sqrt{2} (\tilde\alpha^{\dagger}_{k+} + \tau \tilde\alpha^{\dagger}_{k-}) \vac,\\
 \Psi^{b}_{\tau,k}\vac &= 1/\sqrt{2} (\tilde\beta^{\dagger}_{k+} +\tau \tilde\beta^{\dagger}_{k-}) \vac,
\end{align}
with $\tau= \pm$. Abbreviating $f(t)=  \exp[i \kappa \sin(\omega t)]$ these states read in the original basis as
\begin{align}\label{eq:app_lattice_eigenstates}
 \Phi^{a,m}_{\tau,k}(t) &= 1/2 \left[ \left(f(t) + \tau e^{i \omega t}  \bar{f}(t)\right) a^{\dagger}_k \right.\\
                       & \left. \quad \quad + \left(f(t) -\tau e^{i \omega t}  \bar{f}(t)\right) a^{\dagger}_{k+\pi} \right] e^{i m \omega t}\vac , \\
 \Phi^{b,m}_{\tau,k}(t) &= 1/2 \left[ \left(f(t) +\tau e^{i \omega t}  \bar{f}(t)\right) b^{\dagger}_k \right.\\
                       & \left. \quad \quad + \left(f(t) -\tau e^{i \omega t}  \bar{f}(t)\right) b^{\dagger}_{k+\pi} \right] e^{i m \omega t} \vac,
\end{align}
and the corresponding quasi-energies are
\begin{align}
\epsilon^{a,0}_{\tau,k} &= \tau \epsilon^{a}_k J_{-1}(\kappa) ,\\
\epsilon^{b,0}_{\tau,k} &= \tau \epsilon^{b}_k J_{-1}(\kappa) + \Delta_g.
\end{align}
Again turning to the discussion of the nearest neighbour tight-binding dispersion  $\epsilon(k)=-{\rm t} \cos(k)$ with hopping strength ${\rm t}$ where the subbands were gapped by $V_2$ and the hopping was suppressed by a factor of ${\rm t}/V_2$ in the case of an undriven lattice, we note that the hopping is now modified by $J_{-1}(\kappa)$ instead and the gap between the subbands is closed at $k=\pm \pi/2$.
\section{Matrix elements for the modulated lattice}\label{app:matrixelements}
In this appendix we provide the explicit expressions for the matrix element
\begin{equation}
 I^m_{\tau_{\rm i},\tau_1,\tau_2;k,q}= \braket{\Phi^{0}_{\tau_1,q;\tau_2,-q}}{H_{\text{int}}/g}{ \Phi^{0}_{\rm i}}.
\end{equation}
appearing in the FFGR for the scattering in the modulated lattice, see equation~(\ref{eq:FFGRmatrixelement}) and the following discussion for details.

As a first step we compute
\begin{equation}
\begin{aligned}
  &I^{m}_{\tau_{\rm i},\tau_1,\tau_2;k,q} =\brakett{\Phi^{-m}_{\tau_1,q;\tau_2,-q}}{H_{\text{int}}/g}{ \Phi^{0}_{\tau_{\rm i},k;\tau_{\rm i},-k}}\\
                        &\quad=\frac{1}{T} \int_0^{T} dt \, \braket{\Phi^{-m}_{\tau_1,q;\tau_2,-q;}}{H_{\text{int}}/g}{ \Phi^{0}_{\tau_{\rm i},k;\tau_{\rm i},-k}}\\
		        &\quad=\frac{1}{T} \int_0^{T} dt \, e^{i m \omega t}\braket{\Phi^{0}_{\tau_1,q;\tau_2,-q}}{H_{\text{int}}/g}{ \Phi^{0}_{\tau_{\rm i},k;\tau_{\rm i},-k}}\\
		        &\quad= \frac{1}{T} \int_0^{T} dt \,e^{i m \omega t}  I^{0}_{\tau_{\rm i},\tau_1,\tau_2;k,q}  ,
  \end{aligned}
\end{equation}
where
\begin{equation} \label{eq:FFGRmatrixelement1}
 I_{\tau_{\rm i},\tau_1,\tau_2;k,q}= \braket{\Phi^{0}_{\tau_1,q;\tau_2,-q;0}}{H_{\text{int}}/g}{ \Phi^{0}_{\tau_{\rm i},k;\tau_{\rm i},-k}}
\end{equation}
was defined. This implies that $I^{m}_{\tau_{\rm i},\tau_1,\tau_2;k,q}$ is just the Fourier component of $I_{\tau_{\rm i},\tau_1,\tau_2;k,q}$ oscillating at $\exp[-i m \omega t]$ and $I_{\tau_{\rm i},\tau_1,\tau_2;k,q}$ contains all the relevant information.

We expand the single particle Floquet mode $\Phi^{n}_{i,k}$ in the basis of Bloch functions $\psi^k_{n}(x) $ of band $n$ as
\begin{equation}
 \Phi^{n}_{i,k}(x,t) = c^{n}_{i,k}(t) \psi^k_{n}(x).
\end{equation}
By equation~(\ref{eq:app_lattice_eigenstates}) $c^{n}_{i;k}$ do not depend either on the band ($a$ or $b$) or on the momentum $k$, but only on which subband ($i=\pm$) the particles are in. 
However, the interaction matrix elements $W_{bbaa}^{k_1;k_2;k_3;k_4}$ do depend on the momenta of the particles.

With the definition of the following abbreviations
\begin{align}
 M_1&=W_{bbaa}^{q,-q,k,-k} \\
 M_2 &=W_{bbaa}^{q,-q,k+\pi,-k+\pi} \\
 M_3 &=W_{bbaa}^{q+\pi,-q+\pi,k,-k} \\
 M_4 &=W_{bbaa}^{q+\pi,-q+\pi,k+\pi,-k+\pi} \\
 M_5 &= W_{bbaa}^{q,-q+\pi,k,-k+\pi}\\
 M_6 &=W_{bbaa}^{q,-q+\pi,k+\pi,-k} \\
 M_7 &= W_{bbaa}^{q+\pi,-q,k,-k+\pi} \\
 M_8 &=W_{bbaa}^{q+\pi,-q,k+\pi,-k}
\end{align}
the matrix-elements for transitions from an initial state with particles starting in the lower band $a$ in subband $\tau_{rm i}=+$ with momentum $k$ and $ -k$ into the state with two particles in the upper-band $b$ in subbands $\tau_1$ and $\tau_2$ with momenta $q$ and $-q$, abbreviated as $(a,k,+;a,-k,+)\rightarrow(b,q,\tau_1;b,-q,\tau_2)$, are given in table~\ref{tab:lattice_matrixelements}.
%%%%%%%%%%%%%%%%%%%%%%%%%%%%%%%%%%%%%%%
\begin{table*}
\begin{ruledtabular}
\caption{Matrix-elements $I_{\tau_{\rm i},\tau_1,\tau_2;k,q}$ as defined in equation~(\ref{eq:FFGRmatrixelement1}) for transitions from an initial state with particles starting in the lower band $a$ in subband $\tau_{rm i}=+$ with momentum $k$ and $ -k$ into the state with two particles in the upper-band $b$ in subbands $\tau_1$ and $\tau_2$ with momenta $q$ and $-q$, abbreviated as $(a,k,+;a,-k,+)\rightarrow(b,q,\tau_1;b,-q,\tau_2)$ \label{tab:lattice_matrixelements}}
\begin{tabular}{ll} 
$(a,k,+;a,-k,+)\rightarrow(b,q,+;b,-q,+)$  &$4 M_1 \cos\left[1/2 (\omega t - 2 V_{\omega}/({\hbar \omega}) \sin(\omega t))\right]^4 $\\
                                           &$+ 2 (-M_2 - M_3 + M_4 + M_5 + M_6 + M_7 + M_8)\sin\left[1/2 (\omega t - 2 V_{\omega}/({\hbar \omega}) \sin(\omega t))\right]^2  $ \\
                                           &$+ 2 (-M_2 - M_3 - M_4 + M_5 + M_6 + M_7 + M_8) \cos\left[\omega t - 2 V_{\omega}/({\hbar \omega}) \sin(\omega t)\right])$ \\
                                           & $\quad \times \sin\left[1/2 (t - 2 V_{\omega}/({\hbar \omega}) \sin(\omega t))\right]^2$\\
$(a,k,+;a,-k,+)\rightarrow(b,q,-;b,-q,-)$  & $1/2 (-M_1 + 3 M_3 - M_4 + M_5 + M_6 + M_7 + M_8) $\\
                                           &$ + 4 M_2 \cos\left[1/2 (\omega t - 2 V_{\omega}/({\hbar \omega}) \sin(\omega t))\right]^4-  8 M_3 \cos\left[\omega t - 2 V_{\omega}/({\hbar \omega}) \sin(\omega t)\right]$ \\
                                           & $ + 1/2(M_1 + M_3 + M_4 - M_5 - M_6 - M_7 - M_8) \cos\left[2 (\omega t - 2 V_{\omega}/({\hbar \omega}) \sin(\omega t))\right]$\\
$(a,k,+;a,-k,+)\rightarrow(b,q,+;b,-q,-)$  &$(M_1 + M_2 - M_3 - M_4 - M_5 + M_6 - M_7 +  M_8 )\sin\left[\omega t - 2 \kappa \sin(\omega t)\right]$\\
                                           &$+ 1/2 (M_1 + M_2 + M_3 + M_4 - M_5 - M_6 - M_7 - M_8) \sin\left[2(\omega t - 2 \kappa \sin(\omega t))\right]) $ \\
\end{tabular}
\end{ruledtabular}
\end{table*}
%%%%%%%%%%%%%%%%%%%%%%%%
To better understand the general behaviour of these matrix elements with regard to their Fourier-structure and justify the statements made in their discussion, we will consider more closely the $(a,k,+;a,-k,+)\rightarrow(b,q,+;b,-q,-)$ element given by
\begin{equation}\label{eq:I_pm}
\begin{aligned}
 I_{+,-;k,q} &= f_1(\{M_i\})\sin\left[\omega t - 2 \kappa \sin(\omega t)\right]\\
             &\quad + f_2(\{M_i\}) \sin\left[2(\omega t - 2 \kappa \sin(\omega t))\right])/2 .
\end{aligned}
\end{equation}
with
\begin{equation}
\begin{aligned}
 f_1(\{M_i\})&=M_1+M_2-M_3 - M_4 - M_5 + M_6 - M_7 +  M_8  \, ,\\
 f_2(\{M_i\})&=M_1 + M_2 + M_3 + M_4 - M_5 - M_6 - M_7 - M_8\, .
\end{aligned}
\end{equation}
If the interaction matrix elements are momentum-independent, $M_i=M $, we have that $f_1(\{M_i\})=f_2(\{M_i\})=0$ and the time-dependent terms vanish, and therefore no inelastic scattering occurs.

Using the usual expansion in terms of Besselfunctions, $\exp[i z \sin(\omega t))]=\sum_n J_n(z) e^{i n \omega t} $, the term given in Eq.~(\ref{eq:I_pm}) is seen to in fact contain all frequency components allowing the absorption of an arbitrary integer number of energy quanta $\hbar \omega$. Moreover, for small $\kappa=V_{\omega}/(\hbar \omega)$ higher order processes are suppressed by powers of $\kappa$. Specifically, for this matrix element, in a given order $(\kappa)^n$, frequency components $\exp[i m \omega t]$ from $m=-n-2$ up to $m=n+2$ are present. Or put differently, an $m$-photon transition is at least suppressed by a power of $(\kappa)^{n_0}$ with $n_0=\max\left[|m|-2,0\right]$.

\bibliography{mybib-final}{}

\end{document}